\documentclass[12pt]{article}
\usepackage{graphics,cite,amssymb,epsfig,float}
\usepackage[usenames,dvips]{color}
\usepackage{rotating}

\begin{document} 

\newcommand{\lsim}{\raisebox{-0.13cm}{~\shortstack{$<$ \\[-0.07cm] $\sim$}}~} 
\newcommand{\gsim}{\raisebox{-0.13cm}{~\shortstack{$>$ \\[-0.07cm] $\sim$}}~} 
\newcommand{\ra}{\rightarrow} 
\newcommand{\lra}{\longrightarrow} 
\newcommand{\ee}{e^+e^-} 
\newcommand{\gam}{\gamma \gamma} 
\newcommand{\nn}{\noindent} 
\newcommand{\non}{\nonumber} 
\newcommand{\beq}{\begin{eqnarray}} 
\newcommand{\eeq}{\end{eqnarray}} 
\newcommand{\s}{\smallskip} 
\newcommand{\tb}{\tan\beta}

\def\thefootnote{\fnsymbol{footnote}}
\setcounter{footnote}{1}

\begin{center} 

~~~ 

\vspace{1.2cm} 

{\Large\bf The Higgs sector of supersymmetric theories}

\vspace{0.3cm} 

{\Large\bf and the implications for high--energy colliders\footnote{ 
{\large  In memoriam of Julius Wess, 1934--2007.} \\ \vspace*{-3mm}   

\noindent To be published in "Supersymmetry on the Eve of the LHC" a special
volume of European Physical Journal C, Particles and Fields (EPJC) in memory of
Julius Wess.} }

\vspace{1.cm} 

{\sc\large Abdelhak Djouadi}

\vspace{0.7cm} 
{Laboratoire de Physique Th\'eorique, Universit\'e Paris--Sud and CNRS}


Batiment 210, F--91405 Orsay Cedex, France 

\vspace{0.3cm}

Physikalisches Institut, University of Bonn, Nussallee 12, 53115 Bonn,
Germany.

\vspace{1.5cm} 

{\bf Abstract}

\end{center}

\vspace{0.7cm}

\begin{tabular}{lll}
\begin{minipage}{2cm}
\end{minipage}
& 
\begin{minipage}{14.5cm}

\noindent  One of the main motivations for low energy supersymmetric theories is
their  ability to address the hierarchy and naturalness problems in the Higgs
sector  of the Standard Model. In these theories, at least two doublets of
scalar fields are required to break the electroweak symmetry and to generate 
the masses of the elementary particles, resulting in a rather rich Higgs
spectrum. The  search for the Higgs bosons of Supersymmetry and the
determination of their basic properties is one of the  major goals of
high--energy colliders and, in particular, the LHC which will soon start
operation. We review the salient features of the Higgs sector of the Minimal
Supersymmetric  Standard Model and of some of its extensions and summarize the
prospects for probing them at the LHC and at the future ILC. 

\end{minipage}
&
\begin{minipage}{1cm}
\end{minipage}
\end{tabular}

\thispagestyle{empty} 
\def\thefootnote{\arabic{footnote}}
\setcounter{footnote}{0}
\newpage

\setcounter{page}{1}
\section{Introduction} 

It was known relatively soon after the introduction of the Standard Model (SM)
of the  electroweak interactions \cite{GSW}, which makes use of one Higgs
doublet of complex scalar fields to spontaneously break the ${\rm SU(2)_L
\times  U(1)_Y}$ symmetry to generate in a gauge invariant way the masses of
the  $W^\pm, Z$ gauge bosons and the fermions \cite{Higgs}, that the model
suffers from  a severe flaw: the so--called naturalness or fine--tuning problem
\cite{Hierarchy-SM}. Indeed, when attempting to calculate the quantum
corrections to the squared mass of the single Higgs boson of the theory, one
encounters divergences that are quadratic in the cut--off scale $\Lambda$ beyond
which the theory ceases to be valid and new physics should appear.  If one
chooses the cut--off $\Lambda$ to be the Grand Unification (GUT) scale $M_{\rm
GUT} \simeq 2 \cdot 10^{16}$ GeV or the Planck scale $M_{\rm Pl} \sim  10^{18}$
GeV, the mass of the Higgs particle, which is expected for consistency reasons
to lie in the range of the electroweak symmetry breaking scale $v \sim 250$ GeV,
will prefer to be close to the very high scale unless an unnatural fine
adjustment of parameters is performed. A related issue, called the hierarchy
problem, is why these two scales are so widely different, $\Lambda \gg v$, a
question that has no satisfactory answer in the SM.\s

Supersymmetry (SUSY), introduced in the early seventies by Julius Wess and Bruno
Zumino \cite{Wess-Zumino,WessBagger} among others \cite{Early-SUSY} mainly for
aesthetical reasons, is  presently widely considered as the most attractive
extension of the SM. The  main reason is that  it solves, at least technically,
the hierarchy and naturalness problems \cite{Hierarchy-SUSY}. Indeed, this new
symmetry prevents the Higgs boson mass from acquiring large radiative
corrections: the quadratic divergent loop contributions of the SM particles  are
exactly canceled by the corresponding loop contributions of  their
supersymmetric partners which differ in spin by $\frac12$.  This cancellation
thus stabilizes the huge hierarchy between the GUT and the electroweak scales
and no extreme fine-tuning is required. Later on, two other main motivations for
introducing low energy supersymmetry in particle physics were  recognized: the
satisfactory unification of the  gauge couplings of the electromagnetic,
weak and strong interactions  at the GUT scale \cite{Gauge-Unif} and the
presence of a particle  that is massive, electrically neutral, weakly
interacting, absolutely stable, which is the  ideal candidate for the dark
matter in the universe \cite{DM-first}.\s

The most intensively  studied low energy supersymmetric extension of the SM is
the  most economical one, the so--called MSSM \cite{MSSM,MSSM-rev,Martin}. In
this minimal model,  one assumes the SM  gauge group (and associates a 
spin--$\frac12$ gaugino to each  gauge boson of the model), the minimal particle
content (in particular, three generations of fermions without right--handed
neutrinos and their spin--zero partners, the sfermions) and the conservation of
a discrete symmetry called R--parity  which makes the lightest SUSY particle
absolutely stable. In order to explicitly break SUSY, a collection of soft terms
(i.e. which do not reintroduce quadratic divergences) is added to the Lagrangian
\cite{Soft,mSUGRA}: mass terms for  the spin $\frac12$ gauginos and the spin--0 
sfermions, mass and bilinear terms for the Higgs bosons and trilinear couplings
between sfermions and Higgs bosons. Although incomplete (e.g. it does not have
right--handed (s)neutrinos and has a problem with the $\mu$ parameter), it
serves as a benchmark scenario for the possible phenomenology of SUSY theories.\s

The MSSM requires the existence of two isodoublets of complex scalar fields of
opposite hypercharge to cancel chiral anomalies and to give masses separately to
isospin up--type and down--type fermions \cite{Hierarchy-SUSY}. Three of the
original eight degrees of freedom of the scalar fields are absorbed by the
$W^\pm$ and $Z$ bosons to build their longitudinal polarizations and to acquire
masses. The remaining degrees of freedom will correspond to five scalar Higgs
bosons. In the absence of CP--violation, two CP--even neutral Higgs bosons $h$
and $H$, a pseudoscalar $A$ boson and a pair of charged scalar particles $H^\pm$
are thus introduced by this extension of the Higgs sector 
\cite{HHG,HaberGunion,HaberGunion2,Anatomy1,Anatomy2}. Besides the four masses,
two additional parameters define the properties of these particles at
tree--level: a mixing angle $\alpha$ in the neutral CP--even sector and the
ratio of the two vacuum expectation values $\tb$, which, from GUT restrictions,
is assumed in the range $1 \lsim \tb \lsim m_t/m_b$ with the lower and upper
ranges being favored if the Yukawa couplings are to be unified at the GUT scale
\cite{YukawaUnif}.
Supersymmetry leads to several relations among these parameters and only two of
them, taken in general to be the pseudoscalar Higgs mass $M_A$ and $\tb$, are in
fact independent. These relations impose a strong hierarchical structure on the
mass spectrum, $M_h<M_Z, M_h <M_A<M_H$ and $M_W<M_{H^\pm}$, which is, however,
broken by radiative corrections
\cite{RC-leading,RC-oneloop,RC-all,RC-review,RC-higher}.  
These radiative corrections  turn out to be very large and, for instance, they
shift the upper bound on the mass of the lighter $h$ boson from the tree--level
value $M_Z$ up to $M_h \sim 140$ GeV \cite{RC-all,RC-review}. Thus, in the MSSM,
one Higgs particle is expected to be relatively light, while the masses of the
heavier neutral and charged Higgs particles are expected to be in the range of
the electroweak  scale. \s

The Higgs sector in SUSY models may be more complicated if some basic
assumptions of the CP--conserving MSSM, such as the absence of new sources of CP
violation, the presence of only two Higgs doublets, or R--parity conservation,
are relaxed. For instance, if CP--violation is present in the SUSY sector (which
is required if baryogenesis is to be explained at the weak scale), the new
phases will enter the MSSM Higgs sector through  the large radiative corrections
and alter the Higgs masses and couplings; in particular, the three neutral Higgs
states will not have definite CP quantum numbers and will mix with each other to
produce the physical states \cite{SUSY-CPV,CPHmasses}.  Another interesting
extension is the next--to--minimal supersymmetric SM, the NMSSM, which consists
of simply introducing a complex iso-scalar field which naturally generates a
weak scale value for the supersymmetric Higgs--higgsino parameter $\mu$ (thus
solving the so--called $\mu$ problem) \cite{NMSSMt,NMSSMp}. The model  includes
an additional CP--even and CP--odd Higgs particles compared to the MSSM
\cite{NMSSMp,NMSSMb}.\s 

A large variety of theories,  string theories, Grand Unified theories, 
left--right symmetric models, etc., suggest an additional gauge symmetry which
may be broken only at the TeV scale, leading to an extended particle spectrum
and, in particular, to additional Higgs fields beyond the minimal set of the
MSSM \cite{H-GUTs,H++th,H:higheR}. These extensions also predict extra matter
fields and would lead to a very interesting phenomenology and new collider
signatures in the Higgs sector. In a general SUSY model with an arbitrary number
of singlet and doublet scalar fields (as well as a matter content which  allows
for the unification of the gauge couplings), a linear combination of Higgs
fields has to generate  the $W^\pm/Z$ masses and, from the requirement that all
couplings stay perturbative up to $M_{\rm GUT}$, a Higgs particle should have
significant  couplings to gauge bosons and a mass below 200 GeV
\cite{HSUSY-thbound}. This sets an upper bound on the lighter Higgs particle
mass in SUSY theories.\s

The phenomenology of the SUSY Higgs sector  is thus much richer than the one of
the SM with its unique Higgs boson. The study of the properties of the Higgs
bosons and of those of the supersymmetric particles is one of the most active
fields of elementary particle physics. The search for these new particles and,
if discovered, the determination of their fundamental properties, is one of the
major goals of  high--energy colliders. In this context, the probing of the
Higgs sector has a double importance since, at the same time, it provides the
clue of the electroweak symmetry breaking mechanism and it sheds light on the
SUSY--breaking mechanism. Moreover, while SUSY particles are allowed to be
relatively heavy unless one invokes fine--tuning arguments, the existence of a
light Higgs boson is a generic prediction of low energy SUSY. This particle
should therefore manifest itself at the next round of high--energy experiments,
in particular at the LHC
\cite{CMSTDR,ATLASTDR,LHC,Houches,Houches-last,Weiglein:2004hn}, which will
start operation rather soon, and at the future ILC 
\cite{Weiglein:2004hn,TESLA,H-Desch,DCR}.   We are thus in a situation where
either SUSY with its extended Higgs sector is discovered soon or, in the absence
of a light Higgs boson, the whole SUSY edifice, at least in the way it is
presently viewed, collapses. \s

This review summarizes the salient features of the Higgs sector of SUSY
theories. In the two next sections, we present the Higgs spectrum of the MSSM
and some of its extensions,  and summarize the decays of and into the Higgs 
bosons.  In sections 4 and 5, we discuss the production, the detection and the
study of the properties of  the Higgs particles at the LHC and at the future
ILC. A very brief conclusion is given in Section 6. A short Appendix collects
some basic formulae. 

\section{The Higgs spectrum in SUSY models}

\subsection{The Higgs potential of the MSSM}

In the MSSM, two doublets of complex scalar fields of opposite
hypercharge are required
\beq
\label{Hcomponents}
H_1= \left( \begin{array}{cc} H^0_1 \\ H^-_1 \end{array} \right) \ {\rm 
with} \ Y_{H_1} = -1 \ \ , \ \ 
H_2= \left( \begin{array}{cc}  H^+_2 \\ H^0_2 \end{array} \right) \ {\rm
with} \ Y_{H_2}=+1 \ , 
\eeq
to break spontaneously the electroweak symmetry. There are several reasons for
this  requirement.\s

The first reason is that in the SM,  one generates the masses of the fermions of
a given isospin by using the same scalar field $\Phi$ that also generates the
$W$ and $Z$ boson masses, the isodoublet  $\tilde{\Phi}= i\tau_2 \Phi^*$ with
opposite hypercharge generating the masses of the opposite isospin--type
fermions.  However, in a SUSY theory, the Superpotential should involve only the
superfields and not their conjugate fields. Therefore, we must introduce a
second doublet with the same hypercharge as the conjugate $\tilde{\Phi}$ field
to generate the masses of both isospin--type fermions 
\cite{Hierarchy-SUSY,MSSM}.\s

A second reason is that in the SM,  chiral anomalies which spoil the
renormalizability of  the theory, disappear because the sum of the hypercharges
or charges of all the 15 chiral fermions  of one generation is zero, ${\rm
Tr}(Y_f) = {\rm Tr}(Q_f)=0$. In the SUSY case, if we use only one doublet of
Higgs fields as in  the SM, we will have one additional charged spin $\frac12$
particle, the higgsino corresponding to the SUSY partner of the charged
component of the scalar field, which will spoil this cancellation. With two
doublets of Higgs fields with opposite hypercharge, the cancellation of chiral
anomalies still takes place \cite{Anomaly-SUSY} and the renormalizability  of
the theory is preserved.\s

Finally, a higher number of Higgs doublets would  spoil  the unification of the
electromagnetic, weak and strong coupling constants at the GUT energy scale if
no additional matter particles are added to the spectrum; see for instance Ref.~
\cite{HSUSY-thbound}.\s

In the MSSM, the terms contributing to the scalar Higgs potential $V_H$ come 
from various  sources; see the Appendix. The potential can be written as
\cite{HHG,HaberGunion,Martin}: 
\beq
V_H&=& ( |\mu |^2 +m_{H_1}^2)|H_1|^2 +(|\mu|^2+m_{H_2}^2)|H_2|^2
-\mu B \epsilon_{ij} (H_1^i H_2^j+{\rm h.c.}) \non \\
&& +{g_2^2+g_1^{2}\over 8} (|H_1|^2 - |H_2|^2)^2 +{1\over 2} 
g_2^2 |H_1^\dagger H_2|^2 
\eeq
where $m_{H_1}, m_{H_2}$ are the soft--SUSY breaking terms for the Higgs boson
masses and $B\mu$ is the one of the bilinear term  $\mu H_1 H_2$ of the SUSY
Lagrangian; $g_2$ and $g_1$ are the ${\rm SU(2)_L}$ and ${\rm U(1)_Y}$ 
couplings and  $ \epsilon_{12}=1 = - \epsilon_{21}$. Defining the mass squared 
terms  
\beq
\overline{m}_1^2=|\mu |^2 +m_{H_1}^2 \, , \, \ \overline{m}_2^2=|\mu |^2 
+m_{H_2}^2\, , \, \  \overline{m}_3^2=B\mu
\eeq
one obtains, using the decomposition of the $H_{1,2}$ fields into neutral and 
charged components eq.~(\ref{Hcomponents})
\beq
V_H &=& \overline{m}_1^2 (|H_1^0|^2 +|H^-_1|^2) +\overline{m}_2^2(|H_2^0|^2
+|H_2^+|^2) -\overline{m}_3^2(H_1^-H_2^+ -H_1^0 H_2^0+{\rm h.c.}) 
\hspace*{-1cm} \non \\
&& +{g_2^2+g_1^{2}\over 8} (|H_1^0|^2 + |H_1^-|^2 -|H_2^0|^2 -|H_2^+|^2)^2 
+{g_2^2\over 2} |H_1^{-*} H_1^0 + H_2^{0*}H_2^+|^2 
\eeq
One can then require that the minimum of the potential $V_H$  breaks the  ${\rm
SU(2)_L \times U(1)_Y}$ group while preserving the electromagnetic symmetry 
U(1)$_{\rm Q}$. At the minimum of the potential,  one can  always choose the
vacuum expectation value of the field $H_1^-$ to be zero,  $\langle
H^-_1\rangle$=0, because of SU(2) symmetry.  At  $\partial V/\partial H^-_1$=0,
one obtains then automatically $\langle H^+_2 \rangle$=0. There is therefore no
breaking in the charged directions and the QED symmetry is preserved. Some
interesting and important remarks on the potential $V_H$ can be made  
\cite{HHG,HaberGunion,Martin}: \s

$\bullet$ The quartic Higgs couplings are fixed in terms of the  ${\rm
SU(2)_L\times U(1)_Y}$ gauge couplings. Contrary to a general  two--Higgs
doublet model where the scalar potential   has 6 free parameters and a phase,
in  the MSSM  we have only three free parameters: $\overline{m}^2_{1}, 
\overline{m}^2_{2}$ and $\overline{m}^2_{3}$.\s

$\bullet$ The two combinations  $m_{H_1,H_2}^2+|\mu|^2$ are real and, thus,
only $B\mu$ can be complex. However, any phase in $B\mu$ can be absorbed into 
the phases of the fields $H_1$ and $H_2$. Thus, the scalar potential of the MSSM
is  CP conserving at the tree--level.\s

$\bullet$ To have electroweak symmetry breaking, one needs a combination of the 
$H_1^0$ and $H_2^0$ fields to have a negative squared mass term. This occurs if
$\overline{m}_3^2 > \overline{m}_2^2 \overline{m}_2^2$.   If not,  $\langle
H_1^0 \rangle = \langle H_2^0 \rangle$ will be a stable minimum of the potential
and there is no electroweak symmetry breaking (EWSB).\s  

$\bullet$ In the direction $|H_1^0|$=$|H_2^0|$, there is no quartic
term. $V_H$  is  bounded from below for large values of the field $H_i$ only 
if the condition $\overline{m}_1^2+\overline{m}_2^2 >2|\overline{m}_3^2|$ 
is satisfied. \s

$\bullet$ To have explicit electroweak symmetry breaking and, thus, a  negative
squared term in the Lagrangian, the potential at the minimum  should have a
saddle point which implies $\overline{m}_1^2  \overline{m}_2^2 <
\overline{m}_3^4$.\s

$\bullet$ The two above conditions on the masses $\overline{m}_i$ are 
not satisfied if $\overline{m}_1^2= \overline{m}_2^2$ and, thus, we must have
non--vanishing soft SUSY--breaking scalar masses: $\overline{m}_1^2 \neq 
\overline{m}_2^2$ meaning $m_{H_1}^2 \neq m_{H_2}^2$. \s

Therefore, to break the electroweak symmetry, we need also to break SUSY. This
provides a close connection between gauge symmetry breaking and SUSY--breaking.
In constrained models such as the minimal supergravity model \cite{mSUGRA}, the
soft SUSY--breaking scalar Higgs masses are equal at high--energy, $m_{H_1} =
m_{H_2}$ [and their squares positive], but the running to lower  energies via
the contributions of top/bottom quarks and their SUSY partners in the 
renormalization group evolution (RGE) makes that this degeneracy is lifted at
the weak scale, thus satisfying the relation $m_{H_1}^2 \neq m_{H_2}^2$ above.
In the running one obtains $m_{H_2}^2<0$ or $m_{H_2}^2 \ll m_{H_1}^2$ which thus
triggers EWSB: this is the radiative breaking of the symmetry \cite{REWSB}.
Thus, EWSB is more natural and elegant in the MSSM than in the SM since, in the
latter case, one needs to make the ad hoc choice of a negative mass squared term
for the scalar field  in the Higgs potential while, in the MSSM, this comes
simply from radiative corrections.

\subsection{The masses of the MSSM Higgs bosons}

Let us now determine the Higgs spectrum in the CP--conserving MSSM, following 
Refs.~\cite{HHG,HaberGunion,Martin}. The neutral components
of the two Higgs fields develop vacuum expectations values
\beq
\langle H_1^0\rangle =  v_1/\sqrt 2 \ \ , \ \  
\langle H_2^0 \rangle = v_2/\sqrt 2 
\eeq
Minimizing the scalar potential at the electroweak minimum, $\partial V_H/
\partial H_1^0=\partial V_H/\partial H_2^0=0$, using 
\beq
(v_1^2+v_2)^2=v^2= 4M_Z^2/(g_2^2+g_1^2)=(246~{\rm GeV})^2
\eeq
with $v$ the SM vacuum expectation value, and defining the important parameter
\beq
\tb= v_2/v_1 = (v \sin \beta)/(v \cos \beta)
\eeq 
one obtains two minimization conditions that can be written in the 
following way:
\beq
2B\mu &=& (m_{H_1}^2 -m_{H_2}^2) \tan 2\beta+M_Z^2 \sin2\beta \nonumber \\ 
\mu^2 \cos\beta &=& (m_{H_2}^2\sin^2\beta - m_{H_1}^2 \cos^2 \beta)
- M_Z^2\cos2\beta/2 
\label{min-conditions}
\eeq
These relations show explicitly what we have already mentioned: if $m_{H_1}$ 
and $m_{H_2}$ are known (e.g. from RGEs once fixed at the scale $M_{\rm GUT}$)
and $\tb$ is fixed at the weak scale,  $B$ and $\mu^2$ are fixed while  the sign
of $\mu$ stays undetermined. These relations are very important as the
requirement of radiative EWSB  leads to additional constraints  and lowers the
number of free parameters.  \s

To obtain the Higgs physical fields and their masses, one has to develop the 
two doublet complex scalar fields $H_1$ and $H_2$ around the  vacuum, into  
real and imaginary parts 
\beq
H_1=(H_1^0,H_1^-)=  \frac{1}{\sqrt{2}} \left( v_1+ H_1^0+ i P_1^0
, H_1^- \right) \ ,\
H_2=(H_2^+,H_2^0)= \frac{1}{\sqrt{2}} \left( H_2^+ , v_2+ H_2^0+ i P_2^0 
\right)  
\eeq
where the real parts correspond to the CP--even Higgses and the imaginary 
parts  to the CP--odd Higgs and  Goldstone bosons, and then 
diagonalize the mass matrices evaluated at the vacuum  
\beq
{\cal M}_{ij}^2=\frac{1}{2} \left. \frac{\partial^2 V_H}{\partial H_i 
\partial H_j} \right|_{\langle H_1^0\rangle= v_1/\sqrt{2}, \langle H_2^0 
\rangle=v_2/\sqrt{2},\langle H^\pm_{1,2} \rangle=0} 
\eeq
In the case of the CP--even Higgs bosons, one obtains the following mass matrix
\begin{eqnarray}
{\cal M}_R^2 = \left[ \begin{array}{cc} -\bar{m}_3^2 \tb + M_Z^2 \cos^2 \beta  
&  \bar{m}_3^2 -M_Z^2 \sin \beta \cos \beta \\
\bar{m}_3^2 - M_Z^2 \sin \beta \cos \beta & -\bar{m}_3^2 {\rm cot} \beta +
M_Z^2 \sin^2 \beta \end{array} \right] 
\label{CPeven:matrix-tree}
\end{eqnarray}
while for the neutral Goldstone and CP--odd Higgs bosons, one has the mass 
matrix 
\begin{eqnarray}
{\cal M}_I^2 = \left[ \begin{array}{cc} -\bar{m}_3^2 \tb  &  \bar{m}_3^2 \\
\bar{m}_3^2 & -\bar{m}_3^2 {\rm cot} \beta  \end{array} \right]
\end{eqnarray}  
In the latter case, since Det$({\cal M}_I^2)=0$, one eigenvalue is zero and 
corresponds to the Goldstone boson mass,  while the other corresponds to 
the pseudoscalar Higgs mass and is given by 
\beq
M_A^2= - \bar{m}_3^2 (\tb + {\rm cot} \beta) = - 2 \bar{m}_3^2/
\sin 2\beta
\label{Amass:tree}
\eeq
The mixing angle $\theta$ which gives the physical fields is
in fact simply the angle $\beta$ 
\beq
\left( \begin{array}{c}   G^0 \\ A \end{array} \right) 
&=& {\cal R}_\beta \left( \begin{array}{c}   P_1^0 \\ P_2^0 \end{array} \right)
=\left( \begin{array}{cc} \cos \beta & \sin \beta \\
- \sin \beta & \cos \beta \end{array} \right) \ 
\left( \begin{array}{c}   P_1^0 \\ P_2^0 \end{array} \right)
\eeq
In the charged Higgs case, one can make the same
exercise  and obtain the charged fields,
$(^{G^\pm}_{ H^\pm}) =  {\cal R}_\beta (^{H_1^\pm}_ {H_2^\pm})$,   with a
massless charged Goldstone and a charged Higgs boson with a mass  
\beq
M_{H^\pm}^2= M_A^2 + M_W^2 
\label{H+mass:tree}
\eeq
Coming back  to the CP--even Higgs case, one obtains then for the Higgs boson 
masses 
\beq
M_{h,H}^2= \frac{1}{2} \left[ M_A^2+M_Z^2 \mp \sqrt{ (M_A^2+M_Z^2)^2 -4M_A^2
M_Z^2 \cos^2 2\beta } \right] 
\label{Hmasses:tree}
\eeq
The physical Higgs bosons are obtained from the rotation
of angle $\alpha$, $(^H_h)=  {\cal R}_\alpha (^{H_1^0}_{H_2^0})$,  
where the mixing angle $\alpha$ is given in compact form by
\beq
\alpha  = \frac{1}{2} {\rm arctan} \bigg({\rm tan} 2\beta \, \frac{M_A^2 
+ M_Z^2}{ M_A^2-M_Z^2} \bigg)\ , \ \ - \frac{\pi}{2} \leq \alpha \leq 0 
\label{alpha:tree}
\eeq
Thus, the supersymmetric structure of the theory has imposed very strong 
constraints on the Higgs spectrum. Out of the six parameters which describe
the MSSM Higgs sector, $M_h, M_H, M_A, M_{H^\pm}, \beta$ and $\alpha$, only
two parameters, which can be taken as $\tb$ and $M_A$, are free parameters at
the  tree--level. In addition, a strong hierarchy is imposed on the mass
spectrum and, besides the relations $M_H > {\rm max} (M_A,M_Z)$ and $M_{H\pm} 
>M_W$, we have the very important constraint on the lightest $h$ boson mass 
at  the tree--level which is maximal for large $\tb$ values for which
$\cos2\beta=1$, 
\beq
M_h & \leq  &{\rm min} (M_A, M_Z) \cdot |\cos2\beta|  \leq M_Z 
\label{treelevelMh}
\eeq

\subsection{The couplings of the MSSM Higgs bosons}

The Higgs boson couplings to the gauge bosons \cite{HHG,HaberGunion} are obtained 
from the kinetic terms of the fields $H_1$ and $H_2$ in the  Lagrangian
\beq
{\cal L}_{\rm kin.}=  (D^\mu H_1)^\dagger (D_\mu H_1) + (D^\mu H_2)^\dagger 
(D_\mu H_2) 
\eeq
Expanding the covariant derivative $D_\mu= -ig_2 \frac12 \tau_a W^a_\mu -i g_1
\frac{Y}{2} B_\mu$ and performing the usual transformations on the gauge and
scalar fields to obtain the physical fields, one can identify the trilinear
couplings $V_\mu V_\nu H_i$ among one Higgs and two gauge bosons and $V_\mu H_i
H_j$ among one gauge boson and two Higgs bosons, as well as the couplings
between two gauge and two Higgs bosons $V_\mu V_\nu H_i H_j$. The Feynman rules
for the important couplings of the neutral  Higgs bosons are given below, where
we have used the abbreviated couplings  $g_W=g_2$ and $g_Z=g_2/c_W$
[$c_W^2=1-s_W^2 \equiv \cos^2\theta_W$]:
\beq
Z_\mu Z_\nu h \ : \  ig_Z M_Z \sin(\beta-\alpha)g_{\mu\nu} &,& 
Z_\mu Z_\nu H \ \ \ : \ ig_Z M_Z \cos(\beta-\alpha) g_{\mu\nu}
\nonumber \\
W^+_\mu W^+_\nu h \ \ : \ ig_W M_W \sin(\beta-\alpha)g_{\mu\nu} &,&   
W^+_\mu W^-_\nu H \ : \ ig_W M_W \cos(\beta-\alpha)g_{\mu\nu}
\nonumber \\
Z_\mu hA: +{g_Z \over 2}  \cos(\beta-\alpha) (p_h+p_A)_\mu &,& 
Z_\mu HA: -{g_Z \over 2} \sin(\beta-\alpha) (p_H+p_A)_\mu 
\eeq
with $p_i$ the (entering the vertex) momenta of the Higgs bosons. A few remarks 
are to be made:\s

$\bullet$ The couplings of the charged Higgs bosons follow closely those of the  $A$
boson.\s

$\bullet$ Since the photon is massless, there are no Higgs--$\gamma \gamma$ and
Higgs--$Z\gamma$  couplings at  tree--level (there is no Higgs--gluon-gluon 
coupling as well as the Higgs is colorless) but the couplings can be generated
at the loop level.  CP--invariance also forbids $WWA,ZZA$ and $WZH^\pm$ 
couplings. \s  

$\bullet$ For the $H_i H_j V$ couplings, CP--invariance implies that $H_i$ and $H_j$ 
must have opposite parity;  there are no $Zhh,ZHh, ZHH,ZAA$ couplings and only
the $ZhA$ and $ZHA$ couplings are allowed.\s

$\bullet$ There are many quartic couplings between two Higgs and  two gauge
bosons; they are  proportional to $g_{\mu \nu}$ and involve two powers of the
electroweak  coupling which make them small.\s

$\bullet$ The couplings of the $h$ and $H$ bosons to $VV$ states  are proportional to
either $\sin(\beta-\alpha)$ or $\cos(\beta- \alpha)$; they are thus
complementary and the sum of  their squares is just the square of the SM Higgs
boson coupling $g_{H_{\rm  SM}VV}$. This complementarity  will have very
important consequences. For large $M_A$ values, one can expand the Higgs--VV 
couplings in powers of $M_Z/M_A$ to obtain
\beq
g_{HVV} & \propto  & \cos(\beta-\alpha)  \stackrel{\small M_A \gg M_Z} \lra \ \
\frac{M_Z^2} {2M_A^2} \sin4 \beta \quad \ \stackrel{\small \tb \gg 1} \lra \quad
- \frac{2 M_Z^2} {M_A^2 \tb} \ \ra 0 \non \\
g_{hVV} & \propto & \sin(\beta-\alpha)   \stackrel{\small M_A \gg M_Z} \lra
1- \frac{M_Z^4} {8M_A^4} \sin^2 4\beta  \stackrel{\small \tb \gg 1} \lra 
1- \frac{2 M_Z^4} {M_A^4 \tan^2 \beta} \ra 1 
\label{gHVVdecoup}
\eeq
where we have also displayed the limits at large  $\tb$. One sees that for
$M_A\gg M_Z$, $g_{HVV}$ vanishes while $g_{hVV}$ reaches unity, i.e. the SM
value; this occurs more  quickly if $\tb$ is large. \s

As SUSY imposes that the doublet $H_1 (H_2)$ generates the masses and
couplings of isospin $-\frac12(+\frac{1}{2})$ fermions, Higgs mediated flavor
changing neutral currents are automatically forbidden. The Higgs couplings to
fermions  come from the superpotential;  using  the left-- and right--handed 
projection operators $P_{L/R}\!=\!\frac{1}{2}(1\!\mp \!\gamma_5)$, the Yukawa
Lagrangian  with the first family notation  is 
\beq {\cal L}_{\rm Yuk}&=& - \lambda_u [ \bar u P_L u H_2^0  - \bar u P_L d
H_2^+ ] - \lambda_d [ \bar d P_L d H_1^0  - \bar d P_L u H_1^- ] + {\rm h.c.} 
\label{Yukawa-Lagrangian}  \eeq 
The fermion masses, generated when the Higgs fields acquire their vevs, are
related to the Yukawa couplings by $\lambda_u\!=\!\sqrt{2} m_u / (v \sin\beta)$
and   $\lambda_d\!=\!\sqrt{2} m_d /(v \cos\beta)$. Expressing the $H_1$ and 
$H_2$  fields  in terms of the physical fields, one  obtains the  MSSM Higgs 
couplings to fermions \cite{HHG,HaberGunion}
\beq 
&& G_{huu} = i \frac{m_u}{v} \frac{\cos\alpha}{\sin\beta}  \ , \ \  G_{Huu} =
i \frac{m_u}{v} \frac{\sin\alpha}{\sin\beta} \ , \ \  G_{Auu} =
\frac{m_u}{v} \cot\beta \, \gamma_5 \non \\  
&& G_{hdd} = -i \frac{m_d}{v} \frac{ \sin\alpha}{\cos\beta} \ , \ \ G_{Hdd} =
i \frac{m_d}{v} \frac{ \cos \alpha}{\cos\beta} \ , \ \  G_{Add} =
\frac{m_d}{v} \tan\beta \, \gamma_5 \non  \\ 
&& G_{H^+ \bar u d} =  -  \frac{i}{\sqrt{2} v}  V_{ud}^*  [m_d \tb 
(1+\gamma_5) + m_u{\rm cot}\beta (1-\gamma_5)]  
\eeq  
One notices that the couplings of the $H^\pm$ bosons  have the same $\tb$
dependence as those  of the pseudoscalar $A$ boson and that, for values
$\tb>1$,  the $A$ and $H^\pm$ couplings to down--type (up--type) fermions are
enhanced (suppressed). Thus, for large values of $\tb$, the couplings of these
Higgs bosons to $b$ quarks, $\propto m_b \tb$, become  very strong while those
to the top quark, $\propto m_t/ \tb$, become rather weak.  This is, in fact,
also the case of the couplings of one of the CP--even  Higgs boson $h$ or $H$ to
fermions. depending on the magnitude of $\cos(\beta-\alpha)$.  This can be
viewed in the limit of very large $M_A$ values. In this case, the reduced Higgs
couplings to fermions (normalized to the SM Higgs case) reach the limit:
\beq
M_A \gg M_Z : 
g_{huu}  \to 1  \ , \ 
g_{hdd}  \to 1   \,  \ 
g_{Huu} \to -\cot \beta \ , \ 
g_{Hdd}  \to \tb 
\eeq 
Thus, the couplings of the $h$ boson approach those of the SM Higgs boson, 
while the couplings of the $H$ boson reduce, up to a  sign, to those of the
pseudoscalar Higgs boson. Again, these limits are in general reached more
quickly at large values of $\tb$.. \s

The trilinear and quadrilinear couplings between three or four Higgs fields 
can be obtained from the scalar potential $V_H$ by performing  derivatives
with respect to three or four Higgs fields. Two important  trilinear couplings
among neutral Higgs bosons,  in units of $\lambda_0=-iM_Z^2/v$, are
\cite{HHG,HaberGunion}
\beq
\lambda_{hhh} = 3 \cos2\alpha \sin (\beta+\alpha) \, , \  
\lambda_{Hhh} = 2\sin2 \alpha \sin (\beta+\alpha) -\cos 2\alpha \cos(\beta
+ \alpha) 
\eeq
The numerous quartic Higgs couplings involve two powers of the electroweak
coupling and can be expressed  in units of $\lambda_0/v= M_Z^2/v^2$; they are 
thus very small.\s 

Finally, there are Higgs couplings to SUSY particles. A coupling which 
plays an important role is the $h$ coupling to  top squarks which, in the case 
of the lightest one $\tilde t_1$, 
reads \cite{HaberGunion2}
\begin{eqnarray}
g_{h \tilde{t}_1 \tilde{t}_1 } \propto \cos 2\beta M_Z^2 \left[ \frac{1}{2} \cos^2 
\theta_t - \frac{2}{3} s^2_W \cos 2 \theta_t \right] + m_t^2 + 
\frac{1}{2} \sin 2\theta_t  m_t X_t
\end{eqnarray} 
and involves components which are proportional to $X_t=A_t -\mu \cot \beta$ where
$A_t$ is the stop mixing parameter. For  large values of the parameter $X_t$,
which incidentally make the $\tilde{t}$  mixing angle almost maximal, $|\sin 2
\theta_{t}| \simeq 1$ and  lead to lighter $\tilde{t}_1$ states, the last
components can strongly   enhance the $g_{h\tilde{t}_1 \tilde{t}_1}$ coupling
and make it larger than  the top quark coupling, $g_{htt} \propto m_t/M_Z$. \s

Another class of potentially important couplings of the Higgs bosons are  the
ones to the two charginos $\chi_i^\pm$ and four neutralinos $\chi_i^0$. With 
the notation $\Phi=h,H,A$, they are given by \cite{HaberGunion2}
\beq
g_{\chi^0_i \chi^+_j H^+} &\propto&  \sqrt 2 Z_{j4} V_{i1} +  
\left( Z_{j2} + \tan \theta_W Z_{j1} \right) V_{i2}  \nonumber  \\
g_{\chi^-_i \chi^+_j \Phi} \propto  e_k V_{j1}U_{i2}-d_k V_{j2}U_{i1} & , & 
g_{\chi^0_i \chi^0_j \Phi} \propto  \left( Z_{j2}- \tan\theta_W Z_{j1} \right) 
\left(e_\Phi Z_{i3} + d_\Phi Z_{i4} \right) 
\eeq
where $Z$ and $U/V$ are the $4\times 4 $ and $2 \times 2$ matrices which 
diagonalize the neutralino and chargino matrices and the coefficients $e_\Phi,
d_\Phi$ are sines and cosines of the angles $\alpha$ and $\beta$.  The Higgs
couplings to the $\chi_1^0$ lightest SUSY particle (LSP), for which  $Z_{11},
Z_{12}$ are the gaugino components and $Z_{13},Z_{14}$  the higgsino components,
vanish if the LSP is a pure gaugino or a pure higgsino. This statement can be
generalized to all neutralino and chargino states and the Higgs bosons couple
only to higgsino--gaugino mixtures or states. The couplings of the neutral Higgs
bosons to neutralinos can also accidentally vanish for certain values of $\tb$
and $\alpha$  which enter the coefficients  $d_\Phi, e_\Phi$.

\subsection{Radiative corrections in the MSSM Higgs sector}

It was realized in the early nineties that, as a result of the  large Yukawa
coupling of the top quark, the radiative corrections in the MSSM  Higgs sector
are very important \cite{RC-leading}. The leading part of these corrections 
rise with the fourth power of the top quark mass  and logarithmically with the
stop mass.  These corrections  may push the lighter  Higgs mass well above the
tree--level bound, $M_Z$. In the subsequent years, an impressive theoretical
effort has been devoted to the precise determination of the Higgs boson masses
in the MSSM.  A first step was to provide the full one--loop computation
including the contributions of all SUSY particles \cite{RC-oneloop} and a second
the addition of the dominant two--loop corrections  \cite{RC-all,RC-review}
involving the strongest couplings of the theory, the QCD coupling and the Yukawa
couplings of heavy third generation fermions.  Other small higher--order
corrections have also been calculated \cite{RC-higher}. \s

As seen previously,  at the tree level, the Higgs sector of the MSSM can be
described by two input parameters, which can be taken to be $M_A$ and $\tb$.
The CP--even Higgs mass matrix, given  by eq.~(\ref{CPeven:matrix-tree}),
receives radiative corrections at  higher orders and it can be written as
\begin{eqnarray}
{\cal M}^2 = \left[ \begin{array}{cc} {\cal M}_{11}^2 + \Delta {\cal M}_{11}^2
& {\cal M}_{12}^2 + \Delta {\cal M}_{12}^2 \\ 
  {\cal M}_{12}^2 + \Delta {\cal M}_{12}^2 
&  {\cal M}_{22}^2 + \Delta {\cal M}_{22}^2  
\end{array} \right]
\label{HmatrixRC}
\end{eqnarray}
The leading one--loop radiative corrections $\Delta {\cal M}_{ij}^2$ to the 
mass matrix are controlled by the top Yukawa coupling $\lambda_t$  and one can
obtain a very simple  analytical expression in this case  \cite{RC-leading}
\beq
\Delta {\cal M}_{11}^2& \sim & \Delta {\cal M}_{12}^2 \sim 0 \ , \non \\
\Delta {\cal M}_{22}^2& \sim & \epsilon = \frac{3\, \bar{m}_t^4}{2\pi^2 v^2\sin^
2\beta} \left[ \log \frac{M_S^2}{\bar{m}_t^2} + \frac{X_t^2}{2\,M_S^2} \left( 1 -
\frac{X_t^2}{6\,M_S^2} \right) \right]
\label{higgscorr}
\eeq
where $M_S$ is the arithmetic  average of the stop masses $M_S\!=\!\frac{1}{2}
(m_{\tilde{t}_1}\!+\!m_{\tilde{t}_2})$, $X_{t}\!=\!A_t -\mu /\tan\beta$ where
$A_t$ is the stop mixing parameter and $\bar{m}_t$ is the running ${\rm
\overline{MS}}$ top quark mass to account for the leading two--loop QCD and
electroweak corrections in a renormalization group (RG) improvement.\s

The corrections controlled by the bottom Yukawa coupling $\lambda_b$ are in
general strongly suppressed by powers of the $b$--quark mass $m_b$. However,
this suppression can be compensated by a large value of the sbottom mixing
parameter $X_b= A_b- \mu \tb$, providing a non--negligible correction to ${\cal
M}^2$.  Including these subleading contributions at one--loop, plus the leading
logarithmic contributions at two--loops, provides a rather good approximation of
the bulk of the radiative corrections.  Nevertheless, one needs to include the
full  set of corrections mentioned previously to have precise predictions for
the  Higgs boson masses and couplings to which we turn now. \s

The radiatively corrected CP--even Higgs boson masses are obtained by 
diagonalizing the mass matrix eq.~(\ref{HmatrixRC}). In the approximation 
where only the leading corrections controlled by the top Yukawa coupling, 
eq.~(\ref{higgscorr}), are implemented, the masses are simply given by 
\cite{RC-leading}
\begin{eqnarray}
M_{h,H}^2 = \frac{1}{2} (M_A^2+ M_Z^2+\epsilon) \left[ 1 \mp 
\sqrt{1- 4 \frac{ M_Z^2 M_A^2 \cos^2 2\beta +\epsilon ( M_A^2 \sin^2\beta +
 M_Z^2 \cos^2\beta)} {(M_A^2+ M_Z^2+\epsilon)^2} } \right] \ 
\label{Mhepsilon}
\end{eqnarray}
In this approximation, the charged Higgs mass does not receive radiative 
corrections, the leading contributions being only of ${\cal O} (\alpha m_t^2)$ 
in this case \cite{RC-review}. \s

For large values of the pseudoscalar Higgs mass, $M_A  \gg M_Z$, the lighter
Higgs boson mass reaches its maximum for a given  $\tb$ value and in the
``$\epsilon$ approximation", this value reads
\beq
M_h \stackrel{\small M_A \gg M_Z} \to  \sqrt{M_Z^2\cos^22 \beta + \epsilon 
\sin^2\beta} \ 
\stackrel{\small \tb \gg 1} \to \ \sqrt{M_Z^2 + \epsilon}
\eeq
The radiative corrections are largest and maximize $M_h$ in the so--called 
``maximal mixing" scenario, where the  trilinear stop coupling in the 
$\overline{\rm DR}$ scheme is such that $X_t =A_t - \mu \cot \beta \sim 
\sqrt{6} M_S$, while the radiative corrections are much smaller in the ``no
mixing scenario" where $X_t$ is close to zero.\s

In the limit $M_A \gg M_Z$, the heavier CP--even and charged Higgs bosons
become almost degenerate in mass with the pseudoscalar Higgs boson
\beq
M_H \simeq M_{H^\pm} \simeq M_A
\eeq
This is an aspect of the decoupling limit \cite{Decoupling} which will be
discussed in  more detail later. \s 

The Higgs couplings  are renormalized by the same radiative  corrections which
affect the masses. For instance, in  the $\epsilon$ approximation, the
corrected angle $\bar \alpha$ will be  given by 
\beq
\tan 2 \bar{\alpha} = \tan 2\beta \,  \frac{M_A^2 + M_Z^2} {M_A^2 - 
M_Z^2 + \epsilon /\cos 2 \beta } \ , \hspace*{1cm}- \frac{\pi}{2} \leq \alpha 
\leq 0
\label{alphaCR}
\eeq
The radiatively corrected reduced couplings of the neutral CP--even Higgs 
particles to gauge bosons (i.e. normalized to the SM Higgs coupling) are 
then simply given by
\beq
g_{hVV}= \sin (\beta- \bar \alpha) \ \ , \ \
g_{HVV}= \cos (\beta- \bar \alpha) 
\eeq
where the renormalization of $\alpha$ has been performed in the same
approximation as for the masses.

In the case of the Higgs--fermion couplings, there are additional one--loop 
vertex corrections which modify the tree--level Lagrangian that incorporates 
them \cite{CR-hrs}. In the case of quarks, these corrections involve squarks
and gluino in the loops and can be very large, in particular for the bottom
Yukawa couplings for which they grow as $m_b \tan\beta$, $\Delta_b \simeq 
 \frac{2\alpha_s}{3\pi} \mu m_{\tilde{g}} \tb /{\rm max}(m_{\tilde{g}}^2, 
m_{\tilde{b}_1}^2,m_{\tilde{b}_2}^2)$. 
For instance, the reduced $b\bar b$  couplings of the $H,A$ states [in the 
$\overline{\rm MS}$ scheme and at zero momentum transfer] are given in this 
case by
\beq
g_{Hbb} &\simeq &  \frac{\cos \bar \alpha}{\cos\beta} \bigg[1-\frac{\Delta_b}
{1+\Delta_b}( 1- \tan \bar \alpha \cot \beta  )\bigg] \ , \ 
g_{Abb} \simeq  \tb \bigg[1-\frac{\Delta_b} {1+\Delta_b} \frac{1}
{\sin^2\beta}\bigg] 
\label{ghff:threshold}
\eeq

Finally, the trilinear Higgs couplings are renormalized not  only indirectly by
the renormalization of the angle $\alpha$, but also directly by additional
contributions to the vertices  \cite{CR-hhheps}.  In  the $\epsilon$
approximation,  which here gives only the magnitude of the correction, the
additional shifts  in the Higgs self--couplings $\Delta \lambda = \lambda^{\rm
1-loop} (\bar \alpha)  -\lambda^{\rm  Born} (\alpha \to \bar \alpha) $ are given
by \cite{CR-hhheps}
\beq
\label{Trilinear-RC}
\Delta \lambda_{hhh} = 3 \frac{\epsilon}{M_Z^2} \frac{\cos \alpha}{\sin\beta} 
\cos^2\alpha  \, , \ 
\Delta \lambda_{Hhh} = 3 \frac{\epsilon}{M_Z^2} \frac{\sin \alpha}{\sin\beta}
\cos^2\alpha 
\eeq

\subsection{Summary of Higgs masses, couplings and regimes in the MSSM}

For an accurate determination of the CP--even Higgs boson masses and couplings, 
the $\epsilon$ approach, although transparent and useful for a qualitative
understanding, is not a very good approximation.  The full one--loop
corrections, RGE improvement and  the non--logarithmic two--loop contributions 
due to QCD and the top/bottom Yukawa couplings should also be included. Here,
we will discuss the masses  and couplings of the MSSM Higgs bosons, including
the most important corrections. The Fortran code  {\tt SuSpect} \cite{Suspect}
which  calculates the spectrum of the SUSY and Higgs particles in the MSSM and
which incorporates the set of the dominant radiative corrections (here,
calculated in the on--shell scheme using the  routine {\tt FeynHiggsFast}
\cite{FeynHiggs}),  has been  used.\s

\begin{figure}[h]
\begin{center}
\vspace*{-1.8cm}
\hspace*{-3cm}
\epsfig{file=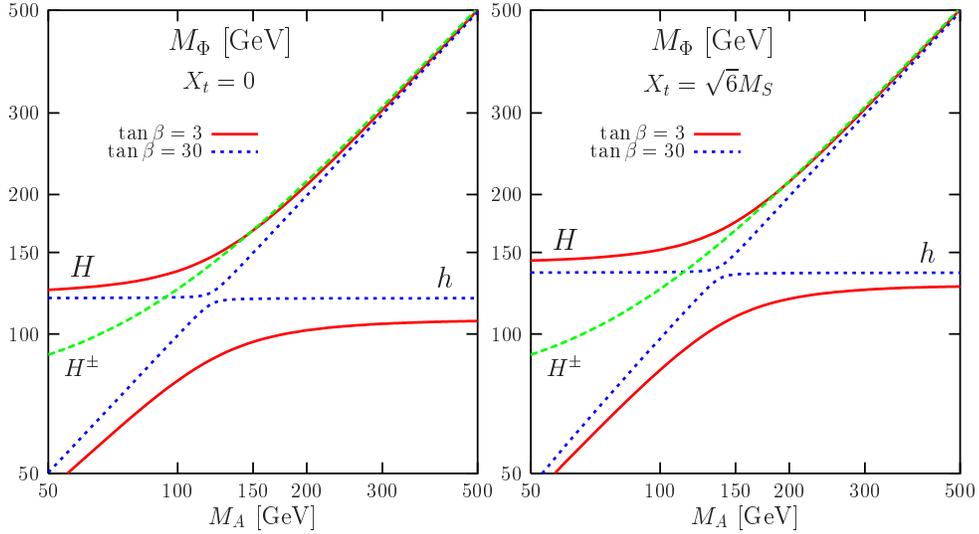,width=14.cm} 
\end{center}
\vspace*{-11.cm}
\caption[]{The masses of the MSSM Higgs bosons as a function of $M_A$ for two
values $\tb=3$ and 30, in the no mixing (left) and maximal mixing  (right)
scenarios with $M_S=2$ TeV and all the other SUSY parameters set to 1 TeV. The
full set of radiative corrections is included with the pole masses $m_t=178$  
GeV, $m_b =4.88$ GeV and with $\alpha_s (M_Z)=0.1172$.} 
\vspace*{-2mm}
\end{figure}

The radiatively corrected masses of the neutral CP--even and the charged Higgs
bosons are displayed in Fig.~1 as functions of $M_A$ for the values  $\tb=3$
and $30$.  The  scenarios of no--mixing with $X_t=0$ (left) and maximal
mixing with $X_t = \sqrt{6} M_S$ (right)  have been assumed.    As can be seen,
a maximal value for the lighter Higgs mass, $M_h \sim 135$ GeV, is obtained for
large $M_A$ values in the maximal mixing scenario with $\tb = 30$; the mass
value is almost constant if $\tb$ is increased.  For no stop mixing, or when
$\tb$ is small, $\tb \lsim 3$, the upper bound on the $h$ boson mass is smaller
by more than 10 GeV in each case and the combined choice $\tb=3$ and $X_t=0$,
leads to a maximal value $M_h^{\rm max}  \sim 110$ GeV. Also for large $M_A$
values, the $A,H$ and $H^\pm$ bosons (the mass of the latter being almost
independent of the stop mixing and $\tb$) become degenerate in mass. In the
opposite case, i.e. for a light pseudoscalar, $M_A \lsim M_h^{\rm max}$, it is
$M_h$ which is very close to $M_A$, and the mass difference is particularly
small for large $\tb$ values.\s  

The squares of the  renormalized Higgs couplings to gauge bosons and to isospin
$\pm \frac12$ fermions are displayed in  Figs.~2,  as functions of $M_A$ in the
no and maximal mixing cases,  respectively; the  SUSY and SM parameters are 
chosen as in Fig.~1.  One notices the very strong variation with $M_A$ and the
different pattern for values above and below the critical value $M_A \simeq
M_h^{\rm max}$. \s

For small $M_A$ values the $hVV$ couplings are suppressed, with the suppression
being stronger with large values of $\tb$. For values $M_A \gsim M_h^{\rm max}$,
the $hVV$ boson couplings tend to unity and reach the values of the SM Higgs
couplings, $g_{hVV}=1$ for $M_A \gg M_h^{\rm max}$; these values are reached
more quickly when $\tb$ is large.  The situation in the case of the heavier
CP--even $H$ boson is just opposite: its couplings are close to unity for $M_A
\lsim M_h^{\rm max}$ [which in fact is very close to the minimal value of $M_H$,
$M_H^{\rm min} \simeq M_h^{\rm max}$, in particular at large $\tb$], while above
this limit, the $H$ couplings to gauge bosons are strongly suppressed.   Note
that the mixing $X_t$ in the stop sector does not alter this pattern, its  main
effect being simply to shift the value of $M_h^{\rm max}$.\s

\begin{figure}[h!]
\begin{center}
\vspace*{-1.7cm}
\hspace*{-6.cm}
\epsfig{file=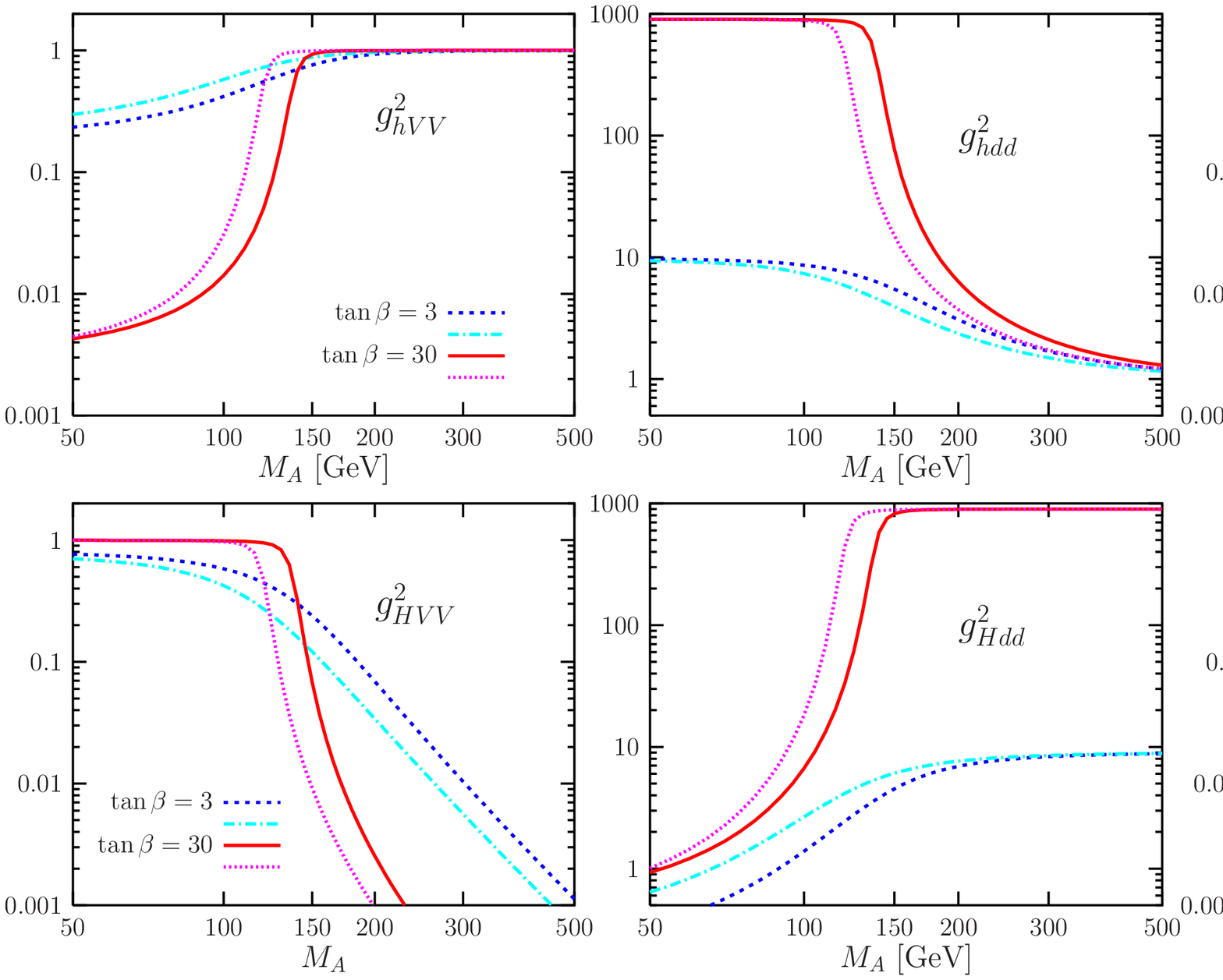,width=12.5cm}  
\end{center}
\vspace*{-6.8cm}
\caption[]{The normalized couplings squared of the CP--even MSSM  neutral Higgs
bosons to gauge bosons and fermions as a function of $M_A$ for  $\tb=3$ and 30 
with the same inputs as in Fig.~1.} 
\end{figure}

As in the case of the $VV$ couplings, there is  a very strong variation  of the
Higgs couplings to fermions with $M_A$ and different behaviors for values above
and below the critical mass $M_A \simeq M_h^{\rm max}$. For  $M_A \lsim M_h^{\rm
max}$ the $h$ couplings to up--type fermions  are suppressed, while those to
down--type fermions are enhanced, with  the suppression/enhancement being
stronger at high $\tb$. For $M_A \gsim  M_h^{\rm max}$, the normalized $h$
couplings tend to unity and reach the  values of the SM Higgs couplings,
$g_{hff}=1$, for $M_A \gg M_h^{\rm max}$;  the limit being reached more quickly
when $\tb$ is large. The situation of the $H$  couplings to fermions is just
opposite: they  are close to unity for $M_A \lsim M_h^{\rm max}$, while for $M_A
\gsim  M_h^{\rm max}$, the couplings to up (down)--type fermions are  suppressed
(enhanced). For $M_H \gg M_h^{\rm max}$, they  become approximately equal to
those of the $A$ boson which couples  to down (up)--type  fermions
proportionally to, respectively, $\tb$ and  $\cot \beta$. In fact,  in this
limit, also the $H$ coupling to gauge bosons  approaches zero, i.e. as in the
case of the $A$ boson.\s

Let us finally summarize the various regimes of the CP--conserving MSSM Higgs
sector \cite{Anatomy2}.\s 

There is first the \underline{decoupling regime} \cite{Decoupling} for large
values of $M_A$, which has been already mentioned. In this regime, which occurs
in practice for $M_A \gsim 300$ GeV for low $\tb$ and $M_A \gsim M_h^{\rm max}$
for $\tb \gsim 10$, the $h$ boson reaches its maximal mass value and its
couplings to fermions and gauge bosons as well as its self--couplings
become SM--like. The heavier $H$ boson has approximately the same mass as the
$A$ boson and its interactions are similar, i.e. its couplings to gauge bosons
almost vanish and the couplings to isospin $-\frac{1}{2}$ ($+\frac{1}{2}$) 
fermions are (inversely) proportional to $\tb$.  The $H^\pm$  boson is also
degenerate in mass with the $A$ boson and its couplings to single $h$ bosons are
suppressed.  Thus, in the decoupling limit, the heavier Higgs bosons decouple
and the MSSM Higgs sector reduces effectively to the SM Higgs sector, but with a
light Higgs with a mass $M_h \lsim 140$ GeV.  This light Higgs particle is
nearly indistinguishable  from the SM Higgs boson. \s

In the \underline{anti--decoupling regime} \cite{Antidecoup}, which occurs for
a  light pseudoscalar Higgs boson, $M_A \ll M_h^{\rm max}$, the situation is
exactly opposite to the one of the decoupling regime. Indeed, in this case, the
lighter tree--level $h$  mass is given by $M_h \simeq M_A |\cos 2\beta|$ while
the tree--level heavier $H$ mass is given by $M_H \simeq M_Z(1 + M_A^2 \sin^2
2\beta /M_Z^2)$. At large values of $\tb$, the $h$ boson is degenerate in mass
with the $A$ boson, $M_h \simeq M_A$, while the $H$ boson has a mass close to
its minimum which is in fact $M_h^{\rm max} \simeq \sqrt{M_Z^2+\epsilon}$. This
is similar to the decoupling regime, except that the roles of the $h$ and $H$
bosons are reversed, and since there is an upper bound on $M_h$, all Higgs
particles are light. Here, it is $\cos(\beta- \alpha)$ which is close to unity 
and  $\sin(\beta- \alpha)$ which is small. Thus, it is the $h$ boson which has 
couplings close to those of the $A$ boson,  while the $H$ boson couplings are
SM--like.\s

The \underline{intense--coupling regime} \cite{ICR,intense} will occur when the mass of the
pseudoscalar $A$ boson is close to $M_h^{\rm max}$. In this case, the three
neutral Higgs bosons $h,H$ and $A$ [and even the charged Higgs particles] will
have comparable masses, $M_h \sim M_H \sim M_A \sim M_h^{\rm max}$. The mass
degeneracy is more effective when $\tb$ is large. In this case both the $h$ and
$H$ bosons have still enhanced couplings to down--type fermions and suppressed
couplings to gauge bosons and up--type fermions.\s

The \underline{intermediate--coupling regime} occurs for low values of $\tb$,
$\tb \lsim 3$--5,  and a not too heavy pseudoscalar Higgs boson, $M_A \lsim
300$--500 GeV \cite{Anatomy2}. Hence, we are not yet in the decoupling regime
and both $\cos^2 (\beta - \alpha)$ and $\sin^2 (\beta - \alpha)$ are sizable,
implying that both CP--even Higgs bosons have significant couplings to gauge
bosons. The couplings between one gauge boson and two Higgs bosons, which are
suppressed by the same mixing angle factors, are also significant. In  addition,
the couplings of the neutral  Higgs bosons to down--type (up--type) fermions are
not strongly enhanced (suppressed) since $\tb$ is not too large. \s

Another possibility  is the \underline{vanishing--coupling regime}.  For
relatively large values of $\tb$ and intermediate to large $M_A$ values, as well
as for specific values of the other MSSM parameters entering the radiative
corrections, there is a possibility of the suppression of the couplings of one
of the CP--even Higgs bosons to fermions or gauge bosons, as a result of the
cancellation between tree--level terms and radiative corrections
\cite{vanishing}. In addition, in the case of the $hbb$ and $hgg$ couplings, a
strong suppression might occur as a result of large direct corrections. 

\subsection{Constraints on the MSSM Higgs sector}

There are various experimental constraints on the MSSM Higgs sector from the 
negative searches that have been performed up to now\footnote{Note that there
are also indirect constraints on the Higgs sector from high-precision
measurements and $B$  physics, but they are more model dependent and not 
very effective in the MSSM; they will not be discussed here.}, mainly at LEP
and  Tevatron. They are  summarized  below.\s

At LEP, which has operated at energies up to 210 GeV, a 95\% confidence level
lower bound $M_{H_{\rm SM}} > 114.4$ GeV has been set on the mass of the SM
Higgs boson, by investigating the Higgs--strahlung process, $\ee \to ZH_{\rm
SM}$ \cite{PDG,LEP2-Higgs}. In the MSSM, this bound is valid for the lighter
CP--even $h$ particle if its coupling to the $Z$ boson is SM--like  $g^2_{ZZh}
\simeq 1$ [i.e. almost in the decoupling regime] or in the case of the heavier
$H$ particle if $g^2_{ZZH} \equiv \cos^2 (\beta- \alpha) \simeq 1$ [i.e. in the
anti--decoupling regime with a rather light $M_A$].  The complementary search of
the neutral Higgs bosons in the associated production processes $\ee \to hA$ and
$HA$, allows to set the following combined 95\% CL limits on the $h$ and $A$
boson  masses\footnote{Note that compared to the SM, there is a $1.7\sigma$
excess of  events at a Higgs mass of $\sim 115$ GeV and a $2.3 \sigma$ excess at
$\sim 98$ GeV; the two can be explained by assuming $M_H \sim 115$ GeV and $M_h
\sim M_A \sim 98$ GeV \cite{Add-ref1}.}  \cite{PDG,LEP2-Higgs}
\beq 
M_h > 91.0~{\rm GeV} \ \ {\rm and} \ \  M_A >91.9 ~{\rm GeV}
\eeq
[which apply only if the $b\bar b$ and $\tau\tau$  couplings of the $h/A$ states
are not suppressed; see Ref.~\cite{Add-ref2} e.g.]
These bounds can be turned into exclusion regions in the MSSM parameter space. 
This is shown for the $\tb$--$M_h$  plane in  Fig.~3 where the no mixing (left)
and maximal--mixing (right)  scenarios are chosen with $M_S=1$ TeV and
$m_t=174.3$ GeV [which is $1\sigma$ higher than the current experimental value
$m_t \simeq 172$ GeV]; $\tb$ is also allowed to be less than unity. As can be
seen, with these specific assumptions, a significant portion of the parameter
space is excluded for the maximal mixing scenario; values $\tb \lsim 2$ are
ruled out at the 95\% CL. The exclusion regions are much larger in the
no--mixing scenario since $M_h^{\rm max}$ is smaller by approximately 20 GeV and
not far from the value that is experimentally excluded at LEP2 in the decoupling
limit, $M_h \gsim 114.4$ GeV; for instance, the range $\tb \lsim 5$ is excluded
at 95\% CL for $m_t=174.3$  GeV. The upper boundaries of the 
parameter space are indicated for other values of the top quark mass.

\begin{figure}[h!]
\begin{center}
\vspace*{-.5cm}
\hspace*{-.9cm}
\mbox{
\epsfig{figure=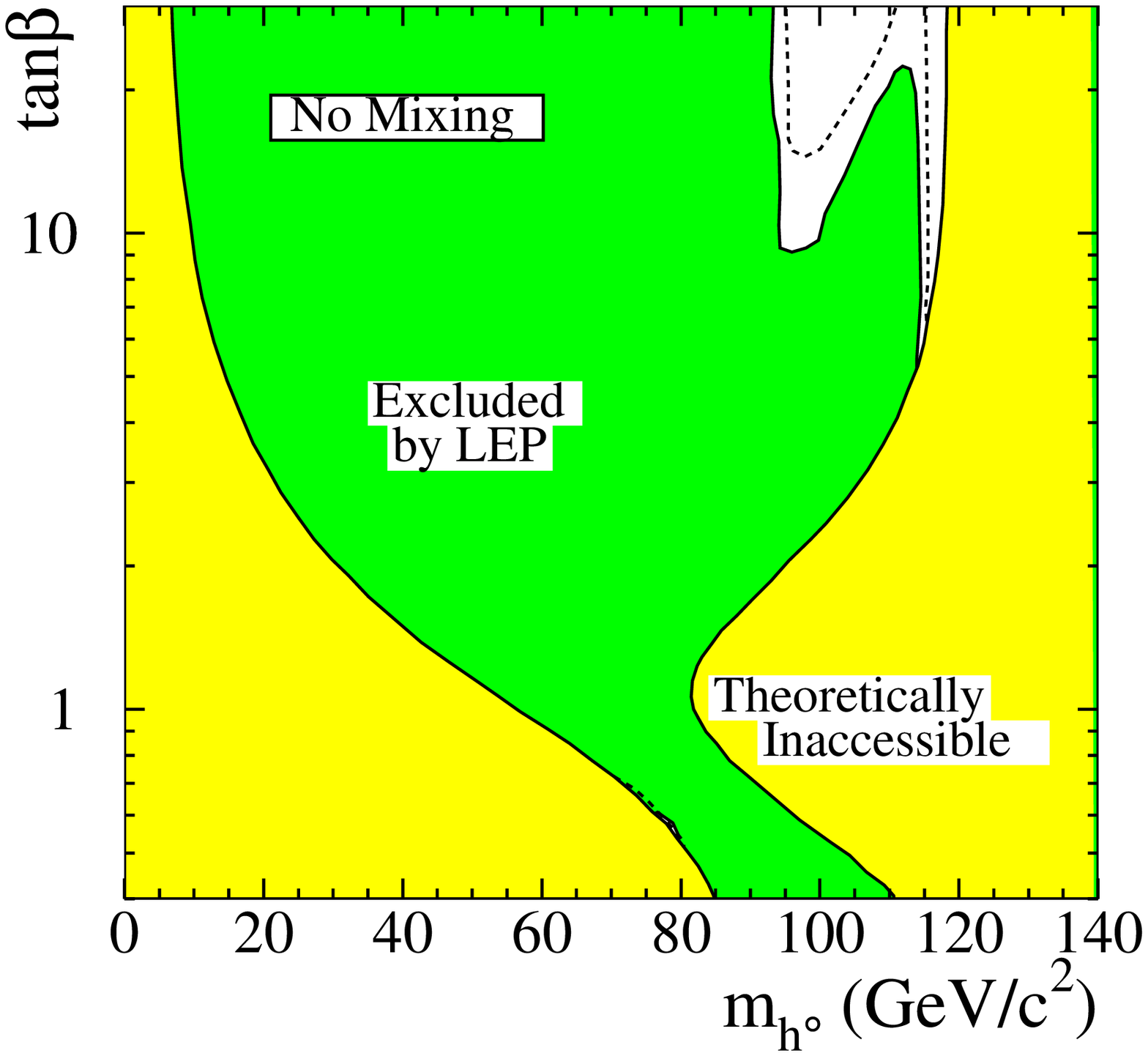,width=8.cm,height=7cm}
\hspace*{-.9cm}
\epsfig{figure=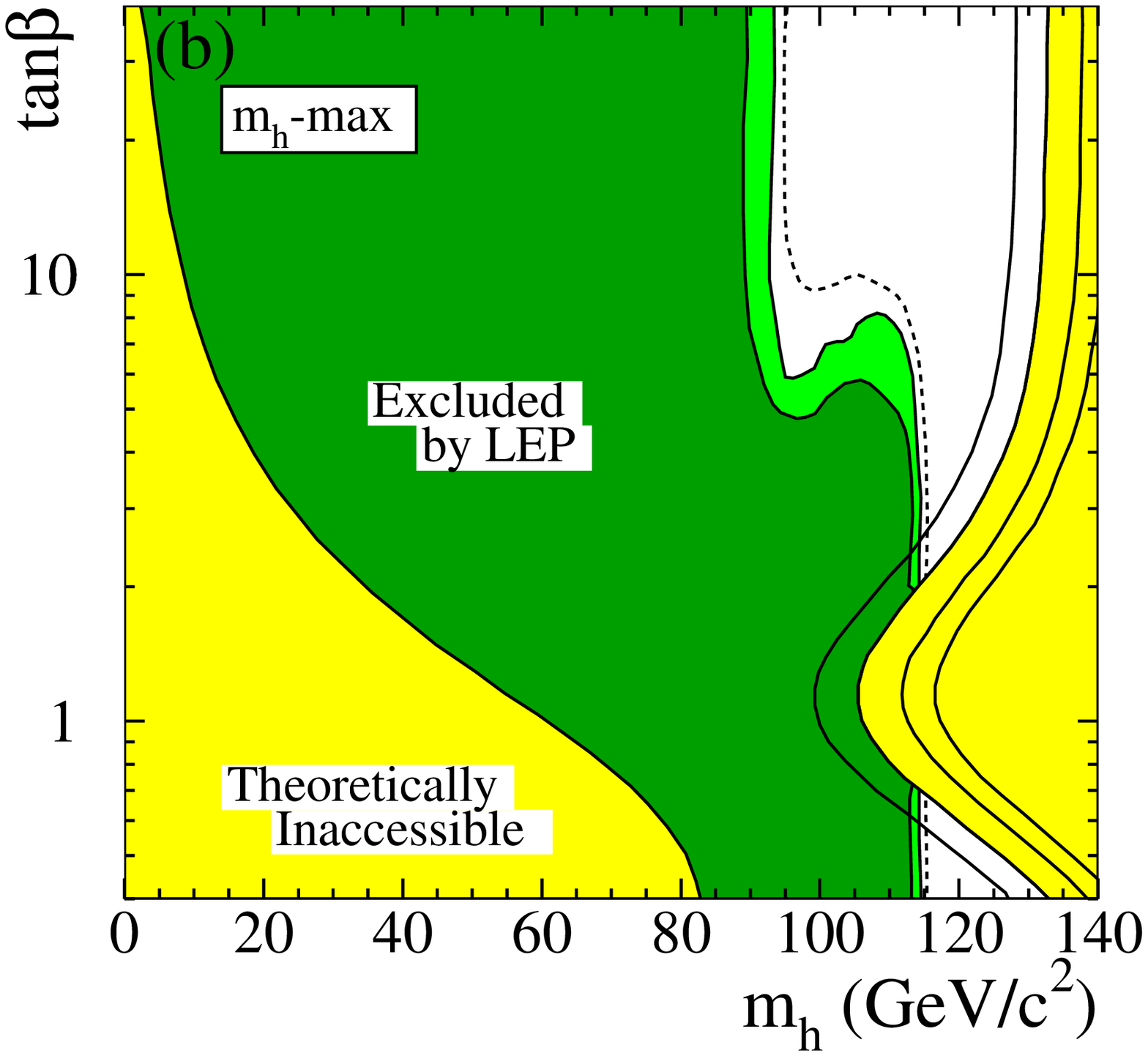,width=8.cm,height=7cm}
}
\end{center}
\vspace*{-.8cm}
\caption[]{95\% CL contours in the  $\tb$--$M_h$  plane excluded by the 
negative searches of MSSM neutral Higgs bosons at LEP2  in the no--mixing 
(left) and maximal mixing (right) scenarios with $M_S=1$ TeV and $m_t=174.3$ GeV. 
The dashed lines indicate the boundaries that are excluded on the basis of 
a simulations in the absence of a signal; the upper boundaries of the 
parameter space are indicated for the values from  left to right: $m_t 
= 169.3, 174.3, 179.3$ and 183 GeV; from \cite{LEP2-Higgs}.}
\vspace*{-3mm}
\end{figure}

In the case of the charged Higgs boson, an absolute bound of $M_{H^\pm} \gsim 
80$ GeV has been set by the LEP collaborations \cite{LEP2-Higgs,LEP-H+ALEPH} by
investigating the pair production $e^+e^- \to H^+ H^-$, with the $H^\pm$ bosons
decaying into either $\nu \tau $ or $cs$ final states (see the next section).
However, since in the MSSM, $M_{H^\pm}$ is constrained to be $M_{H^\pm}=\sqrt{
M_W^2 + M_A^2}$ and in view of the absolute  bound on $M_A$, one should have
$M_{H^\pm} \gsim 120$ GeV. The previous bound does not provide any additional
constraint in the MSSM. A more restrictive bound is obtained from  $H^\pm$
searches at the Tevatron in the decays of the heavy top quark, $t\ra bH^+$
\cite{top-toH+,Tevatron}, if $M_{H^\pm}\lsim m_t-m_b\sim  170$ GeV (see also
next section). However, the branching ratio  compared to the dominant standard
decay  $t \rightarrow bW^+$,  is large only for rather small, $\tb  \lsim 3$, 
and large, $\tb \gsim 30$, values when the $H^\pm tb$ coupling is  strongly
enhanced. The outcome of the search is summarized in the  right-hand side of
Fig.~4 and as can be seen, it is only for $M_{H^\pm}\lsim 140$ GeV and $\tb$
values below unity and above 60 (i.e. outside the theoretically favored $\tb$
range in the  MSSM) that the constraints are obtained \cite{Tevatron-H+}.\s

\begin{figure}[h!]
\vspace*{-.6cm}
\begin{center}
\begin{tabular}{cc}
\begin{minipage}{7cm}
\epsfxsize=6.0cm
\epsfbox[24 147 538 668]{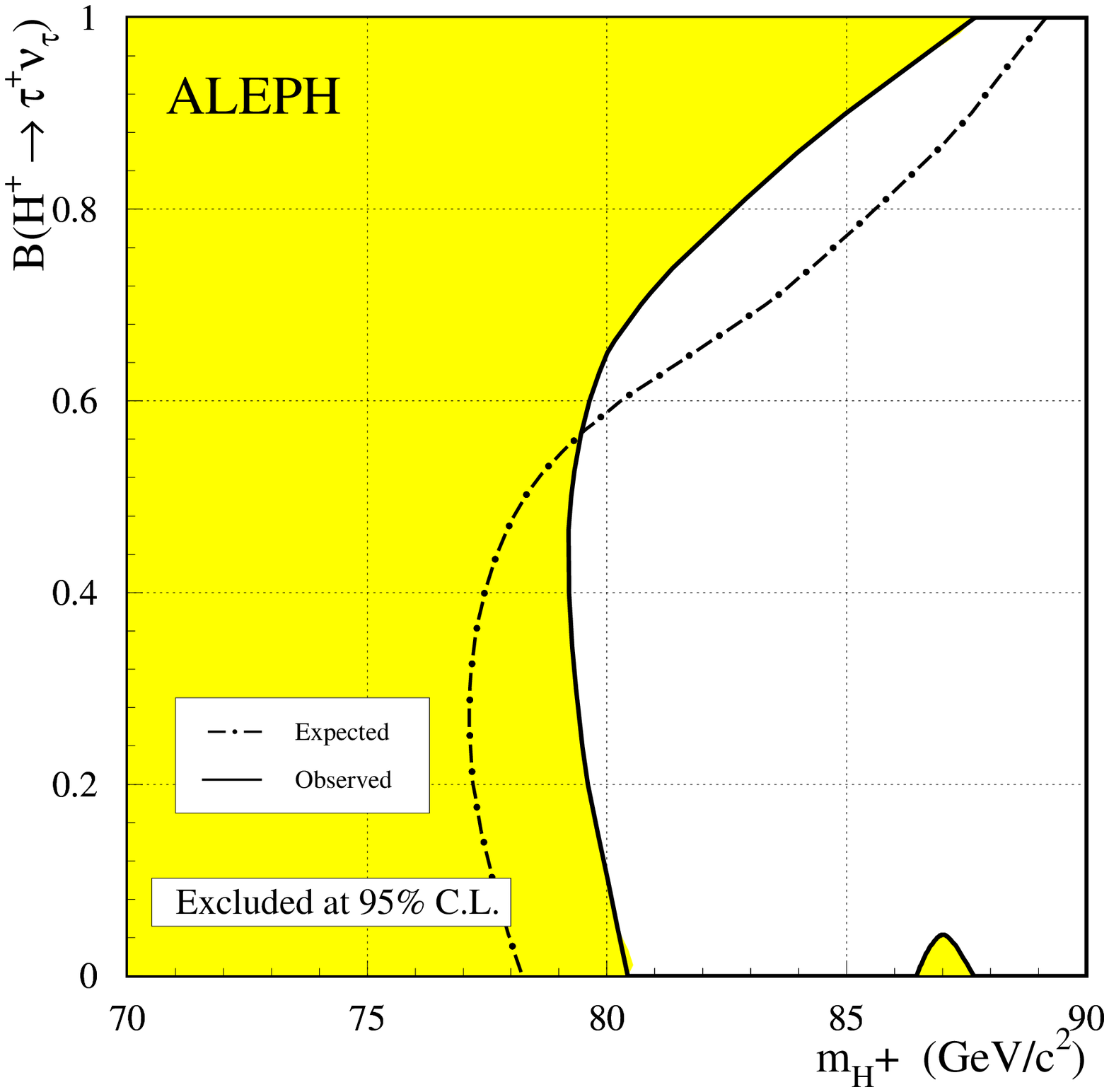} 
\end{minipage}
& \ \ \ 
\begin{minipage}{8cm}
\epsfig{figure=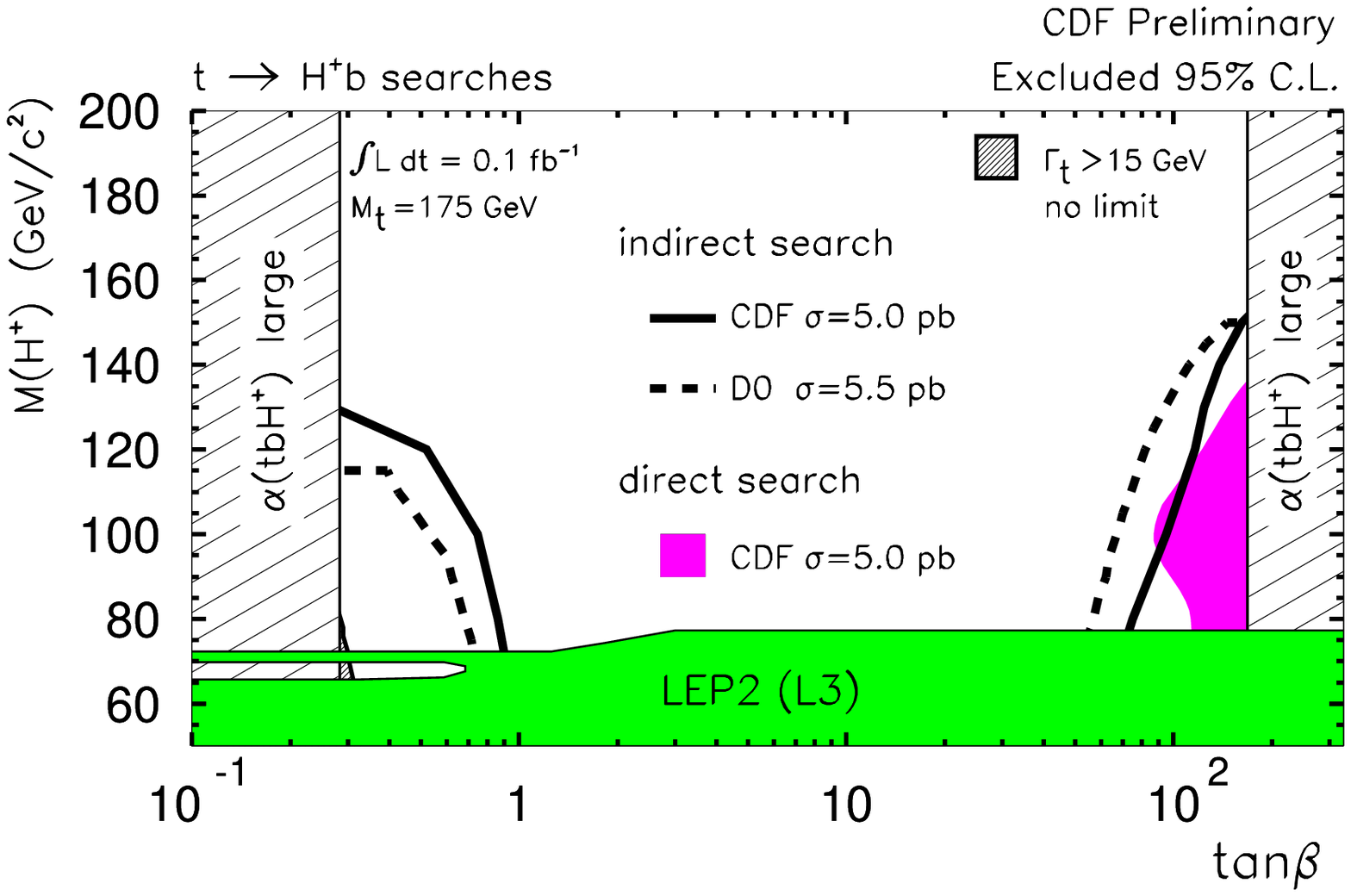,width=7cm,height=7.5cm} 
\end{minipage}
\end{tabular}
\end{center}
\vspace*{-.9cm}
\caption[]{The constraint on $M_{H^\pm}$ as a function of BR$(H^\pm \to \tau 
\nu)$ from the negative searches of $H^\pm$ states by the ALEPH collaboration at
LEP2 \cite{LEP-H+ALEPH} (left)  and the $\tb$--$M_{H^\pm}$ parameter space
excluded at the Tevatron  from the non--observation of the top decay $t \to 
H^+b$ \cite{Tevatron-H+} (right).}
\vspace*{-.3cm}
\end{figure}

\subsection{Higgs bosons in non--minimal SUSY models}

The Higgs sector in SUSY models may be slightly more complicated than the one of
the CP--conserving  MSSM discussed in the previous subsections. In the
following, we briefly discuss the Higgs spectrum in some of these extensions and
highlight the major differences with the  MSSM. \s

In the presence of new sources of \underline{CP--violation in the SUSY sector},
which is required if baryogenesis  is to be explained at the electroweak scale,
the new phases will enter the MSSM Higgs sector (which is CP--conserving at
tree--level as discussed in one of the previous subsections) through the large
radiative corrections which depend, for instance,  on the parameters $A_t$ and
$\mu$ that can involve complex phases in general. These corrections will affect
the masses and the couplings of the neutral and charged Higgs particles. In
particular, the three neutral Higgs bosons will not have definite CP quantum
numbers and will mix with each other to produce the physical states $H_1,H_2$
and $H_3$. The decay and  production properties of the various Higgs particles
can be significantly affected; for reviews, see
e.g.~Refs.~\cite{SUSY-CPV,CPHmasses,cpxbenchmark}.  Note, however, that there is
a sum rule which forces the three $H_i$ bosons to share the coupling of the SM
Higgs boson to gauge bosons, $\sum_i g_{H_iVV}^2 =g^2_{H_{\rm SM}}$;  only the
CP--even component is projected out in these couplings. \s

\begin{figure}[!h]
\begin{center}
\mbox{
\includegraphics[width=0.32\linewidth,height=6.0cm]{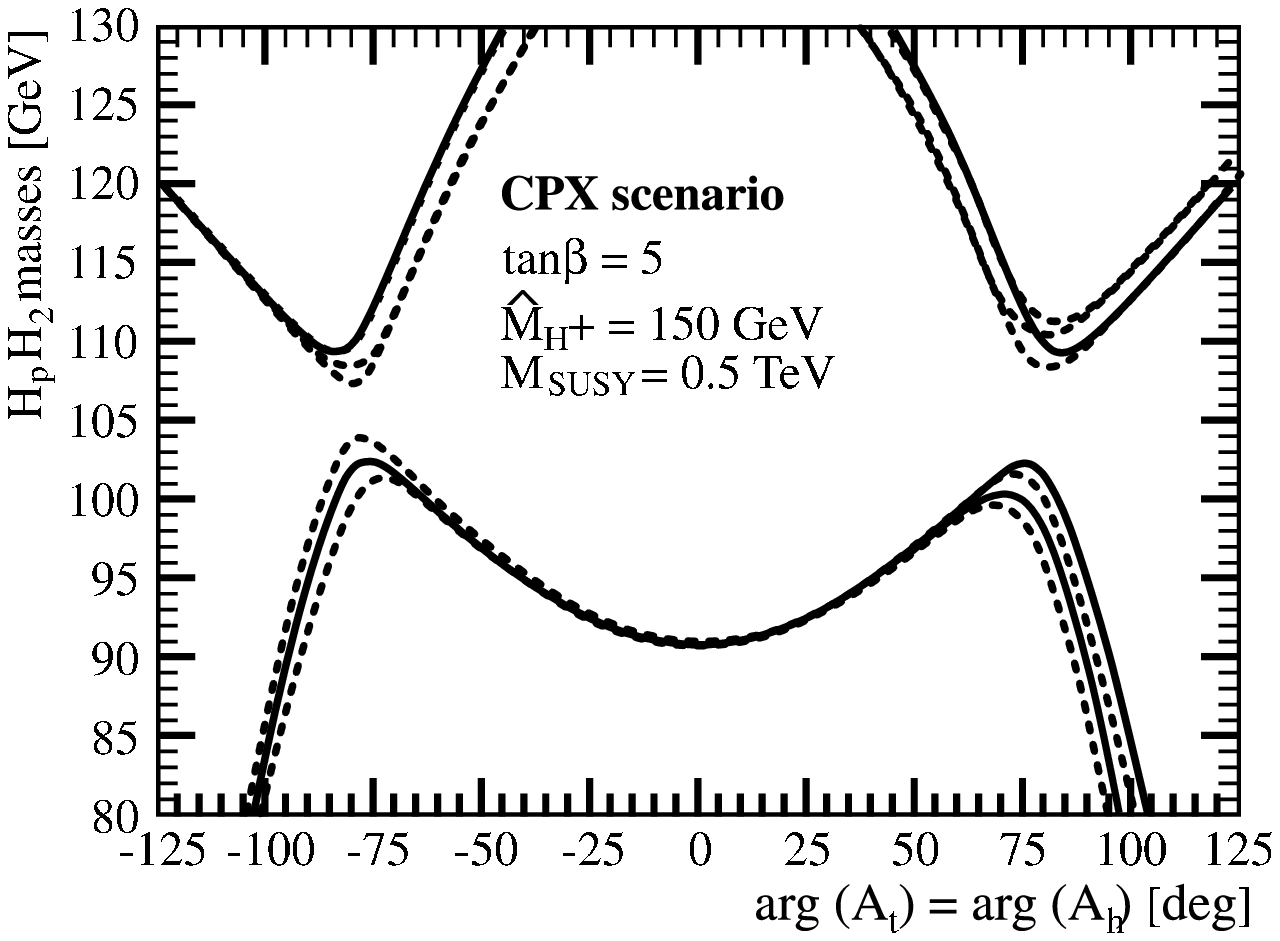}\hspace*{1mm}
\includegraphics[width=0.32\linewidth,height=6.0cm]{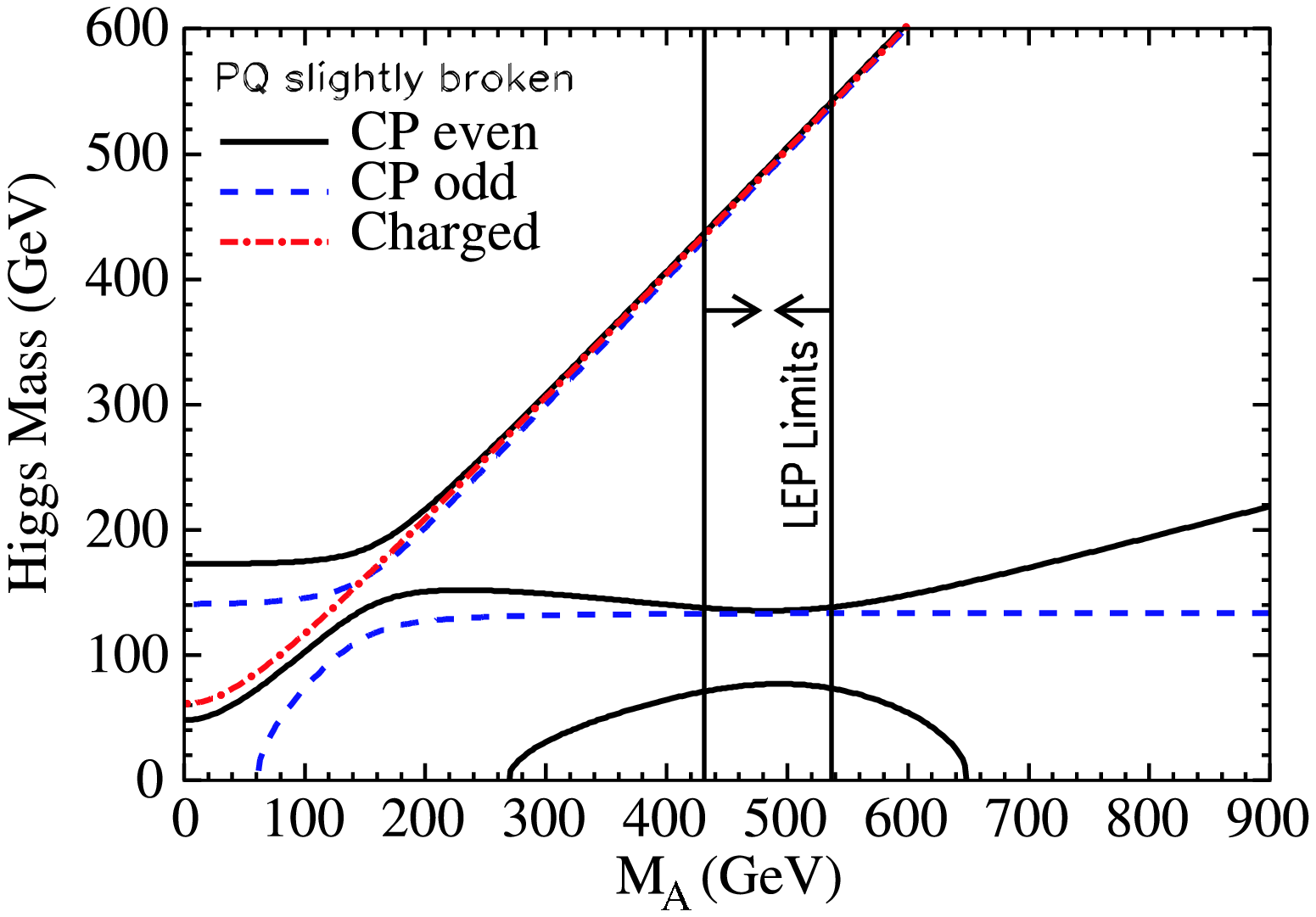}\hspace*{1mm}
\includegraphics[width=0.32\linewidth,height=6.0cm]{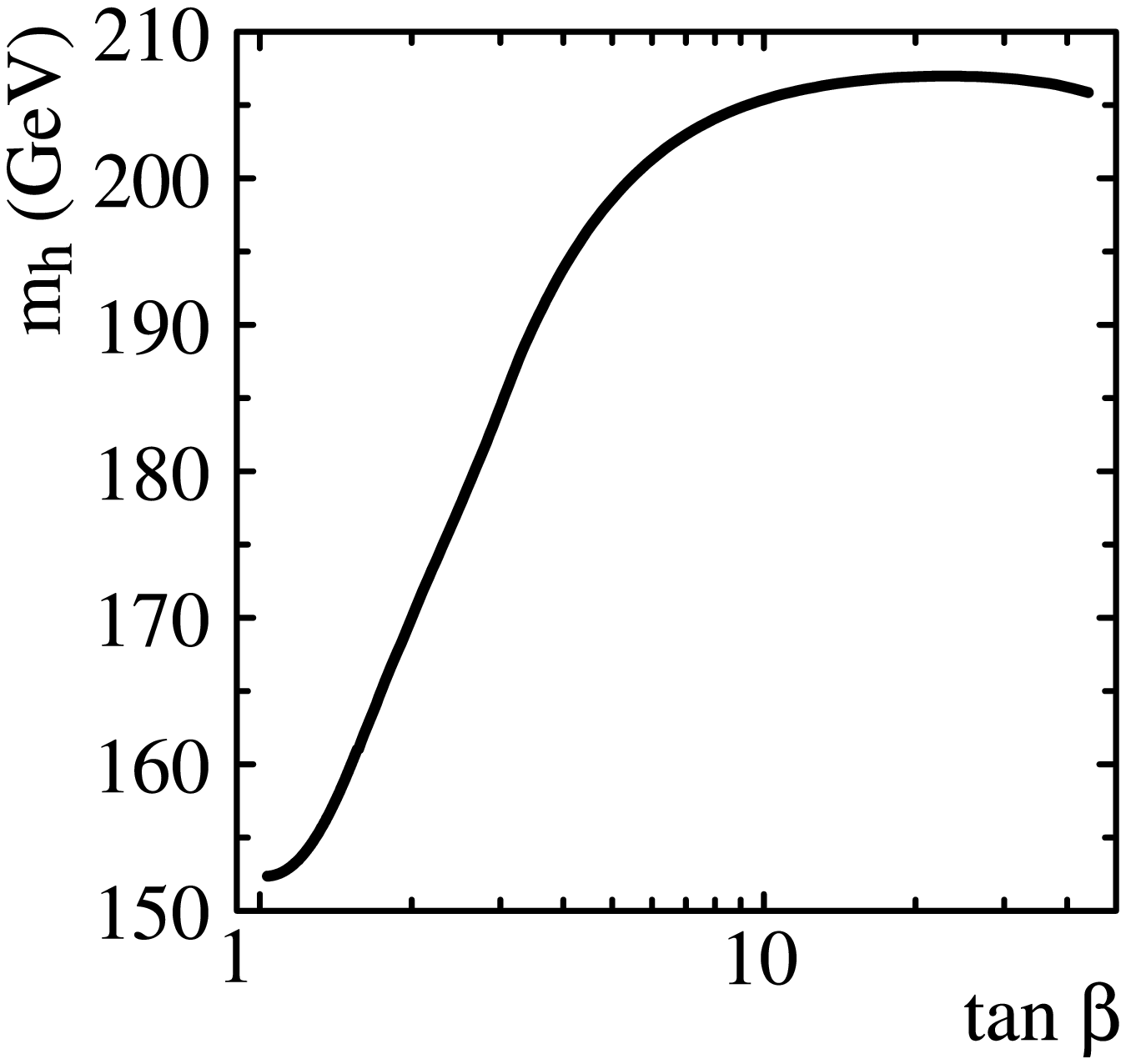}
}
\end{center}
\vspace*{-.5cm}
\caption[The spectrum of neutral Higgs particles in extensions of the MSSM.]
{The spectrum of neutral Higgs particles in a CP--violating MSSM
scenario (for $\tb\!=\!5, M_{H^\pm}\!=\!150$ GeV and $M_S\!=\!0.5$ TeV)
\cite{CPHmasses} (left) typical Higgs mass spectrum in the NMSSM  as a function
of $M_A$ \cite{HNMSSMp} (center) and the upper bound on the lighter Higgs mass
in a general SUSY model with an arbitrary number of doublets as a function of 
$\tb$ \cite{HSUSY-thbound}.}
\label{Hbeyond}
\vspace*{-1mm}
\end{figure}

An illustration  of the Higgs mass spectrum is shown in  Fig.~\ref{Hbeyond}
(left) as a function of the phase of the coupling $A_t$.  As examples of new
features compared to the usual MSSM, we simply mention the possibility of a
relatively light $H_1$ state with very weak couplings to the gauge bosons. In
this case, the cross section for $\ee \to ZH_1$ is very small and if the states
$H_2,H_3$ are heavy, all Higgs particles can escape  detection at LEP2
\cite{cpxopal}. Another interesting feature is the possibility  of resonant
$H/A$ mixing when the two Higgs particles are degenerate in mass
\cite{SUSY-CPV}.   These features have to be proven to be a result of
CP--violation. \s

\underline{The next--to--minimal SUSY extension, the NMSSM}, in which the
spectrum of the MSSM is extended by one singlet superfield, was among the first 
SUSY models based on supergravity-induced SUSY-breaking terms \cite{mSUGRA}. It
has gained a renewed interest in the last decade, since it solves in a natural
and elegant way the so-called $\mu$ problem~\cite{NMSSMt} of the MSSM; in the
NMSSM this parameter is linked to the vev of the singlet Higgs field (see
Appendix), generating a $\mu$ value close to the SUSY-breaking scale.
Furthermore, when the soft--SUSY  breaking terms are assumed to be universal at
the GUT scale, the model is very constrained as one single parameter allows to
fully describe it \cite{cNMSSM}.  The NMSSM  leads to an interesting
phenomenology as the MSSM spectrum is extended to include an additional  CP-even
and CP-odd  Higgs states as well as a fifth neutralino, the singlino. An example
of the Higgs mass spectrum \cite{HNMSSMp} is shown in Fig.~\ref{Hbeyond}
(center).  The upper bound on the mass of the lighter CP--even  particle
slightly exceeds that of the MSSM $h$ boson and the negative searches at LEP2
lead to looser constraints on the mass spectrum.\s

In a large area of the parameter space, the Higgs sector of the NMSSM reduces to
the one of the MSSM but there is a possibility, which is not completely 
excluded, that is,  one of the neutral Higgs particles, in general the lightest 
pseudoscalar $A_1$, is very light with a mass  of a few ten's of GeV. The  light
CP--even Higgs boson, which is SM--like in general, could then decay into pairs
of $A_1$ bosons, $H_1 \to A_1 A_1 \to 4b, 4\tau$, with a large branching
fraction.  The possibility of having the CP--even $H_1$ state to be as light as
$\sim 50$ GeV can also  occur: being singlino--like, it will couple very weakly
to $Z$ bosons and cannot be produced at LEP2. In this case, the SM--like Higgs
boson is $H_2$ which would decay into pairs of $H_1$ states leading mostly to
$4b$ jets,  $H_2 \to H_1 H_1 \to 4b$. \s

\underline{Higgs bosons in GUT theories.} A large variety of theories,  string
theories, grand unified theories,  left--right symmetric models, etc., suggest
an additional gauge symmetry which may be broken only at the TeV scale.  This
leads to an extended particle spectrum and, in particular, to additional Higgs
fields beyond the minimal set of the MSSM \cite{H-GUTs}. Especially common are
new U(1)' symmetries broken by the vev of a singlet field (as in the NMSSM) 
which lead to the presence of a $Z'$ boson and one additional CP--even Higgs
particle compared to the MSSM; this is the  case, for instance, in the
exceptional MSSM based on the string inspired $E_6$ symmetry. The secluded ${\rm
SU(2)\times U(1) \times U(1)'}$ model, in turn, includes four  additional
singlets that are charged under U(1)', leading to  6 CP--even and 4 CP--odd
neutral Higgs states. Other exotic Higgs sectors in SUSY models are, for
instance, Higgs representations that transform as SU(2) triplets or 
bi--doublets under the ${\rm SU(2)_L}$ and  ${\rm SU(2)_R}$ groups in
left--right symmetric models, that are motivated by the seesaw approach to 
explain the small neutrino masses and which lead e.g. to a doubly charged Higgs 
boson $H^{--}$ \cite{H++th,H:higheR}. These extensions, which also predict extra
matter fields, would lead to a very interesting phenomenology and new collider
signatures in the Higgs sector.   \s

In a \underline{general SUSY model}, one can use an arbitrary number of
isosinglet and isodoublet scalar fields to break the electroweak symmetry, while
keeping  the parameter $\rho= M_W^2/ (\cos^2\theta_W M_Z^2)$  naturally equal to
unity at the tree level as it has been verified experimentally \cite{PDG} (this
is not the case of higher representations such as triplets without  finetuning
the vevs). However, in this case, one would need an extended  matter content to 
allow for the unification of the three gauge couplings at the GUT scale. In this
general model, a linear combination of Higgs fields has to generate  the $W/Z$
masses and thus, from the triviality argument (which tells us that in the SM, the
Higgs mass should be small if the model has to be extended to the GUT scale
while leaving the quartic Higgs couplings finite), a Higgs particle should have
a mass below 200 GeV  and significant couplings to gauge bosons
\cite{HSUSY-thbound}.  The upper bound on the mass of the lightest Higgs boson
in this most general SUSY model is displayed in Fig.~\ref{Hbeyond} (right) as a
function of $\tb$.  This tell us that in supersymmetric theories, even in the
most general case, a Higgs boson should be relatively light. \s 

\underline{R--parity violating models.} in which  R--parity is spontaneously
broken (and where one needs to either enlarge the SM symmetry or the spectrum to
include additional gauge singlets),  allow for an explanation of the light
neutrino data \cite{H-RparityV}.  Since R--parity breaking  entails the breaking
of the total lepton number $L$, one of the CP--odd scalars, the Majoron $J$,
remains massless being  the Goldstone boson associated to $L$ breaking. In
these models, the neutral Higgs particles have also reduced couplings to the
gauge bosons. More importantly,  the CP--even Higgs particles can decay into
pairs of invisible Majorons, $H_i \to JJ$, while the CP--odd particle can decay
into a CP--even Higgs and a Majoron, $A_i \to H_i J$, and three Majorons,  $A
\to JJJ$ \cite{H-RparityV}.  In the decoupling regime, only  $H_1$ is light and
one would have only one accessible Higgs boson which decays invisibly.


\section{Decays of and into SUSY Higgs bosons}

In this section, we discuss the various decay modes of the Higgs particles
of the CP--conserving MSSM.
We first assume that the SUSY particles are very heavy and do not affect the
decay patterns and then, summarize the impact of light SUSY particles for both
loop and direct decays. The decays of some SUSY particles  into the MSSM Higgs
bosons and the top quark decay into charged Higgs bosons will also be briefly
discussed. But firstly, let us summarize the decay pattern of the SM Higgs
particle, which can serve as a benchmark to be confronted later with the MSSM.  

\subsection{Decays of the SM Higgs boson}

In the Standard Model, since the mass of the single Higgs boson $H$ is the only
free parameter of the theory, the profile is uniquely determined once this
parameter is fixed. In particular, the Higgs boson partial decay widths into the
various  final states and their branching fractions are fixed as the Higgs
coupling to the particles are simply proportional to their masses. The decay
modes \cite{H-LQT,Hdecay-early}  their branching ratios and the total Higgs
decay width  are summarized in Fig.~6, which is obtained using  the Fortran code
{\tt HDECAY} \cite{hdecay} mainly based on the work of Ref.~\cite{Decays}.  The
pole quark  mass values, $m_t=172$ GeV,  $m_b=4.9$ GeV and $m_c=1.64$ GeV and
$\alpha_S=0.117$  have been used as inputs \cite{PDG}. The most important
radiative corrections have been included, in particular the QCD  corrections to
Higgs decays into quark pairs,  the bulk of which  can be mapped into running
$\overline{\rm MS}$ quark masses defined at the  scale $M_H$;  the generally
small electromagnetic and weak corrections are also incorporated. In addition, 
the QCD corrections to the loop decay modes into gluons and photons are 
included. Finally, below threshold three body decays into $WW^*, ZZ^*$ and $\bar
t t^*$ final states are implemented (in fact,  the double off--shell decays of
the massive gauge bosons which then decay into massless  fermions $H\!\to\! V^*
V^*\! \to\!4f$ are incorporated); see Ref.~\cite{Decays}  \s

In the ``low mass" range, 100 GeV $\lsim M_{H}\lsim 130$ GeV,  the main decay
mode of the SM Higgs boson  is by far $ H \ra b\bar{b}$ with a branching ratio 
of $\sim \, $75--50\% for $M_H = 115$--130 GeV, followed by the decays into 
$\tau^+\tau^-$ and $c\bar{c}$ pairs with branching ratios of the order of 
$\sim$ 7--5\% and $\sim$ 3--2\%, respectively. Also of significance is the 
$H\to gg$ decay with a branching fraction of $\sim \, $7\% for $M_{H} \sim 120$
GeV. The $\gamma \gamma$ and $Z \gamma$ decays are rare, with branching ratios 
at the level of a few per mille, while the decays into pairs of muons and 
strange quarks (where $\bar{m}_s (1~{\rm GeV})=0.2$ GeV is used as input) are 
at the level of a few times $10^{-4}$. The $H \to WW^*$ decays, which are  below
the 1\% level for $M_H \sim 100$ GeV, dramatically increase with $M_H$  to reach
$\sim 30\%$ at $M_H \sim 130$ GeV; for this mass value, the mode $H\to  ZZ^*$
occurs at the percent level.\s 

In the ``intermediate mass"  range, $130 \lsim M_H \lsim 180$ GeV, the Higgs
decays mainly  into $WW$ and $ZZ$ pairs, with one virtual gauge boson below the
$2M_V$  thresholds. The only other decay mode which survives is the $b\bar{b}$
decay  which has a branching ratio that drops from 50\% at $M_H \sim 130$ GeV to
the  level of a few  percent for $M_H \sim 2M_W$. The $WW$ decay starts to
dominate at $M_H \sim 130$ GeV and becomes gradually overwhelming, in particular
for  $2M_W \lsim M_H \lsim 2M_Z$ where the $W$ boson is real (and thus $H \to
WW$ occurs at the two--body level) while the $Z$ boson is still virtual,
strongly suppressing the $H \to ZZ^*$ mode and leading to a $WW$  rate of almost
100\%.\s

In the ``high mass" range, $M_H \gsim 2M_Z$, the Higgs boson decays exclusively
into the massive gauge boson channels with a branching ratio of $\sim 2/3$ for
$WW$ and $\sim 1/3$ for $ZZ$ final states, slightly above the $ZZ$ threshold. 
The opening of the $t\bar{t}$ channel for $M_H \gsim 350$ GeV does not alter
significantly this pattern, in particular for high Higgs masses: the $H \to
t\bar{t}$ branching ratio is at the level of 20\% slightly above the $2m_t$
threshold and starts decreasing for $M_H \sim 500$ GeV to reach a level
below 10\% at $M_H \sim 800$ GeV.  The reason is that while the $H \to
t\bar{t}$ partial decay width grows as $M_H$, the partial decay width
into (longitudinal) gauge bosons increases as $M_H^3$.\s

Finally, for the total decay width, the Higgs boson is very narrow in the low
mass range, $\Gamma_{H} <10$ MeV, but the width becomes rapidly wider for
masses larger than 130 GeV, reaching $\sim 1$ GeV slightly above the $ZZ$
threshold. For larger Higgs masses,  $M_{H} \gsim 500$ GeV, the Higgs boson
becomes obese: its decay width is comparable to its mass because of the
longitudinal gauge boson contributions in the decays $H \to WW,ZZ$. For $M_H
\sim 1$ TeV, one has a total decay width of $\Gamma_H \sim 700$ GeV, resulting 
in a very broad resonant structure. \s

\begin{figure}[!h]
\begin{center}
\vspace*{-2.cm}
\hspace*{-3.cm}
\mbox{
\epsfig{file=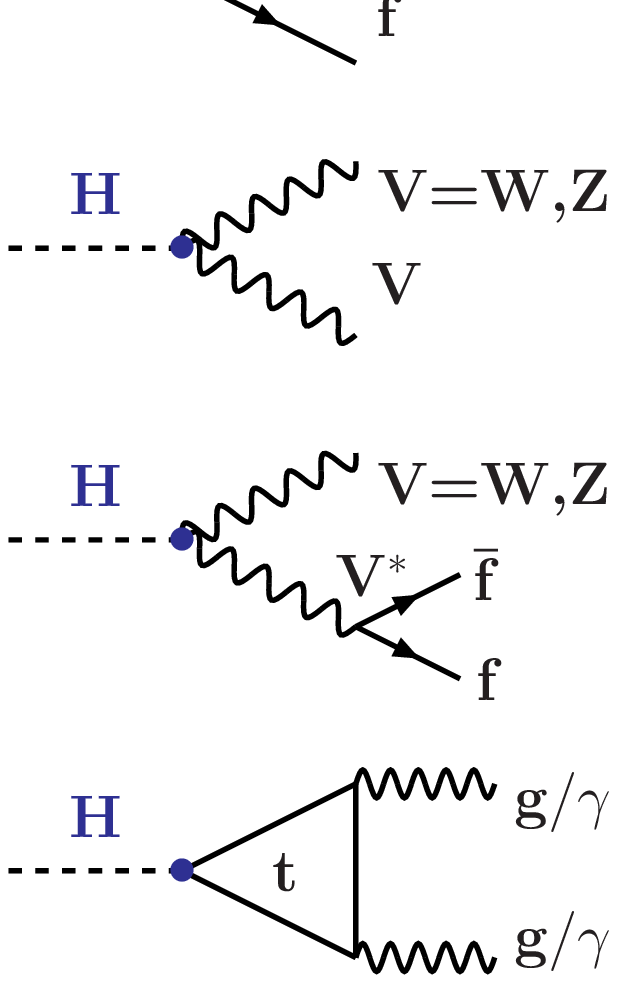,width=8cm,height=14.5cm} \hspace*{-2.7cm} 
\epsfig{file=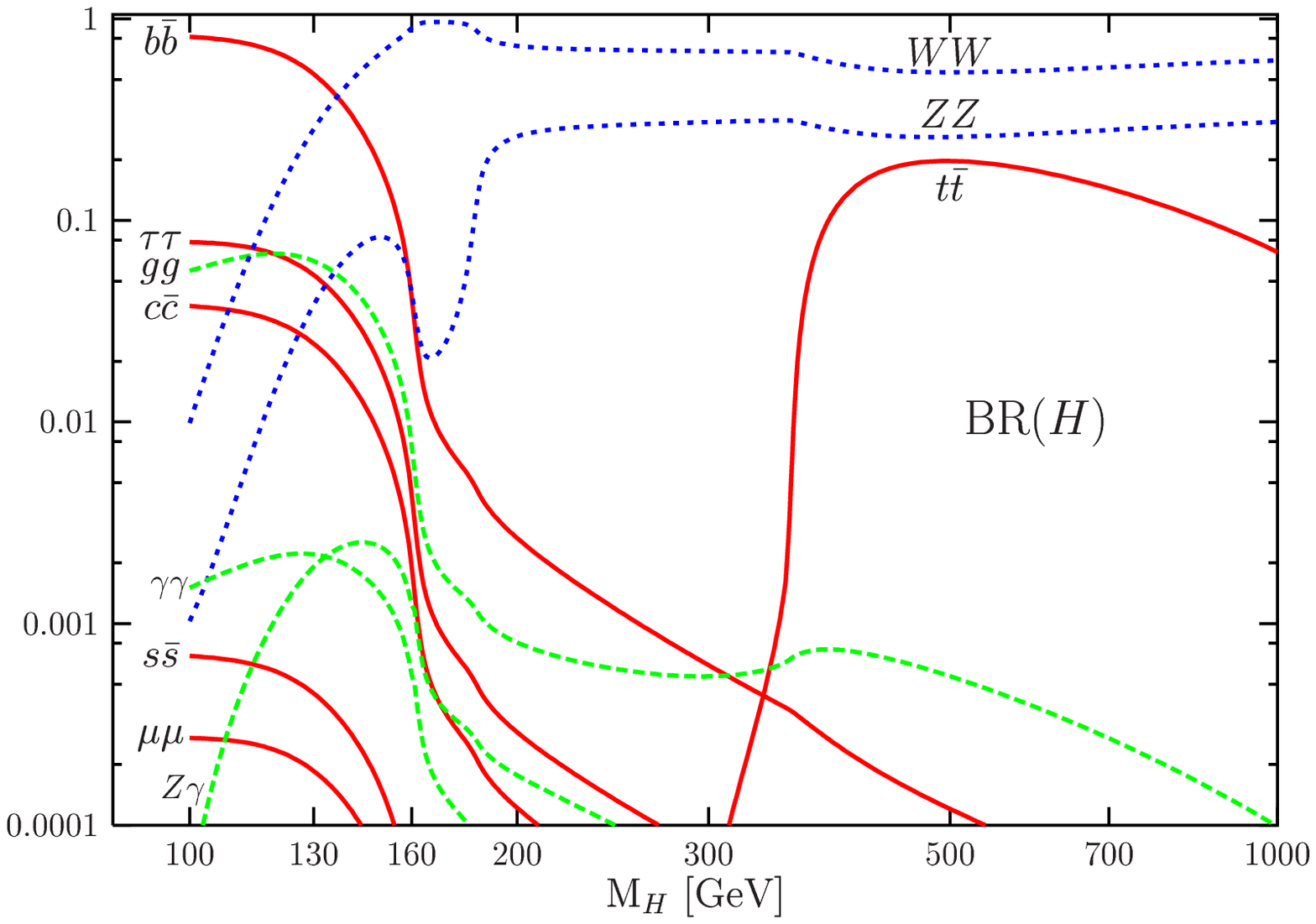,width=8cm,height=17.5cm}  \hspace*{-1.6cm} 
\epsfig{file=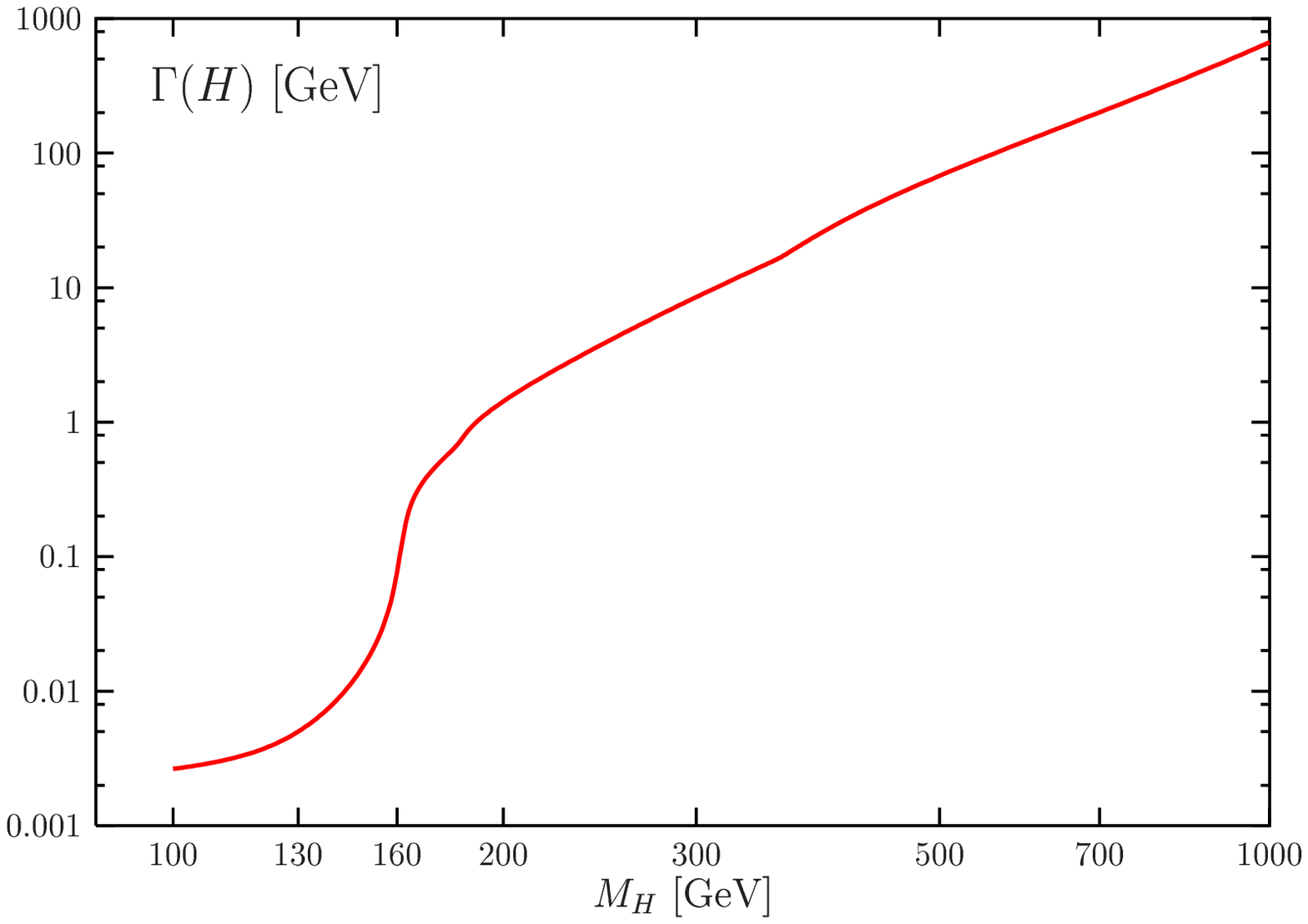,width=8cm,height=17.5cm} 
}
\end{center}
\vspace*{-9.6cm}
\caption{The main decay processes (left), the branching ratios (center) and the
total decay width (right) of the SM Higgs boson as a function of its mass, as 
obtained with {\tt HDECAY} \cite{hdecay}.} 
\vspace*{-.6cm}
\end{figure}

\subsection{Decays of the MSSM Higgs bosons}

In the decoupling regime, $M_A \gsim 150$ GeV for $\tb=30$ and $M_A \gsim 400
$--500 GeV for $\tb=3$, the situation is quite simple; Fig.~7. The lighter $h$
boson reaches its maximal mass value and has SM--like couplings and, thus,
decays as the SM Higgs boson discussed previously.  Since $M_h^{\rm max}\lsim
130$ GeV, the dominant modes are the decays into $b\bar b$ pairs and into $WW^*$
final states, the branching ratios being of the same size in the upper mass
range.   The decays into $\tau^+\tau^-, gg, c\bar c$ and also $ZZ^*$ final
states are at the level of a few percent and the loop induced decays into
$\gamma \gamma$ and $Z\gamma$ at the level of a few per mille. The total decay
width of the $h$ boson is small, $\Gamma (h) \lsim {\cal O}$(10 MeV). \s

For the heavier Higgs bosons, the decay pattern depends on $\tb$. For $\tb \gg
1$, as a result of the strong enhancement of the couplings to down--type
fermions, the  $H$ and $A$ bosons will decay almost exclusively into $b\bar{b}$
($\sim 90\%$) and $\tau^+ \tau^-$ ($\sim 10\%)$ pairs; the $t\bar t$ decay when
kinematically allowed and all other decays, including the $H \to VV^{(*)}$ 
modes, are strongly suppressed.  The $H^\pm$ boson decays mainly into $tb$ pairs
but there is also a a significant fraction  of $\tau \nu_\tau$ final states
($\sim 10\%$). For low values of $\tb$, the decays of the neutral Higgs bosons
into $t\bar t$ pairs and the decays of the charged Higgs boson in $tb$ final
states are by far dominating. For intermediate values, $\tb \sim 10$, the rates
for the $H,A \to b\bar b$ and $t \bar t$ decays are comparable, while the $H^\pm
\to \tau \nu$ decay stays at the 10\% level. For small and large $\tb$ values,
the total decay widths of the four Higgs bosons are, respectively,  of ${\cal
O}$(1 GeV) and of ${\cal O}$(10 GeV) and thus not large. This is because the
decay modes into $W$ and $Z$ bosons are absent or strongly suppressed, contrary
to the SM case.\s

\begin{figure}[!h]
\begin{center}
\vspace*{-1.7cm}
\hspace*{-2.2cm}
\mbox{
\epsfig{file=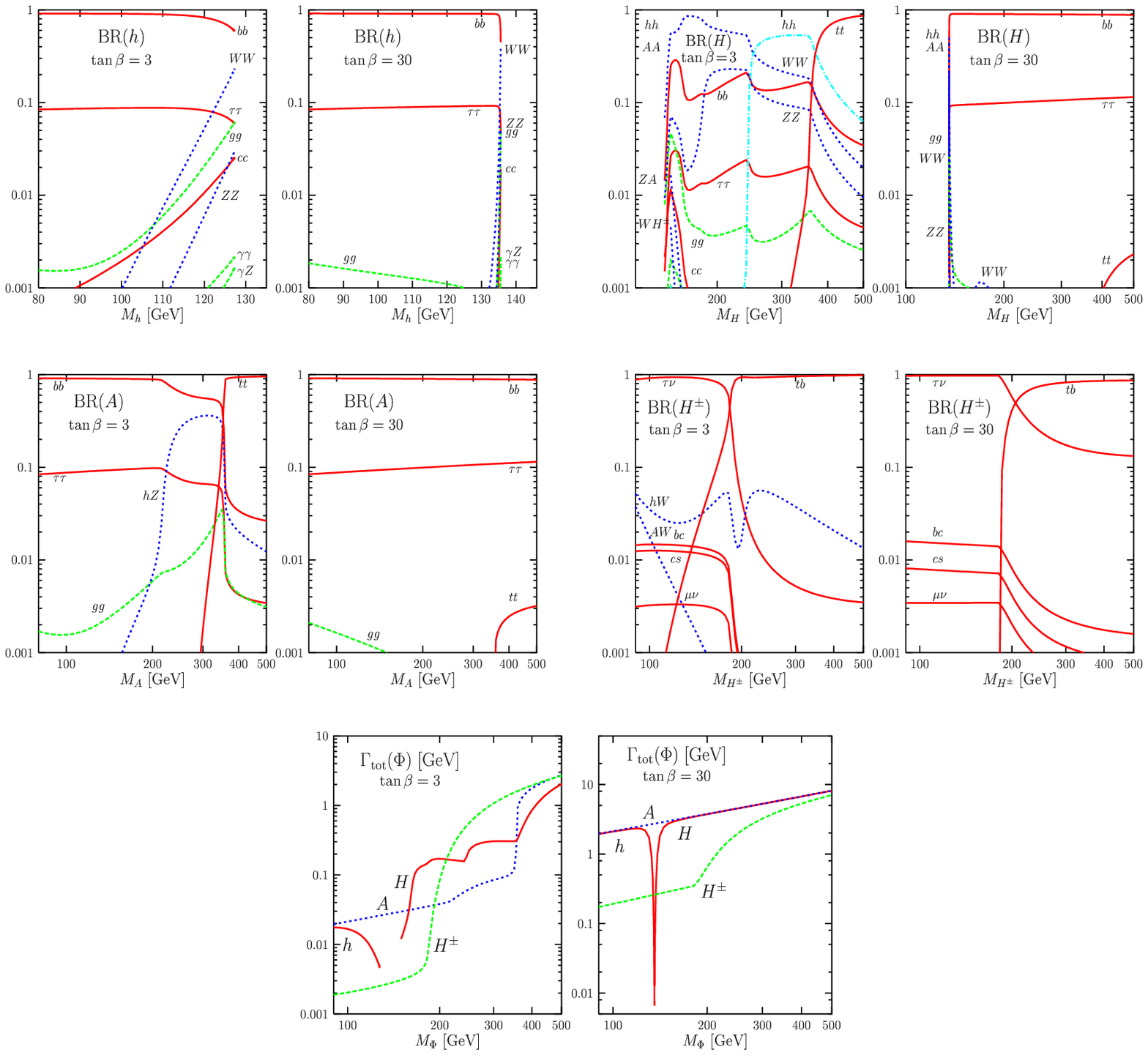,width=19cm} }
\end{center}
\vspace*{-11.3cm}
\caption{The decay branching ratios and total widths of the MSSM Higgs bosons 
as functions of their masses for $\tb=3,30$ in the maximal mixing scenario as 
obtained with {\tt HDECAY} \cite{hdecay} with the inputs of Fig.~1; the
radiative corrections \cite{FeynHiggs} and the three-body decays are included.} 
\vspace*{-.3cm}
\end{figure}




Outside the decoupling regime, the decay pattern can be summarized as follows:\s
 
-- In the anti--decoupling regime, i.e. when $\tb \gsim 10$ and $M_A \lsim M_h^
{\rm max}$, the pattern for the Higgs decays is also rather simple. The $h$ and
$A$ bosons will mainly decay into $b\bar{b}$ ($\sim 90\%$) and $\tau^+ \tau^-$
($\sim 10\%)$ pairs, while the charged $H^\pm$ boson decays almost all the time
into $\tau \nu_\tau$ pairs ($\sim 100\%)$.  All other modes are suppressed down
to a level below $10^{-3}$ except for the gluonic decays of $h$ and $A$  [in
which the $b$--loop contributions are enhanced by the same $\tb$ factor] and
some fermionic decays of $H^\pm$. Although their masses are small, the three
Higgs bosons have relatively large total widths, $ \Gamma (h,A,H^\pm) \sim {\cal
O} $(1 GeV) for $\tb=30$. The heavier $H$ boson will play the role of the SM
Higgs boson, but with one major difference: in the low $M_A$ range (which is now
excluded by LEP2 searches), the $h$ and $A$ particles are light enough for the
two--body decays $H\to hh$ and $H \to AA$ to take place and to dominate with  a
branching fraction of $\sim 50\%$ each. These decays can be very important in
some extensions such  as the CP--violating MSSM and the NMSSM. \s 

-- In the intense--coupling regime, with  $\tb \gsim 10$ and $M_A \sim 100$--140
GeV, the couplings of both $h$ and  $H$  to gauge bosons and up--type fermions
are suppressed and those  to down--type fermions are enhanced. Because of this
enhancement,   the branching ratios of the $h$ and $H$ bosons to $b\bar{b}$ and
$\tau^+\tau^-$ final states are the dominant ones, with values as in the
pseudoscalar Higgs case, i.e. $\sim 90$\% and $\sim 10$\%,  respectively. The
interesting rare decay mode into $\gamma \gamma$  is very strongly suppressed
for the three neutral Higgs particles compared to the SM. The  branching ratios
for the decays into muons, which are not displayed in Fig.~7 are at the  level
of $3 \times 10^{-4}$. The $H^\pm$  boson in this scenario  decays mostly into
$\tau\nu$ final states. \s

-- In the intermediate--coupling regime, i.e. for  $\tb\!\sim\!3$ and $H/A$
masses below the $t\bar t$ threshold,  interesting decays of the $H,A$ and
$H^\pm$ bosons occur. For the pseudoscalar $A$, the decay $A \to hZ$  is
dominant when kinematically accessible, i.e. for $M_A \gsim 200$ GeV, with a
branching ratio exceeding the 50\% level. In the case of $H$, the channel  $H\to
hh$ is very important, reaching the level of 60\% in a significant $M_H$ range;
the decays into weak vector bosons and $b\bar b$ pairs are also significant. 
For the $H^\pm$ boson, the interesting decay $H^\pm \to hW^\pm$ is at the level
of a few percent while the other decay $H^\pm \to AW^\pm$  is kinematically
challenged and occurs at the three--body level. \s

-- Finally, for the choice of input SUSY parameters of Fig.~7, the vanishing
coupling regime does not occur. However, when Higgs couplings to bottom quarks
and $\tau$ leptons accidentally vanish, the outcome is rather clear. For the $h$
boson for instance,  the $WW^*$ mode becomes the dominant one, followed by the
loop induced $h \to gg$ decay; the interesting $h\to \gamma \gamma$ decay mode 
is enhanced but stays below the permille level.

\subsection{The impact of light SUSY particles}

In the preceeding  discussion, we have assumed that the SUSY particles are too
heavy to substantially contribute to the loop induced decays of the neutral 
Higgs bosons and to the radiative corrections to the tree--level decays. In
addition, we have ignored the Higgs decay channels into sparticles which were
considered as being kinematically shut.  However, some SUSY particles such as
the charginos, neutralinos and possibly sleptons and third generation squarks,
could be light enough to play a significant role in this context. We thus
summarize their possible impact. \s

In the case of Higgs decays into $b$ quarks, besides the radiative corrections
to the Higgs masses and the angle $\alpha$, there are large direct corrections,
eq.~(\ref{ghff:threshold}). The corrections generate a strong variation of the
$b\bar b$ partial widths of the three neutral Higgs bosons which can reach the
level of 50\% for large $\mu$ and $\tb$ values, and not too heavy squarks and
gluinos.  However, they have only a small impact on the $b\bar b$ rates since
these decays dominate in general.  In turn, they can have a large influence on
the rates for the other decay modes, in particular, on the $\tau^+ \tau^-$
channels. This can be seen in Fig.~8 (left) where the rates of $h,H,A$ decays
into  $b\bar b$ and $\tau^+\tau^-$  are shown for $\tb=30$;  variations of
BR$(\tau^+\tau^-)$ by a factor of two can be noticed. In the case of the $H,A$
bosons with masses above the $t\bar t$ threshold and for intermediate $\tb$
values when the $b\bar b$ and $t\bar t$ channels compete with each other, these
corrections can be felt by both the $H/A \to b\bar b$ and $t\bar t$ rates. The
same  features occur in the case of the $H^\pm$  boson decaying into $tb$ and
$\tau \nu$ final states.\s

\begin{figure}[!h]
\begin{center}
\vspace*{-2.2cm}
\hspace*{-5.cm}
\mbox{ \hspace*{-1.2cm}
\epsfig{file=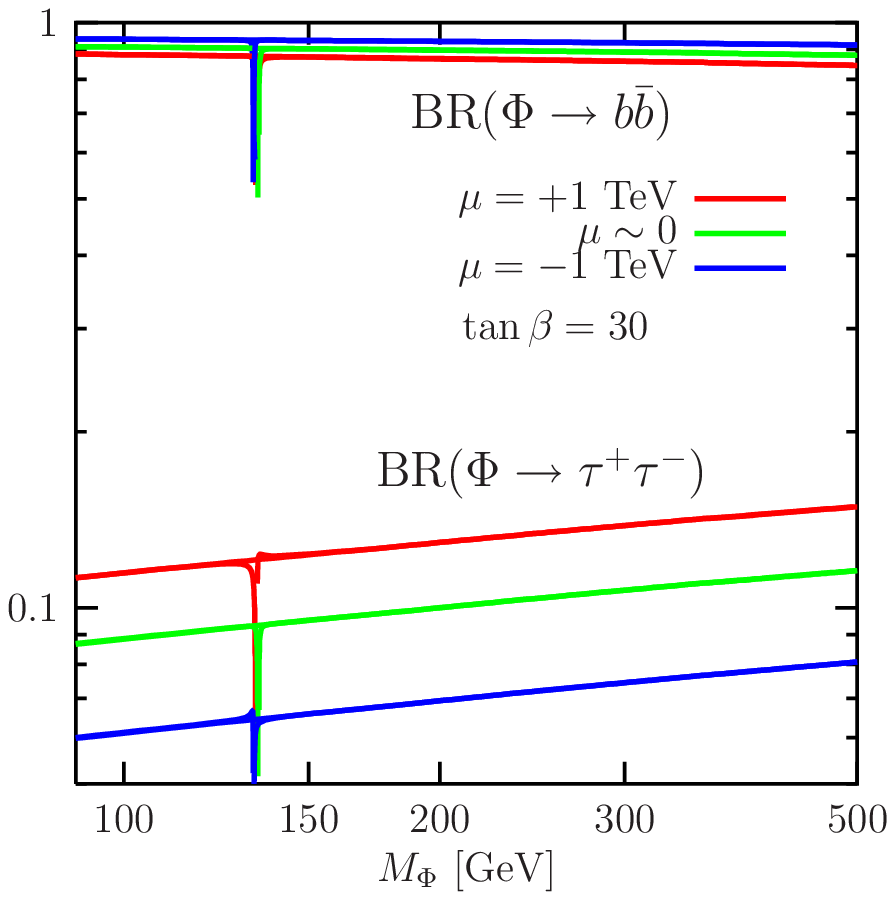,width=13cm} \hspace*{-7.3cm}
\epsfig{file=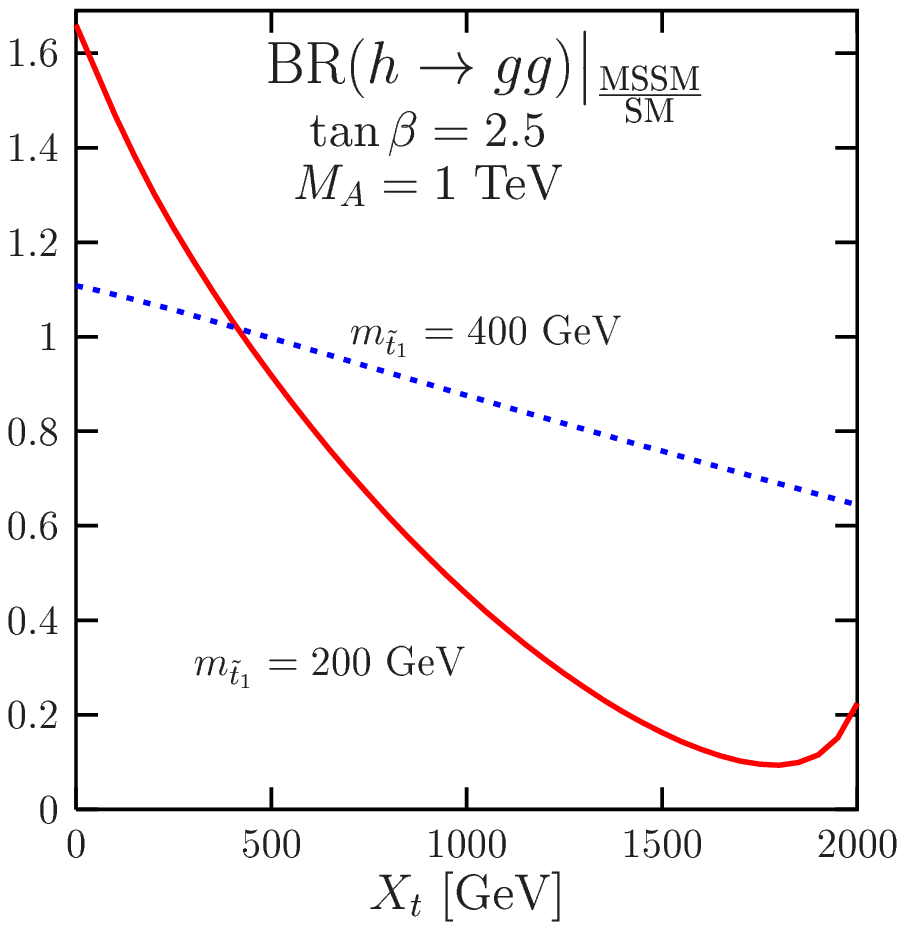,width=13cm}   \hspace*{-7.5cm}
\epsfig{file=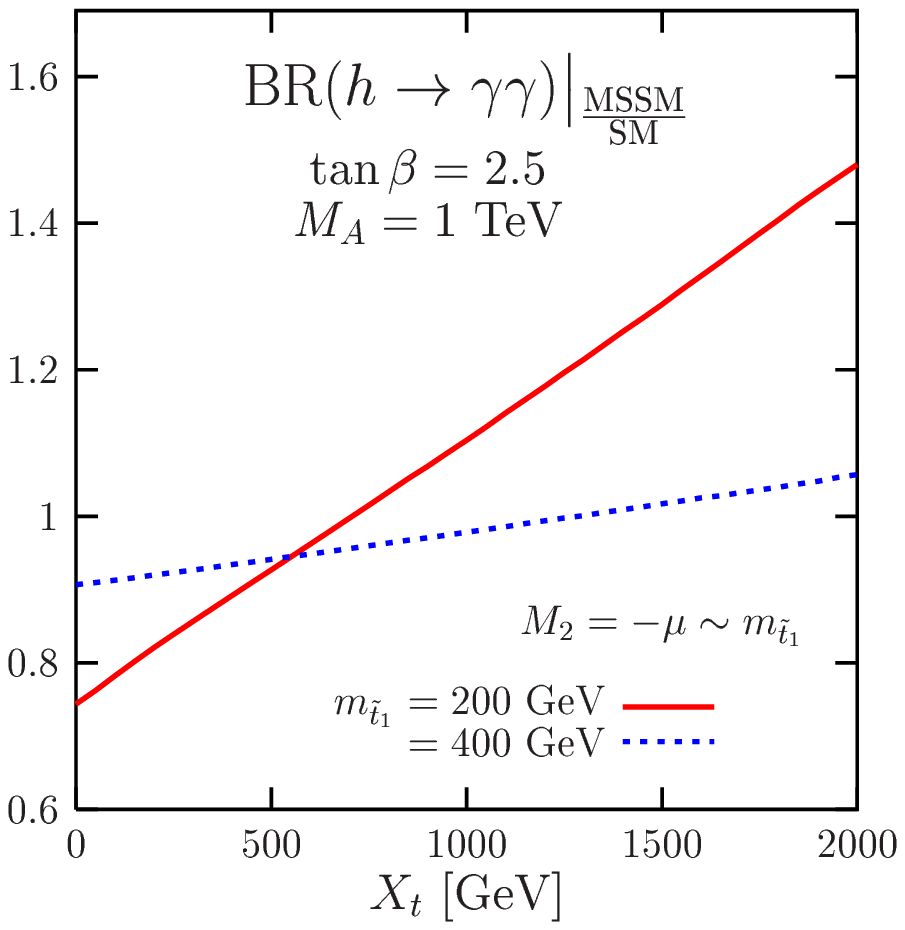,width=13cm}  \hspace*{-5cm}
}
\end{center}
\vspace*{-11.2cm}
\caption[]{The branching ratios: for  $h,A,H \to b\bar b$ and $\tau^+\tau^-$ for 
$\tb=30$  with/without  the SUSY--QCD corrections  \cite{Anatomy2} (left) and  
for the gluonic (center) and photonic (right) decays of the $h$ boson in the
decoupling limit relative to their SM values including SUSY loops with
$\tb=2.5$ \cite{SUSYloops}.}
\vspace*{-.3cm}
\end{figure}

If squarks are relatively light, they can lead to sizable contributions to the 
loop induced decays $h,H \to gg$ and $\gamma \gamma$; due to  CP--invariance
which forbids $A$  couplings to identical $\tilde q_i \tilde q_i$ states, squark
loops do not contribute to $A\to gg, \gamma \gamma$.  Since squarks have Higgs
couplings  that are not proportional to their masses, their contributions are
damped by loop factors $1/m_{\tilde Q}^2$ and, contrary to SM quarks, the
contributions become very small at high $m_{\tilde Q}$ and the sparticles
decouple completely from the vertices. However, when $m_{\tilde Q} \sim
M_{h,H}$,  the contributions can be significant \cite{SUSYloops}.  This is
particularly true in the case of top squarks in the decays $h \to gg$,  the
reason being  two--fold: $(i)$ the  $\tilde t$ mixing, $\propto m_t X_t$,  can
be very large   and could lead to $\tilde{t}_1$ that is much  lighter than all 
other  squarks and even the top quark, and $(ii)$ the coupling of top squarks to
the $h$ boson  involves a component which is proportional to $m_t X_t$ and for
large  $A_t$,  it can be strongly enhanced. Sbottom mixing, $ \propto m_b X_b$,
can also be sizable for large $\tb$ and $\mu$ values and can lead to light
$\tilde b_1$ states with strong couplings to the $h$ boson. Besides, chargino
loops enter also the $h,H,A \to \gamma \gamma$ decays but their contributions is
in general smaller  since the Higgs$\chi \chi$ couplings are not strongly
enhanced. \s 

Figure 8 shows the deviations of the gluonic and photonic  widths of the $h$
boson, relative to their SM values, as a result of $\tilde t$  contributions. 
In the case of $h\to gg$, the partial width can be reduced by an order of
magnitude for light stops and large $X_t$ mixing. For  the $h\gamma \gamma$
coupling, as the interference can be either positive or negative, the  rate can
be  increased by more than 50\% or slightly suppressed. Chargino loops in
$h\gamma \gamma$ contribute less than 10\%.  Note that for the $H gg$ and
$H\gamma \gamma$ couplings, SUSY effects might  be larger  as the $H$ boson  and
loop masses can be comparable; however, in this case, both the photonic and
gluonic branching  ratios are too small. \s

Let us now turn to decays of the MSSM Higgs bosons into SUSY particles 
\cite{HaberGunion2,SUSYdecays,H-chi-inv,H-sfermions} and start with decays into
charginos and neutralinos, collectively called inos.  The sum of the branching
ratios for the Higgs  decays into all possible combinations of ino states are
shown in Fig.~9 as a function of the Higgs masses for the values $\tb\!=\!3,30$
for $H,A$ and $H^\pm$ and $\tb\!=\!10$ for $h$. To allow for such decays, we
have departed from the benchmark of Fig.1, to adopt a scenario in which we have
still $M_S =2$ TeV with maximal stop mixing, but where the parameters in the
gaugino sector are $M_2\!=\!-\mu = 150$ GeV. Here, the universality of the
gaugino masses at the GUT scale, giving  $M_2 \sim 2M_1$ at low scales, is
assumed  while $M_3$ is  still large. This choice leads to rather light ino
states, $m_{\chi_i} \lsim 200$--250 GeV  which still satisfy the LEP  bound,
$m_{\chi_1^\pm} \gsim 100$ GeV \cite{PDG}.\s

\begin{figure}[!h]
\begin{center}
\vspace*{-2.1cm}
\hspace*{-5.cm}
\epsfig{file=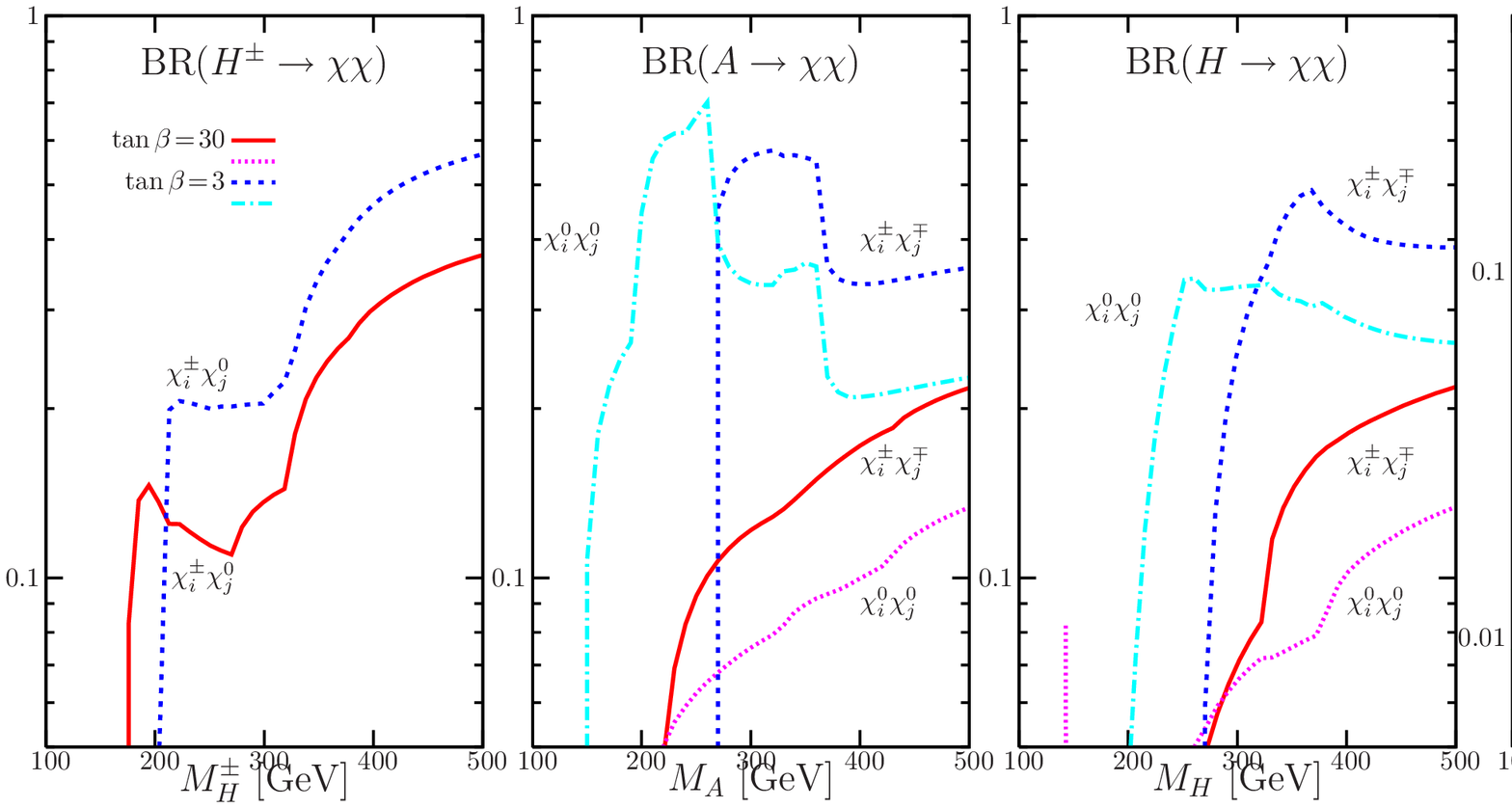,width=14cm} 
\end{center}
\vspace*{-11.4cm}
\caption[]{The branching ratios for the MSSM $H,A,H^\pm\, (h)$ decays into the
sum of charginos and/or neutralinos as a  function of their masses for $\tb\!=
\!3,30(10)$ \cite{Anatomy2}. The relevant SUSY parameters are  $M_S=2$  TeV and
$M_2\!=\!-\! \mu\!=\!150$ GeV and for $h$, the relation $M_2\!\sim\!2M_1$ is
relaxed.}
\vspace*{-.1cm}
\end{figure}

In general, for the heavy $H,A,H^\pm$ states, the sum of these branching ratios
is always large except in a few cases: ($i$) for small $M_A$  when the phase
space is too penalizing and does not allow for the decay into (several) inos to
occur; $(ii$) for the $H$ boson in the mass range $M_H\sim 200$--350 GeV and
small $\tb$ values when the decay $H \ra hh$ is largely dominant; and ($iii$)
for $H^\pm$ just above the $t\bar{b}$ threshold if not all ino  decay channels
are open.  In fact, even above the thresholds of Higgs decays into top quarks 
and/or large $\tb$ values, the decays into inos can be important:  for very
heavy  Higgs bosons, they reach a common value of $ 30\%$ for low $\tb \sim 2$
and large  $\tb \sim 30$ and are dominant for moderate values $\tb \sim  10$
when the Higgs--$b \bar b$  couplings are not yet strongly enhanced. Note that
when kinematically open,   neutral Higgs decays into charginos dominate over
those into neutralinos, as the charged couplings are larger than  the  neutral
ones.\s

The  bound $m_{\chi_1^\pm} \gsim 100$ GeV  does not allow for ino decay modes of
the lightest $h$ boson since $M_h \lsim 140$ GeV, except for the invisible
decays into a pair of the lightest neutralinos, $h \to \chi_1^0\chi_1^0$
\cite{SUSYdecays,H-chi-inv}. This is particularly true when the universality relation 
$M_2 \sim 2M_1$  is relaxed leading to light LSPs while the bound on
$m_{\chi_1^\pm}$ is respected \cite{H-chi-inv}.  In general, when the $h \to
\chi_1^0\chi_1^0$ decay is kinematically allowed, the branching ratio is sizable
only in the decoupling regime (where the $hbb$ couplings are not enhanced) and
for mixed higgsino--gaugino states (which maximizes the $h\chi \chi$ couplings).
Figure 9 (right) shows that the rate can  exceed the 10\% level in this case.
\s 

Another possible decay channel for the heavy $H,A,H^\pm$ bosons is into
sfermions; for the $h$ boson, these decays are kinematically closed as
$m_{\tilde f}\gsim 100$ GeV from LEP and  Tevatron searches. The decays into 
first/second generation sfermions are  marginal, as the Higgs  couplings to
these states are small, while those into third generation sfermions, can be more
important \cite{H-sfermions}. For instance,  $H$ decays  into light top squarks
can be significant and even dominant if the $H\tilde t_1 \tilde t_1$ coupling is
enhanced. Mixed $H,A \to \tilde t_1 \tilde t_2$ and $H^+ \tilde t_1 \tilde b_1$
decays can also be significant when phase space allowed. Decays of the Higgs
bosons into tau sleptons, which are more favored by phase space, can also  be
sizeable  but they have to compete with decays into $b\bar b$ which are strongly
enhanced at large $\tb$. This is also the case for the decays involving $\tilde
b$ squarks in the final state. 

\subsection{Decays of the sparticles and the top quark into Higgs bosons }

Let us now briefly comment on a related issue which is the decays of SUSY
particles into Higgs bosons \cite{Cascade0pp,cascade}. If the mass splitting
between the heavier $\chi_{3,4}^0, \chi_2^\pm$ chargino/neutralino states and
the lighter $\chi_{1,2}^0, \chi_1^\pm$ states is substantial, the heavier inos
can decay into the lighter ones and neutral and/or charged Higgs bosons, 
$\chi_2^\pm, \chi_3^0, \chi_4^0 \to  \chi_1^\pm, \chi_2^0, \chi_1^0 +  h , H, A
, H^\pm$. In fact, even the next--to--lightest neutralino can decay into the LSP
neutralino and a neutral Higgs boson and the lighter chargino into the LSP and a
charged Higgs boson, $\chi_2^0 \to   \chi_1^0 + h , H, A$ and $\chi_1^\pm \to  
\chi_1^0 + H^\pm$. \s

These decay processes will be in direct competition with decays into  gauge
bosons and, if sleptons/squarks are light, decays into sfermions and fermion
partners. The decay branching ratios of the heavier  $\chi_2^\pm$ and
$\chi_{3}^0$ states into the lighter ones $\chi_1^\pm$ and $\chi_{1,2}^0$  and 
Higgs bosons are shown in Fig.~10 for $\tb=10$ and $M_A=180$  GeV with $\mu=150$
GeV,  which means that the lighter inos are higgsino like. The other parameter
$M_2$ is varied with the mass of the  decaying ino. Sleptons and squarks are
assumed to be too heavy to  play a role here. Since the Higgs bosons couple
preferentially to mixtures of gauginos and higgsinos, the couplings to
mixed heavy and light chargino/neutralino states are maximal. To the contrary, 
the gauge boson couplings to inos are important only for higgsino-- or
gaugino--like states.  Thus, in principle, the (higgsino or gaugino--like)
heavier inos $\chi_2^\pm$  and $\chi_{3,4}^0$ will dominantly decay, if phase
space allowed, into Higgs  bosons and the lighter $\chi$ states. As is usually
the case, the charged current decay modes will be more important than the
neutral modes.  A similar pattern occurs for large values of $\mu$ compared to
$M_2$ in which case the light (heavy) inos are gauginos (higgsinos). \s

\begin{figure}[!h]
\begin{center}
\vspace*{-.9cm}
\hspace*{-3.cm}
\epsfig{file=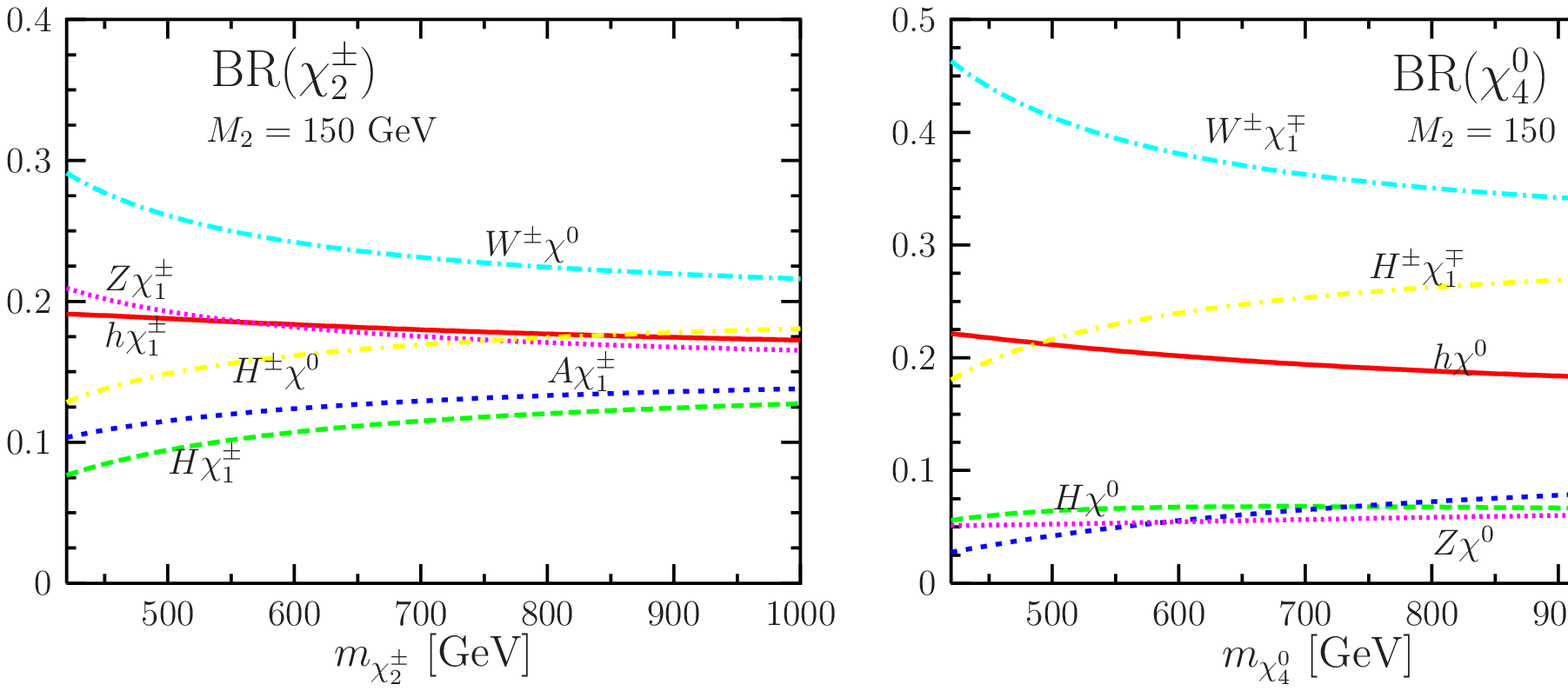,width=15cm} 
\end{center}
\vspace*{-15.3cm}
\caption[]{The branching ratios for the decays of $\chi_2^\pm$ and $\chi_4^0$
into Higgs and gauge bosons for $\tb=10, M_A=180$  GeV and $\mu=150$ GeV, as a
function of the ino masses and hence $M_2$ \cite{cascade}. Similar features
occur for the decays of $\chi_3^0$  and when $\mu$ and $M_2$ are interchanged.}
\vspace*{-.2cm}
\end{figure}

Another potentially large source of Higgs bosons comes from the decays of
sfermions \cite{H-sfermions}. If the mass splitting between two squarks of the
same generation is large enough, as is generally the case of the
$(\tilde{t},\tilde{b})$ isodoublet, the heavier squark can decay into the
lighter one plus a neutral or charged Higgs boson, a channel which will compete
with the usually dominant modes into  quarks and charginos or neutralinos. This
is particularly the case  for the $\tilde t_2 \to \tilde t_1 + h/H/A$ decays
which can have a  substantial rate for moderate to large $X_t$ values which
enhance the Higgs--$\tilde{t}_1\tilde{t}_2$ coupling.\s

Finally, another important source of relatively light charged Higgs bosons, 
$M_{H^\pm} \lsim m_t$,  comes for the decays of the heavy top  quark, $t \to H^+
b$ \cite{top-toH+}.  The couplings of the $H^\pm$ bosons to $tb$ states  are
proportional to the  combinations $m_b \tb (1+\gamma_5) + m_t {\rm cot}\beta 
(1-\gamma_5)$.  They are thus strong enough for small $\tb \sim 1$ or large $\tb
\gsim 30$  values to make this decay compete with the standard $t\to bW^+$
channel, the only relevant mode otherwise. For intermediate values of $\tb$, the
$t\, (b)$--quark component of the coupling is suppressed (not too strongly 
enhanced yet) and the overall couplings is small; the minimal value  occurs at 
$\tb= \sqrt{m_t m_b} \sim 6$. The $t \to b H^+$ branching ratio is shown in 
Fig.~11 as a function of the  $H^\pm$ mass for three values, $\tb=3,10$ and 30. 
One notices the small value of the rate at intermediate $\tb$, while it  exceeds
the level of a few percent for $\tb=3$ and 30. There also a clear suppression
near the threshold: for $M_{H^\pm} \gsim 160$ GeV, the branching ratio being
below the per mille level even for $\tb=3$ and $30$.  \s

\begin{figure}[!h]
\begin{center}
\vspace*{-1.4cm}
\hspace*{3.cm}
\epsfig{file=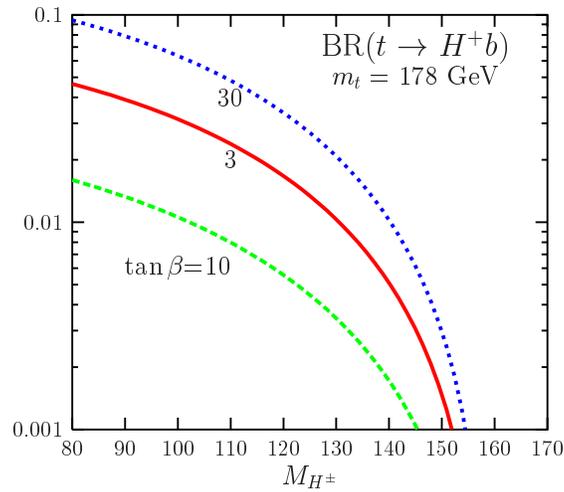,width= 15.cm} 
\end{center}
\vspace*{-14.2cm}
\caption[]{The branching ratio for the decay of the top quark into
a charged Higgs boson and a bottom quark as a function of $M_{H^\pm}$ for three
values $\tb=3,10$ and 30. } 
\vspace*{-.3cm}
\end{figure}

\section{SUSY Higgs production at the LHC}

As in the previous section, we will first summarize the salient features of SM
Higgs  production at the LHC and then then discuss the main differences for MSSM
Higgs production. We first assume  that the SUSY particles are heavy and  then
emphasize impact of light SUSY particles. 

\subsection{Production of the SM Higgs particle}

There are essentially four mechanisms for the single production of the SM Higgs
boson at hadron colliders\footnote{Another possibility would be diffractive
Higgs production; see Ref.~\cite{Diffractive} for a recent and detailed review.}
\cite{P1:Hgg,P1:HV,P1:VVH,P1:ttH}, the Feynman
diagrams of which are shown in the left-hand side of Fig.~12. The total
production cross sections, as obtained with the Fortran programs of
Ref.~\cite{Michael} and the SM inputs used for the Higgs decays in the SM, are
displayed in the center of Fig.~12 for the LHC with a center of mass energy
$\sqrt{s}=14$ TeV as a function of the Higgs mass.  The MRST parton
distributions functions \cite{MRST} have been adopted and the next-to--leading
order (NLO), and eventually the next-to-NLO (NNLO), radiative corrections  have
been implemented \cite{Anatomy1,Spira,Harlander} as will be summarized later
when the main features of each production channel will be discussed. The
significance for detecting the Higgs particle in the various production and
decay channels is shown in  the  right-hand side of Fig.~12, assuming a 100
fb$^{-1}$ integrated luminosity. \s

\begin{figure}[!h]
\begin{tabular}{ll} 
\begin{minipage}{10cm}
\vspace*{-2.2cm}
\centerline{\hspace*{-1.cm}
\includegraphics[scale=0.48]{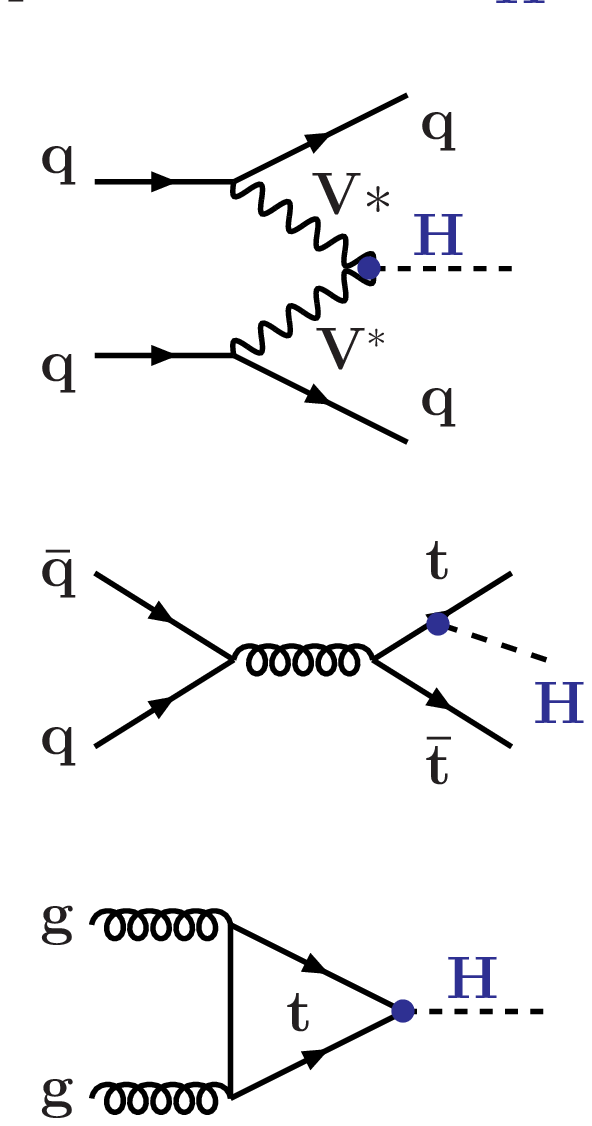} \hspace*{-6.6cm}
\includegraphics[scale=0.6]{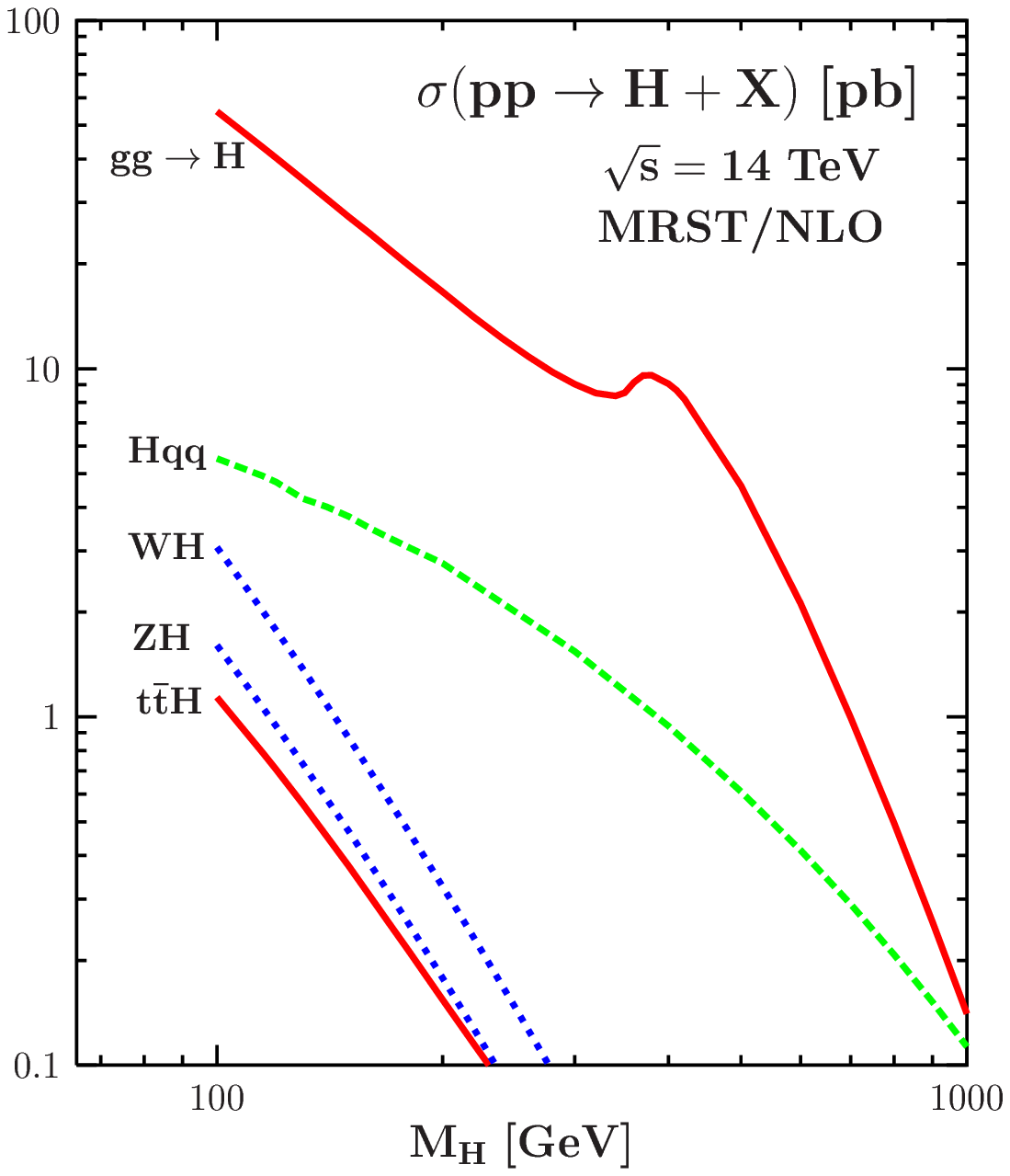} \hspace*{.6cm}
}
\vspace*{-8.7cm}
\end{minipage}
&
\begin{minipage}{6cm}
\centerline{ 
\includegraphics[scale=0.42]{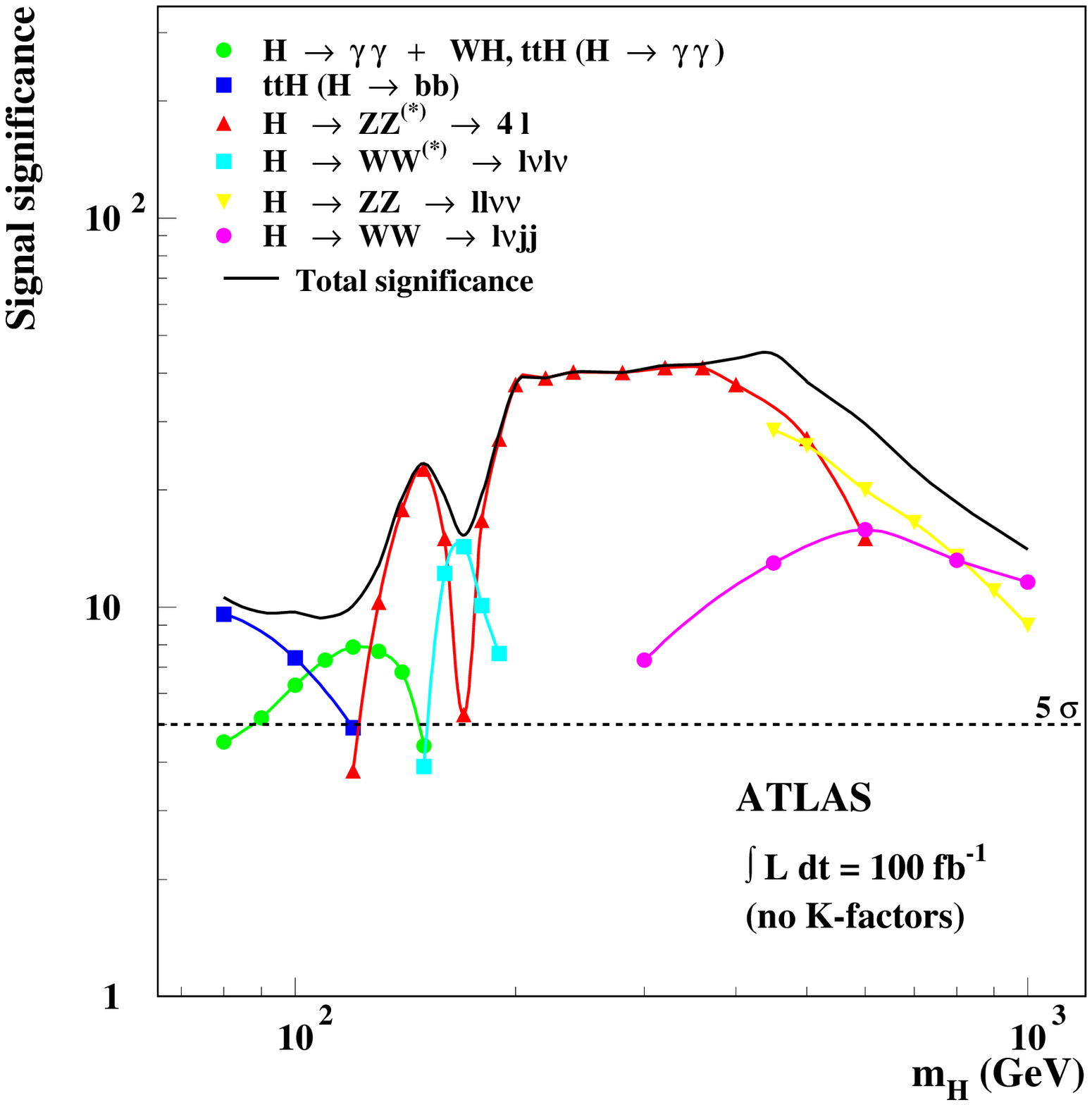}}
\end{minipage}
\end{tabular}
\caption{The dominant production mechanisms (left), the total production 
cross sections (center) and the significance for the experimental detection 
\cite{LHC} (right) of the SM Higgs boson at the LHC.}
\end{figure}

\underline{The gluon--gluon fusion process $gg \to H$} \cite{P1:Hgg}, which
proceeds almost exclusively through a heavy top quark loop (the $b$ quark
contribution is at the few percent level),  is by far the dominant Higgs
production mechanism at the LHC. For a relatively light Higgs boson, $M_H \lsim
200$ GeV,  the production cross section is more than one order of magnitude
larger than those of the other processes  and it dominates for masses up to $M_H
\approx 1$ TeV. At the  LHC, the most promising detection channels  are
\cite{gg-detection} the clean but rare $H \to \gamma \gamma$ signature for $M_H
\lsim 130$ GeV and, slightly above this mass value, the mode $H\to ZZ^* \to
4\ell^\pm$  and/or $H\to WW^{(*)}\to \ell \ell \nu \nu$ with $\ell=e,\mu$ for
Higgs masses below  $2M_Z$. For higher Higgs masses, $M_H \gsim 2M_Z$, the main
signature is the golden mode $H \ra ZZ \ra 4\ell^\pm$ which, from $M_H \gsim
500$ GeV on,  can be complemented by $H \to ZZ \to \nu\bar{\nu} \ell^+ \ell^-$
and $H \to WW \to \nu \ell jj$ to increase the statistics; see Ref.~\cite{LHC}
for details.  \s

The next--to--leading order (NLO) QCD corrections have been calculated in both
the limit where the internal top quark has been integrated out \cite{P2inf}, an
approximation which should be valid in the Higgs mass range $M_H \lsim 300$ GeV,
and in the case where the full quark mass dependence has  been taken into
account \cite{P2}. The corrections lead to an increase of the  cross sections by
a factor of $\sim 1.7$. The challenge of deriving the three--loop corrections
has been performed in the infinite top--quark mass limit; these NNLO corrections
lead to the increase of the rate by an additional 30\% \cite{P3} [see also
Refs.~\cite{P3-add,P3-add2} for recent further improvements]. This results in a
nice convergence of the perturbative series and a strong reduction  of the scale
uncertainty, which is the measure of unknown higher order effects. The
resummation of the soft and collinear corrections, performed at
next--to--next--to--leading logarithm accuracy,  leads to another increase of
the rate by $\sim 5\%$ and a decrease of the scale uncertainty \cite{SG-resum}. 
The QCD corrections to  the Higgs transverse momentum and rapidity
distributions, have also been  calculated at NLO [with a resummation for the
former] and shown to be rather large \cite{Pt-eta-distrib}. The dominant
components of the electroweak corrections, some of which have been derived only
recently, are comparatively very small \cite{EW-CR}. \s 

\underline{The Higgs-strahlung process $q\bar q \to HV$} \cite{P1:HV} where the
Higgs boson is produced in  association with gauge bosons, with $H \to b\bar{b}$ and
possibly $H \to WW^* \to \ell^+ \nu jj$, is the most relevant mechanism at the
Tevatron \cite{Tevatron}, since the dominant $gg$ mechanism has too large a QCD
background. At the LHC, this process plays only a marginal role; however, the
channels $HW \to \ell \nu \gamma \gamma$ and eventually $ \ell \nu b\bar b$
could be useful for the measurement of Higgs couplings. The QCD corrections,
which at NLO \cite{HV-NLO,Spira}, can be inferred from Drell--Yan production,
have been calculated  at NNLO \cite{HV-NNLO}; they are of about 30\% in
total. The ${\cal O} (\alpha)$ electroweak corrections have been also derived
 \cite{HV-EW} and decrease the rate by $5$ to 10\%. The remaining scale
dependence is very small, making this process the theoretically cleanest of all
Higgs production processes.\s

\underline{The vector boson fusion mechanism} \cite{P1:VVH} which leads to $pp
\to Hqq$ final states  has the second largest cross section at the LHC. The QCD
\cite{VV-NLO,Spira}, electroweak \cite{VV-NLO-EW} and SUSY  \cite{SUSY-QCD2} 
radiative corrections are known and are at the level of a few percent. The QCD
corrections including cuts, and in particular those to the $p_T$ and $\eta$
distributions, have also been calculated and implemented into a parton--level
Monte--Carlo program \cite{VV-MC}.  The process has a large enough cross section
[a few picobarns for $M_H \lsim 250$ GeV] and the use of cuts, forward--jet
tagging, mini--jet veto for low luminosity as well as triggering on the central
Higgs decay products \cite{WWfusion0}, lead to small backgrounds, thus allowing
precision measurements. A variety of final states,  $H \to \tau^+ \tau^-, ZZ^*,
WW^*$ and $\gamma \gamma$, can be detected and could allow for measurements of
ratios of couplings \cite{Houches,LHC,Dieter}. The interesting signatures  $H
\to b\bar{b}, \mu^+\mu^-$ and  $H \to $ invisible are more challenging 
\cite{WWfusion-others}.\s

\underline{Higgs production in association with top quarks} \cite{P1:ttH}, $pp
\to t\bar tH$  with $H \to \gamma \gamma$ or $b\bar{b}$, can in principle be
observed at the LHC and would allow for the direct measurement of the top Yukawa
coupling (a CMS analysis has shown that $pp \to t\bar t H\to t\bar t b\bar b$
might be subject to a too large jet background \cite{CMSTDR}). As at
tree--level, the process is at the three--body level, the calculation of the NLO
corrections was a real challenge which was met a few  years ago
\cite{ttH,ttH-SUSY}. The $K$--factors turned out to be rather small, $K\sim 1.2$
but the scale dependence is drastically reduced from a factor of two at LO to
the level of 10--20\% at NLO. Note that the NLO corrections to the $q\bar q /gg
\to b\bar b H$  process, which are more relevant in the MSSM, increases the rate
at the 50\% level if the scale is chosen properly \cite{bbH,bbH-SUSY}. Compared
with the NLO rate for the $bg \to bH$ process where the  initial $b$-quark is
treated as a parton \cite{bg-bH}, the calculations agree  within the scale
uncertainties \cite{bbH-comp}. Note that a similar situation occur for $H^\pm$
production in the $gb$ process: the $K$--factor is moderate $\sim\!1.2$--1.5 if
the cross section is evaluated at scales $\mu \sim \frac12 (m_t+M_{H^\pm})$
\cite{Kfac-H+}.\s

Note that besides the uncertainties due to higher order corrections, an 
additional error on the rates for these processes would be the one due the
parton distribution functions which range from 5\% to 15\% depending on the
considered process and on the Higgs boson mass \cite{Samir-PDFs}.

\subsection{Production of the MSSM Higgs bosons}
 
In the MSSM, the production processes for the CP--even $h,H$ bosons are
practically the same as for the SM Higgs and the ones depicted in Fig.~12 (left)
are all relevant. However, the $b$ quark will play an important role for
moderate to large $\tb$ values as its Higgs couplings are enhanced.  First, one
has to take into account the $b$ loop contribution  in the $gg \to h,H$  process
which becomes the dominant component in the MSSM [here, the  QCD corrections are
available only at NLO  where they have been calculated  in the full massive
case \cite{P2}; they increase the rate by a factor $\sim 1.5$].  Moreover, in associated
Higgs production with heavy quarks, $b\bar{b}$ final states must be considered,
$pp \to b \bar b + h/H$, and this process for either $h$ or $H$ becomes the
dominant one in the MSSM [here, the QCD corrections are available in both the
$gg$ and $gb \to b\Phi, b\bar b \to \Phi$ pictures \cite{bbH,bg-bH,bbH-comp}
depending on how many $b$--quarks are to be tagged, and which are equivalent if
the renormalization and factorization scales are chosen to be small, $\mu  \sim
\frac14 M_\Phi$]. The cross sections for the associated production with
$t\bar{t}$ pairs and with $W/Z$ bosons as well as the $WW/ZZ$ fusion processes,
are suppressed for at least one of the particles as a result of the VV coupling
reduction.\s  

Because of CP invariance which forbids $AVV$ couplings, the $A$ boson cannot be
produced in the Higgs-strahlung and vector boson fusion processes; the rate  for
the $pp \to t\bar t A$ process is suppressed by the small $At\bar t$ couplings
for $\tb \gsim 3$. Hence, only the $gg\to A$ fusion with the $b$--quark loops
included [and where the QCD corrections are also available only at NLO and are
approximately the same as for the CP--even Higgs boson with enhanced  $b$--quark
couplings] and associated production with $b\bar b$ pairs, $pp \to b \bar b +A$
[where the QCD corrections are the same as for one of the CP--even Higgs bosons
as a result of chiral symmetry] provide large cross sections.  However, the
one--loop induced processes  $gg \to AZ, gg\to Ag$ [which hold also for CP--even
Higgses] and associated production with other Higgs particles, $pp \to
A+h/H/H^+$  are possible but the rates are much smaller in general, in
particular for $M_A \gsim 200$ GeV \cite{gg-others}. \s

For the charged Higgs boson, the dominant channel is the production from top
quark decays, $t \to H^+ b$, for masses not too close to $M_{H^\pm}=m_t\!-\!
m_b$; this is particularly true at low or large $\tb$ when the $t\to H^+b$
branching ratio is significant. For higher masses \cite{pp-H+},  the processes
to be considered are the fusion process $gg \to H^\pm tb$ supplemented by $gb
\to H^\pm t$. The two processes have to be properly combined and the NLO
corrections  for both  processes have been derived \cite{Kfac-H+} and are
moderate, increasing the cross sections by 20 to 50\%  if they are  evaluated at
low scales, $\mu \sim \frac12 (m_t+M_{H^\pm})$].  Additional sources
\cite{pp-H+-others} of $H^\pm$ states for masses below $M_{H^\pm} \approx 250$
GeV are provided by pair and associated production with neutral Higgs bosons in
$q\bar q$ annihilation as well as $H^+H^-$ pair and associated $H^\pm W^\mp$
production in $gg$ and/or $b\bar b$ fusion but the cross sections are not as
large,  in particular for $M_{H^\pm} \gsim m_t$. \s

The cross sections for the dominant production mechanisms are shown in Fig.~13
as a function of the Higgs masses for $\tb=3$ and $30$ for the same set of input
parameters as Fig.~7. The NLO QCD corrections are included, except  for the $pp
\to Q \bar Q\,$Higgs processes where, however, the scales have  been chosen as
to approach the NLO results;  the MRST NLO structure functions have been
adopted.  As can be seen, at high $\tb$, the largest cross sections are by far
those of the $gg \to \Phi_A/A$ and $q\bar q/ gg \to b\bar b+ \Phi_A/A$
processes, where $\Phi_A=H\, (h)$ in the (anti--)decoupling regimes $M_A > (<)
M_h^{\rm max}$: the other processes involving these two Higgs bosons have cross
sections that are several orders of magnitude smaller. The production cross
sections for the other CP--even Higgs boson, that is $\Phi_H=h\,(H)$ in the
(anti--)decoupling regime when $M_{\Phi_H} \simeq M_h^{\rm max}$, are similar to
those of the SM Higgs boson with the same mass and are substantial in all the
channels which have been displayed. At small $\tb$, the $gg$ fusion and $b\bar
b$--Higgs cross sections are not strongly enhanced as before and all production
channels [except for $b\bar b$--Higgs which is only slightly enhanced] have
cross sections that are smaller than in the SM Higgs case, except for $h$ in the
decoupling regime.\s

\begin{figure}[!h]
\begin{center}
\vspace*{-1.5cm}
\hspace*{-5.cm}
\epsfig{file=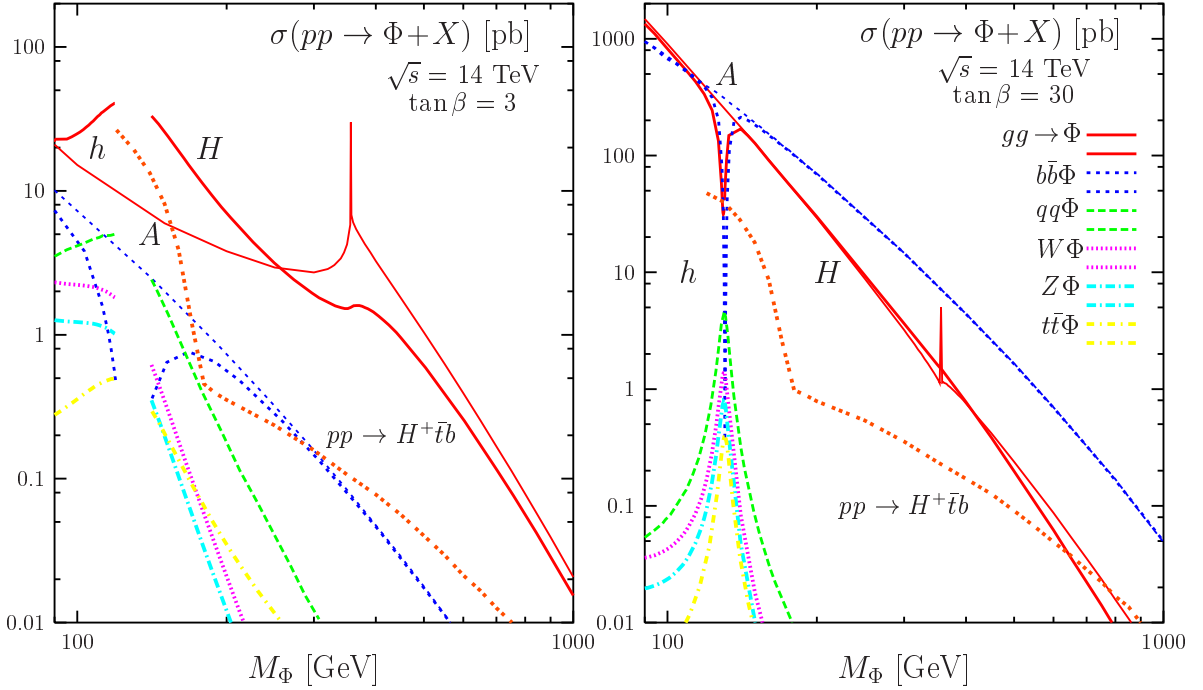,width=15.cm} 
\end{center}
\vspace*{-10.4cm}
\caption{The cross section for the neutral and charged MSSM Higgs production in
the main channels at the LHC as a function of their respective masses for 
$\tb=3$ and $30$ in the maximal mixing scenario; the SM and SUSY inputs are
as in Fig.~7.} 
\vspace*{-.3cm}
\end{figure}

The principal detection signals of the neutral Higgs bosons at the LHC, in the 
various regimes of the MSSM, are as follows 
\cite{LHC,Houches,CMSTDR,ATLASTDR,Anatomy2,intense}.  \s

In the \underline{decoupling regime}, i.e. when $M_h \simeq M_h^{\rm max}$, the
lighter $h$ boson is SM--like and has a mass smaller than $\approx 140 $ GeV. 
It can be detected in the $h \to \gamma \gamma$ decays [possibly supplemented
with a lepton in associated $Wh$ and $t\bar t h$ production], and eventually in
$h\to ZZ^*, WW^*$ decays in the upper mass range, and if the vector boson
fusion processes are used, also in the decays $h \to \tau^+ \tau^-$ and
eventually $h \to W W^*$ in the higher mass range $M_{h} \gsim 130$ GeV; see
Fig.~14 (left).  For relatively large values of $\tb$ $(\tb \gsim 10)$, the heavier
CP--even $H$ boson which has enhanced couplings to down--type fermions, as well
as the pseudoscalar Higgs particle, can be observed in the process $pp \to
b\bar b + H/A$ where at least one $b$--jet is tagged and with the Higgs boson
decaying into $\tau^+ \tau^-$, and eventually, $\mu^+ \mu^-$ pairs in the low
mass range. With a luminosity of 30 fb$^{-1}$ (and in some cases lower) a
large part of the $[\tb,M_A]$ space can be covered; Fig.\,14 (right).  \s

In the \underline{anti-decoupling regime}, i.e. when $M_A < M_h^{\rm max}$ and
at high $\tb$ ($\gsim 10$), it is the heavier $H$ boson which will be SM--like
and can be detected as above, while the $h$ boson will behave like the
pseudoscalar Higgs particle and can be observed in $pp \to b\bar b+ h$ with $h
\to \tau^+ \tau^-$ or $\mu^+ \mu^-$ provided its mass is not too close to $M_Z$
not to be swamped by the background from $Z$ production. The part of the
$[\tb,M_A]$ space which can be covered is also shown in Fig.~14 (left).\s

In the \underline{intermediate coupling regime}, that is for not too large $M_A$
values and moderate $\tb \lsim 5$, the interesting decays $H \ra hh$, $A \ra hZ$
and even $H/A \ra t\bar{t}$ [as well as the decays $H^\pm \to Wh$] still have
sizable branching fractions and can be searched for; Fig.~15 (left set).  In
particular, the $gg \to H \to hh \to b\bar b \gamma \gamma$ process (the $4b$
channel is more difficult as a result of the large background) is observable for
$\tb \lsim 3$ and $M_A \lsim 300$ GeV, and would allow to measure the trilinear
$Hhh$ coupling. These regions of parameter space have to be reconsidered in the
light of the new Tevatron value for the top quark mass.  \s

\begin{figure}[!h]
\vspace*{-.4cm}
\begin{center}
\mbox{
\includegraphics[width=8.cm,height=7.cm]{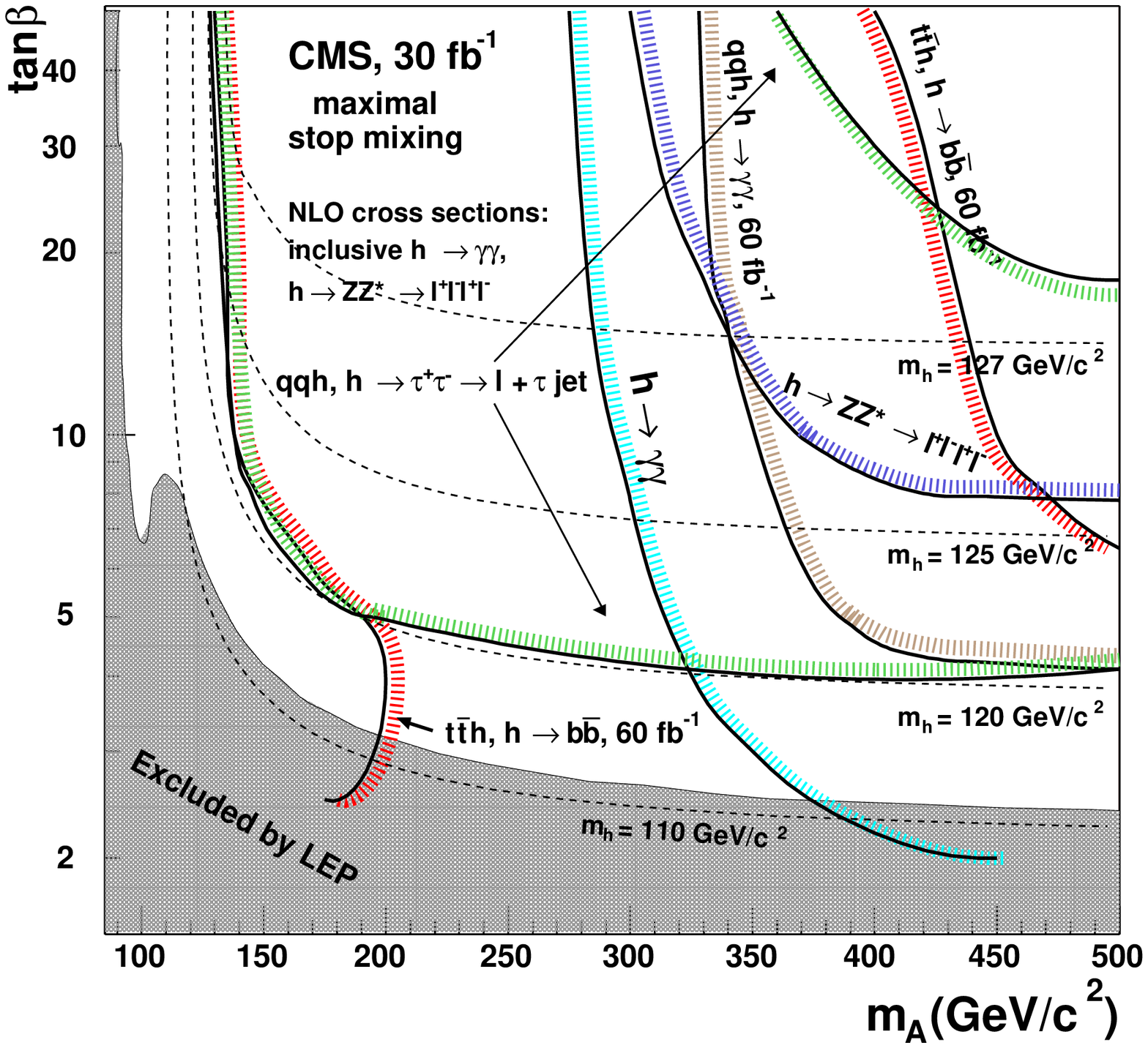} 
\includegraphics[width=8.cm,height=7.cm]{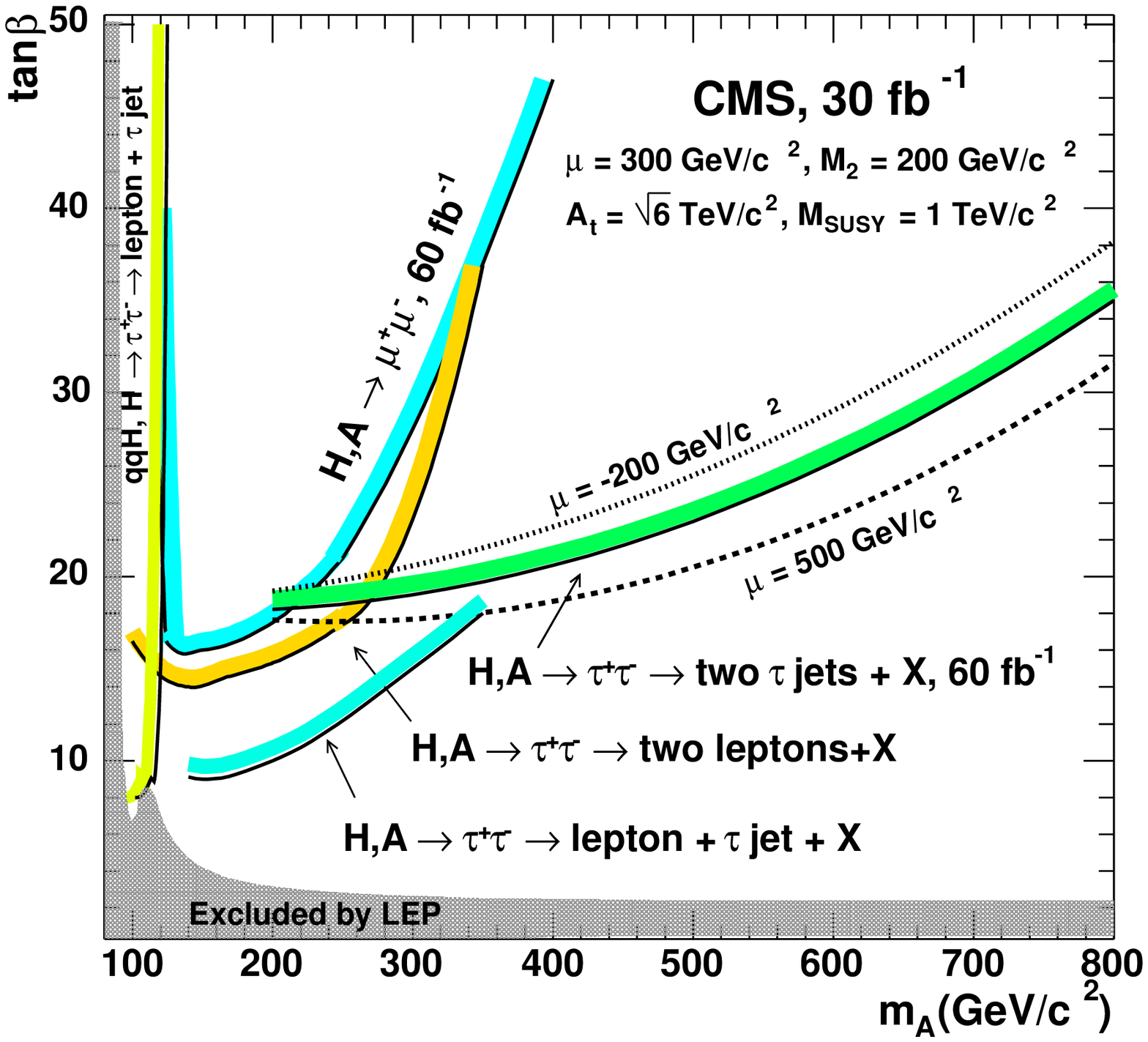} }
\end{center}
\vspace*{-.6cm}
\caption{The areas in the $(M_A, \tb)$ parameter space where the lighter (left)
and heavier (right) MSSM neutral Higgs bosons can be discovered at the LHC with
an integrated luminosity of 30 fb$^{-1}$ in the standard production channels; 
from \cite{CMSTDR}.}
\vspace*{-.3cm}
\end{figure}

In the \underline{intense--coupling regime}, that is for $M_A \sim M_h^{\rm
max}$ and $\tb \gg1$, the three neutral Higgs bosons $\Phi=h,H,A$ have
comparable masses and couple strongly to isospin $-\frac{1}{2}$ fermions leading
to dominant decays into $b\bar b$ and $\tau\tau$ and large total decay widths
\cite{ICR,intense}. The three Higgs bosons can only be produced in the channels
$gg \to \Phi$ and $gg/q\bar q \to b\bar b + \Phi$ with $\Phi \to b\bar b,
\tau^+\tau^-$ as the interesting $\gamma \gamma, ZZ^*$ and $WW^*$ decays of the
CP--even Higgses are suppressed. Because of background and resolution problems,
it is very difficult to resolve between the three particles. A solution
advocated in Ref.~\cite{intense} (see also Ref.~\cite{ggmumu}), would be the
search in the channel $gg/q\bar q \to b\bar b + \Phi$ with the subsequent decay
$\Phi \to \mu^+ \mu^-$ which has a small BR, $\sim 3 \times 10^{-4}$, but for
which the better muon resolution,  $\sim 1\%$, would allow to disentangle
between at least two Higgs particles.  The backgrounds are much larger for the
$gg \to\Phi \to \mu^+ \mu^-$ signals. The simultaneous discovery of the three
Higgs particles is very difficult and in many cases impossible, as exemplified
in Fig.~15 (right) where one observes only one single peak corresponding to $h$
and $A$ production.  \s

\begin{figure}[!h] 
\vspace*{-2mm}
\begin{center}
\begin{tabular}{cc}
\begin{minipage}{7cm}
\hspace*{-5mm}
\mbox{
\includegraphics[width=4cm,height=3cm]{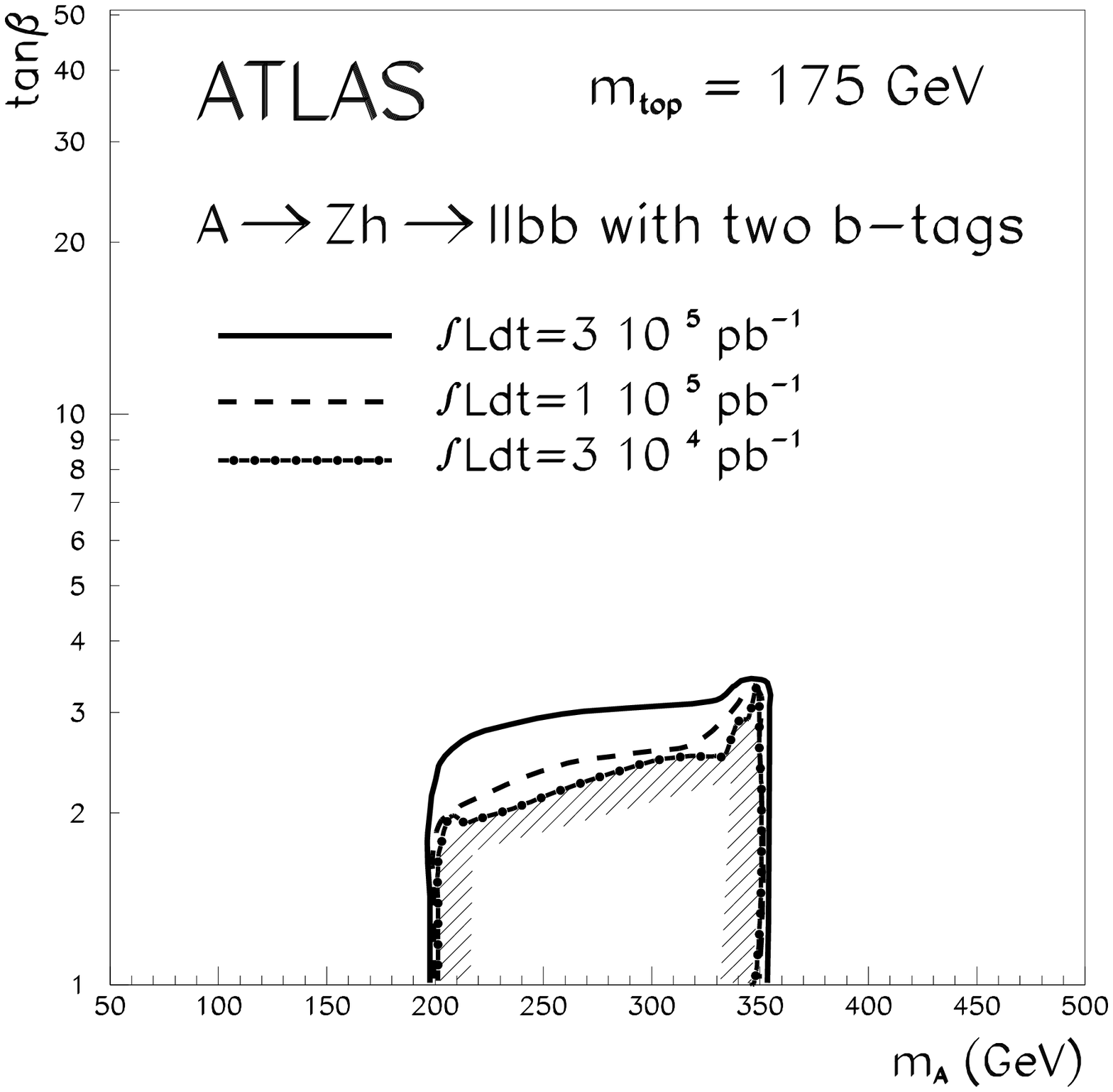} \hspace{-2mm}
\includegraphics[width=4cm,height=3cm]{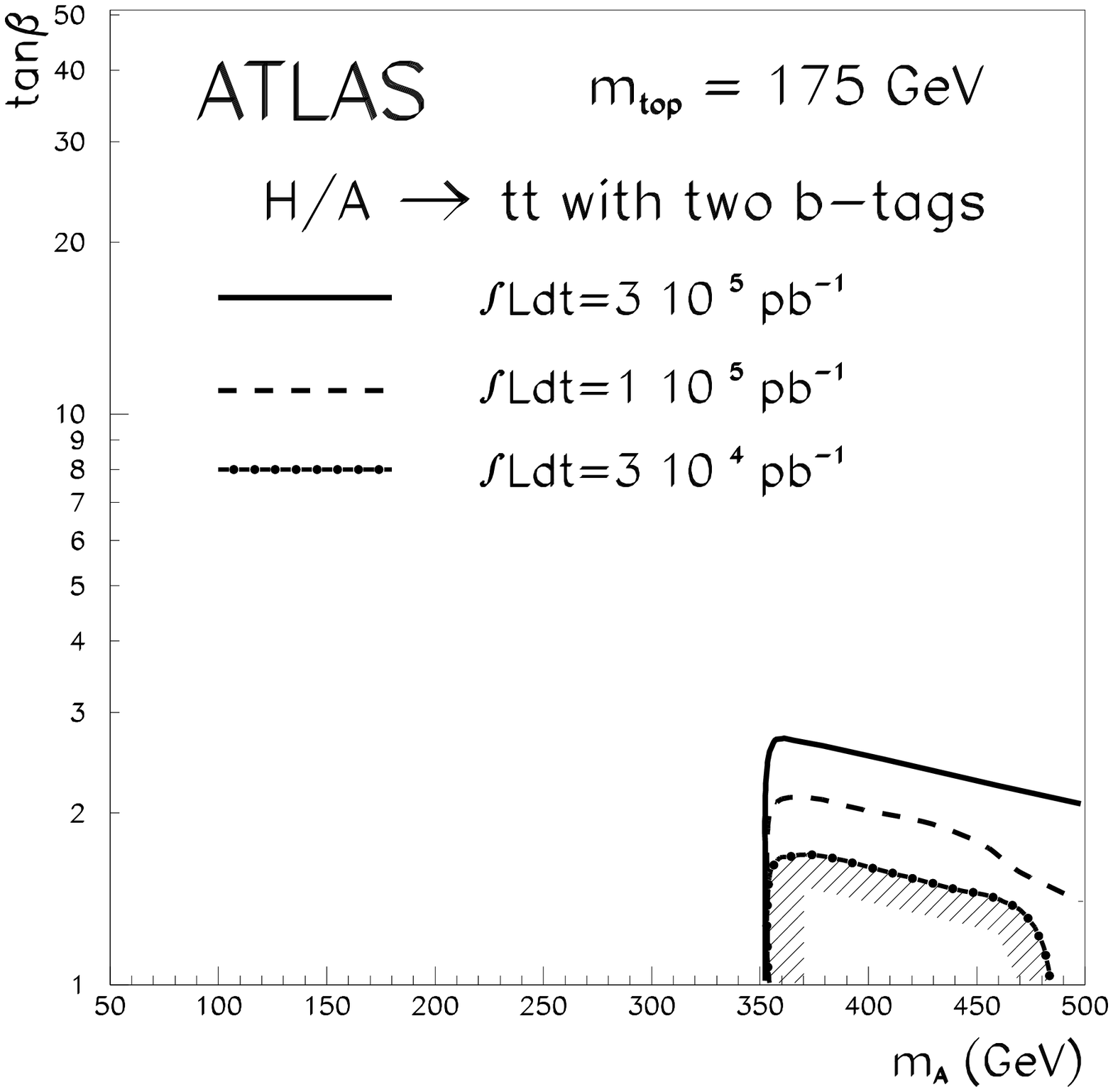}}\\[2mm]
\hspace*{20mm}
\mbox{
\includegraphics[width=5cm,height=3cm]{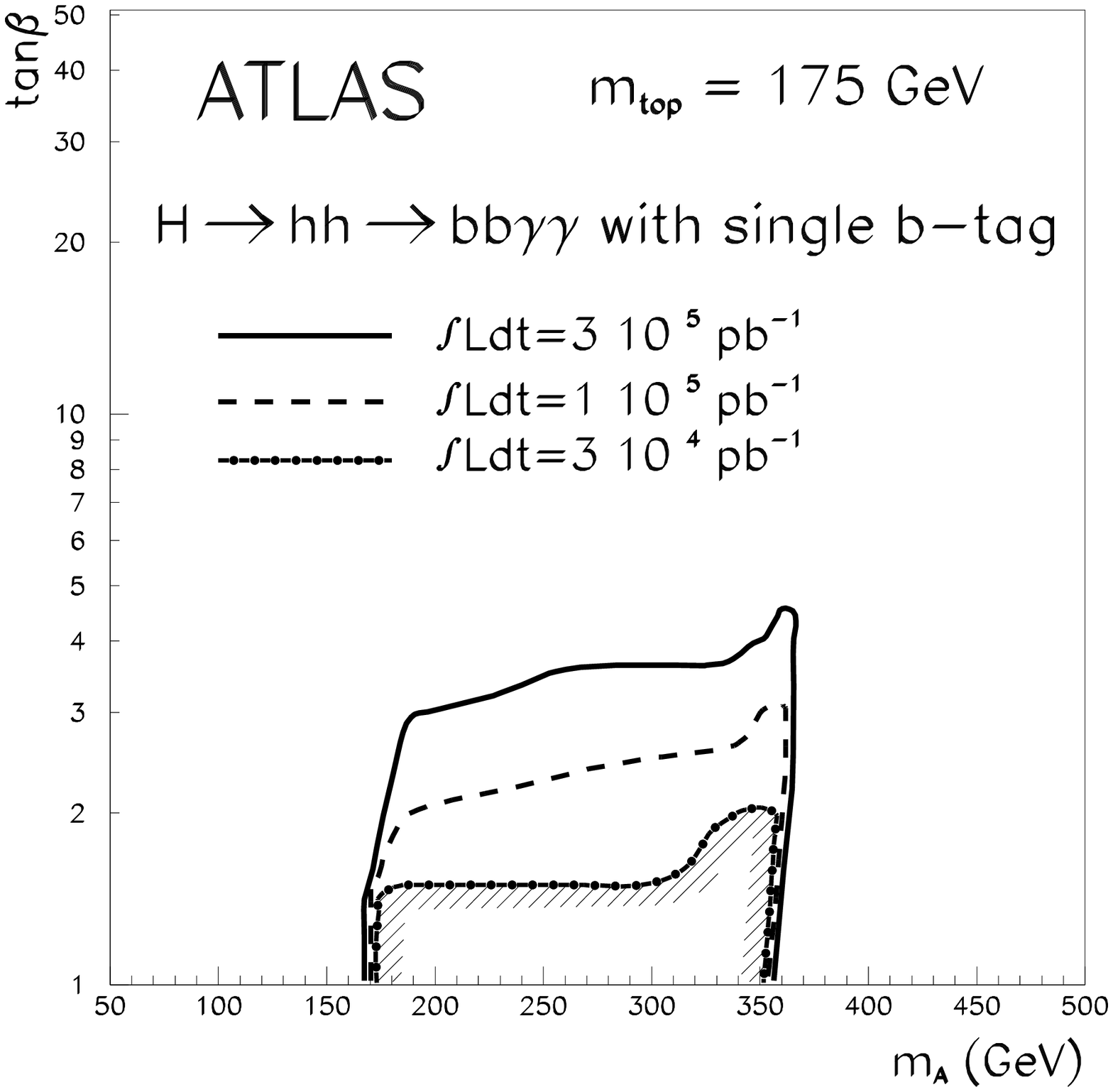} \hspace{-2mm}
}
\end{minipage}
&
\begin{minipage}{7cm}
\hspace*{15mm}
\includegraphics[width=6.5cm,height=6.5cm]{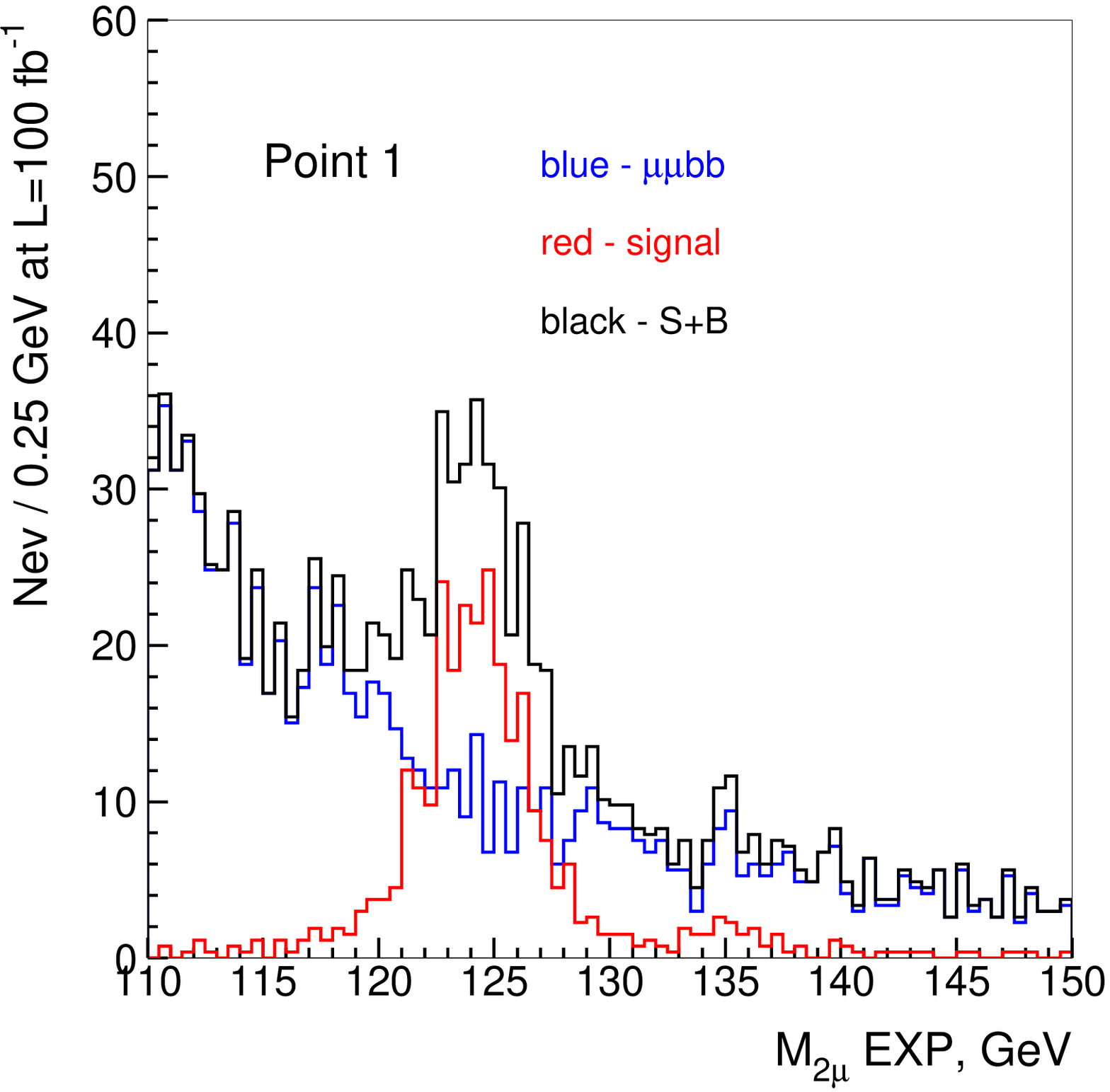} 
\end{minipage}
\end{tabular}
\end{center}
\vspace*{-5mm}
\caption{Left: the regions in the $[\tb,M_A]$ parameter space where the
channel $gg \to H \to hh \to b\bar b \gamma \gamma$, $gg \to A \to hZ \to
b\bar b \ell^+ \ell^-$ and $gg \to H/A \to t\bar t \to \ell \nu jj 
b\bar b $ can be detected at the LHC; from Ref.~\cite{ATLASTDR}. Right: the
$\mu^+ \mu^-$ pair invariant mass distributions for the three Higgs signal 
peaks with $M_A=125$ GeV and $\tb=30$ (leading to $M_h \sim 124$ GeV and $M_H 
\sim 134$ GeV) and backgrounds after detector resolution smearing; from 
Ref.~\cite{intense}.}
\vspace*{-3mm}
\end{figure}

Finally, as mentioned previously, light $H^\pm$ particles with masses below
$M_{H^\pm} \sim m_t$ can be observed in the decays $t \ra H^+b$ with $H^-\ra
\tau \nu_\tau$, and heavier ones can be probed for large enough $\tb$, by
considering the properly combined $gb \to t H^-$ and $gg \ra t \bar{b} H^-$
processes using the decay $H^-\ra \tau \nu_\tau$ and taking advantage of the
$\tau$ polarization to suppress the backgrounds, and eventually the decay $H^-
\to \bar{t}b$  which however, seems more problematic as a result of the large
QCD background. See Ref.~\cite{DP-H+} for more detailed discussions on $H^\pm$
production.  

\subsection{The impact of SUSY particles}

The previous discussion on MSSM Higgs production and detection at the LHC might
be significantly altered  if some supersymmetric particles are relatively light.
Some standard production processes can be affected, new processes can occur and 
the additional detection channels of the Higgs bosons involving SUSY final
states might drastically change the detection strategies of the Higgs bosons.
Let us briefly comment on some possibilities. \s

As discussed in section 3.3, the $Hgg$ and $hgg$ vertices in the MSSM are
mediated not only by heavy $t/b$ loops but also by loops involving squarks [the
NLO QCD corrections are also available \cite{SUSY-QCD} and are moderate]. If the
top and bottom squarks are relatively light, the cross section for the dominant
production mechanism of the lighter $h$ boson in the decoupling regime, $gg \to
h$, can be  significantly altered by their contributions, similarly to the
gluonic decay $h\to gg$.  In addition, in the $h\to \gamma \gamma$ decay which
is one of the most promising detection channels, the same stop and sbottom loops
together with chargino loops, will affect the branching ratio. The cross section
times branching ratio $\sigma( gg \ra h) \times {\rm BR}(h \ra \gamma \gamma)$
for the lighter $h$ boson at the LHC can be thus very different from the SM,
even in the decoupling limit in which the $h$ boson is supposed to be SM--like
\cite{SUSYloops}. This is illustrated  in Fig.~16 (left) where we have simply
adopted the low $\tb$ scenario of Fig.~8 for the $h\to gg$ and $\gamma\gamma$
decays. Here again, for light stops and strong mixing which enhances the
$h\tilde t_1 \tilde t_1$ coupling, the effects can be drastic leading to a
strong suppression of the cross section $\sigma(gg \ra h \ra \gamma \gamma)$
compared to the SM case.\s

\begin{figure}[!h]
\begin{center}
\vspace*{-2.5cm}
\hspace*{-2.7cm}
\mbox{ \hspace*{-3.2cm}
\epsfig{file=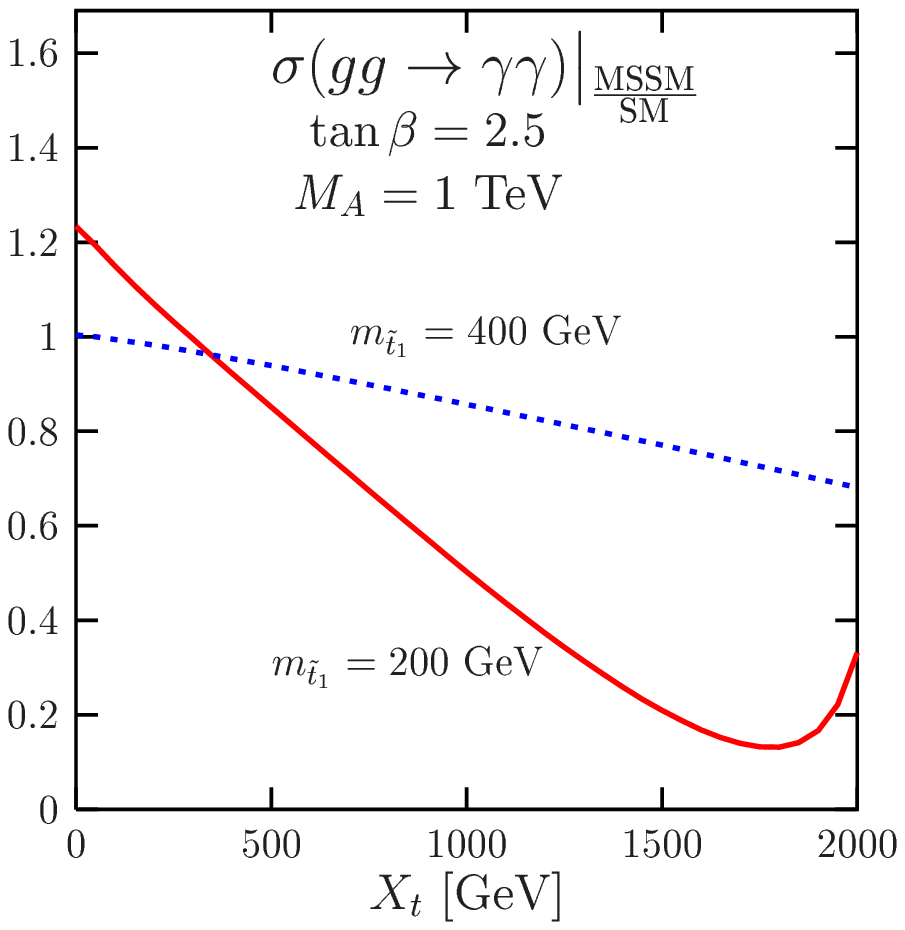,width=15cm}   \hspace*{-7.5cm}
\epsfig{file=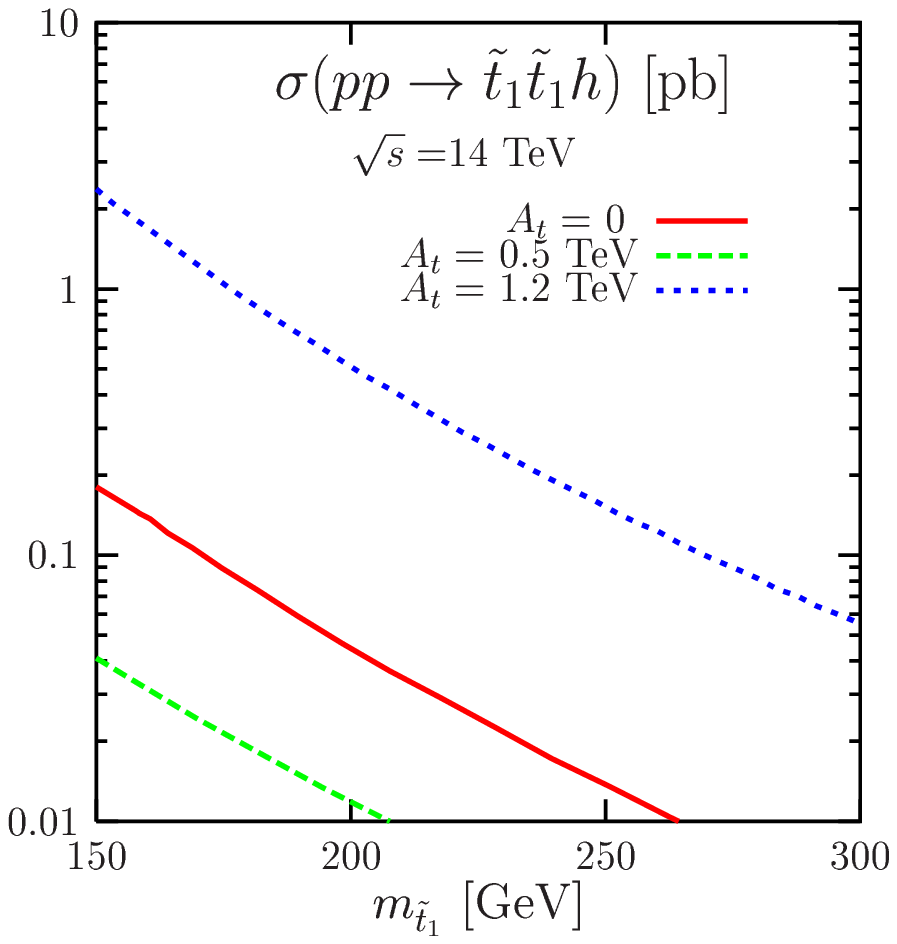,width=15cm}  \hspace*{-5cm}
}
\end{center}
\vspace*{-13.cm}
\caption[]{The $gg$--fusion cross section times the photonic  branching ratio
for the $h$ boson in the MSSM relative to  its SM value with stop contributions 
included \cite{SUSYloops} (left). The cross section for the process $ 
pp  \ra \tilde{t}_1 \tilde{t}_1 h$ for threes scenarios of stop mixing 
\cite{SUSYdirect} (right).}
\vspace*{-.1cm}
\end{figure}

If one of the top squarks is light and its coupling to the $h$ boson is 
enhanced, an additional process might provide a new source for Higgs  particles
in the MSSM: associated production with $\tilde{t}_1$ states \cite{SUSYdirect}, 
$pp \ra gg/ q \bar{q} \ra \tilde{t}_1 \tilde{t}_1 h$. This  process is similar
to the standard $pp \to t\bar t h$ mechanism and in fact, for small masses and
large mixing of the $\tilde t_1$ the cross section can be comparable as shown 
in Fig.~16 (right) where it can reach the picobarn level; in the no or moderate 
mixing cases, the cross sections are much smaller. The stop will mainly decay
into $b \chi_1^+$, with the chargino decaying into $bW^+$ plus missing energy;
this  leads to $\tilde{t}_1 \ra bW^+$ final states which is the same topology as
the decay $t \ra bW^+$ except for the larger  amount of missing energy which 
would help isolating the process if the initial production rates are
significant. Note that final states with the heavier $H,A,H^\pm$ and/or other
squark species than $\tilde t_1$ are less favored by  phase space. \s

Another possible source of  MSSM Higgs bosons would be from the  cascade decays
of strongly interacting sparticles, which have large production rates at the
LHC.  In particular, the lighter $h$ boson and the heavier $A,H$ and $H^\pm$
particles with masses $\lsim 200$--300 GeV, can be produced from the decays of
squarks and gluinos into the heavier charginos/neutralinos, which then decay
into the lighter ones and Higgs bosons. This can occur either in ``little
cascades", $\chi_2^0, \chi_1^\pm \to \chi_1^0+\,$Higgs, or in ``big cascades"
$\chi_{3,4}^0, \chi_2^\pm \to \chi_{1,2}^0, \chi_1^\pm + \,$Higgs. As was shown
in Fig.~10, the rates for ino decays into Higgs bosons can be dominant while
decays of squarks/gluinos into the heavier inos are substantial.  Detailed
studies \cite{Cascade0pp,cascade} have shown  that these processes  can be
isolated in some areas of the SUSY parameter space. In this case, they  can be
complementary to the direct production ones  in some areas of the MSSM parameter
space; see Fig.~17 (left). In particular, one can probe the region $M_A \sim
150$ GeV and $\tb \sim 5$, where only  $h$ can be observed in  standard
searches.\s

One can take advantage of the possibility of light charginos and neutralinos  to
search for the heavier $H,A$ and $H^\pm$ states in regions of the parameter 
space in which they are not accessible in the standard channels [this is the 
case e.g. for $M_A \sim 200$ GeV and moderate $\tb$ values]. There are
situations in which  the signals for Higgs decays into charginos and neutralinos
are clean enough to be detected at the LHC.  One of the possibilities is that
the neutral $H/A$ bosons decay into pairs of the second lightest neutralinos,
$H/A \to \chi_2^0 \chi_2^0$, with the subsequent decays of the latter into the
LSP neutralinos and leptons, $\chi_2^0 \to \tilde \ell^* \ell \to \chi_1^0 \ell
\ell$ with $\ell^\pm=e^\pm,\mu^\pm$, through the exchange of relatively light
sleptons. This leads to four charged leptons and missing energy in the final
state. If the  $H/A$ bosons are produced in the $gg$--fusion processes, there
will be little hadronic activity and the $4\ell^\pm$ final state is clean enough
to be detected. Preliminary analyses show that  the decays can be isolated from
the large (SUSY) background; Fig.~17 (right). Note that  in the scenario in 
which the Higgs bosons, and in particular the lightest one $h$, decay into
invisible lightest neutralinos, the discovery of the particles will be 
challenging but possible  \cite{Hinvisble}.  \s

\begin{figure}[!h]
\vspace*{-1.cm}
\begin{center}
\mbox{
\epsfig{file=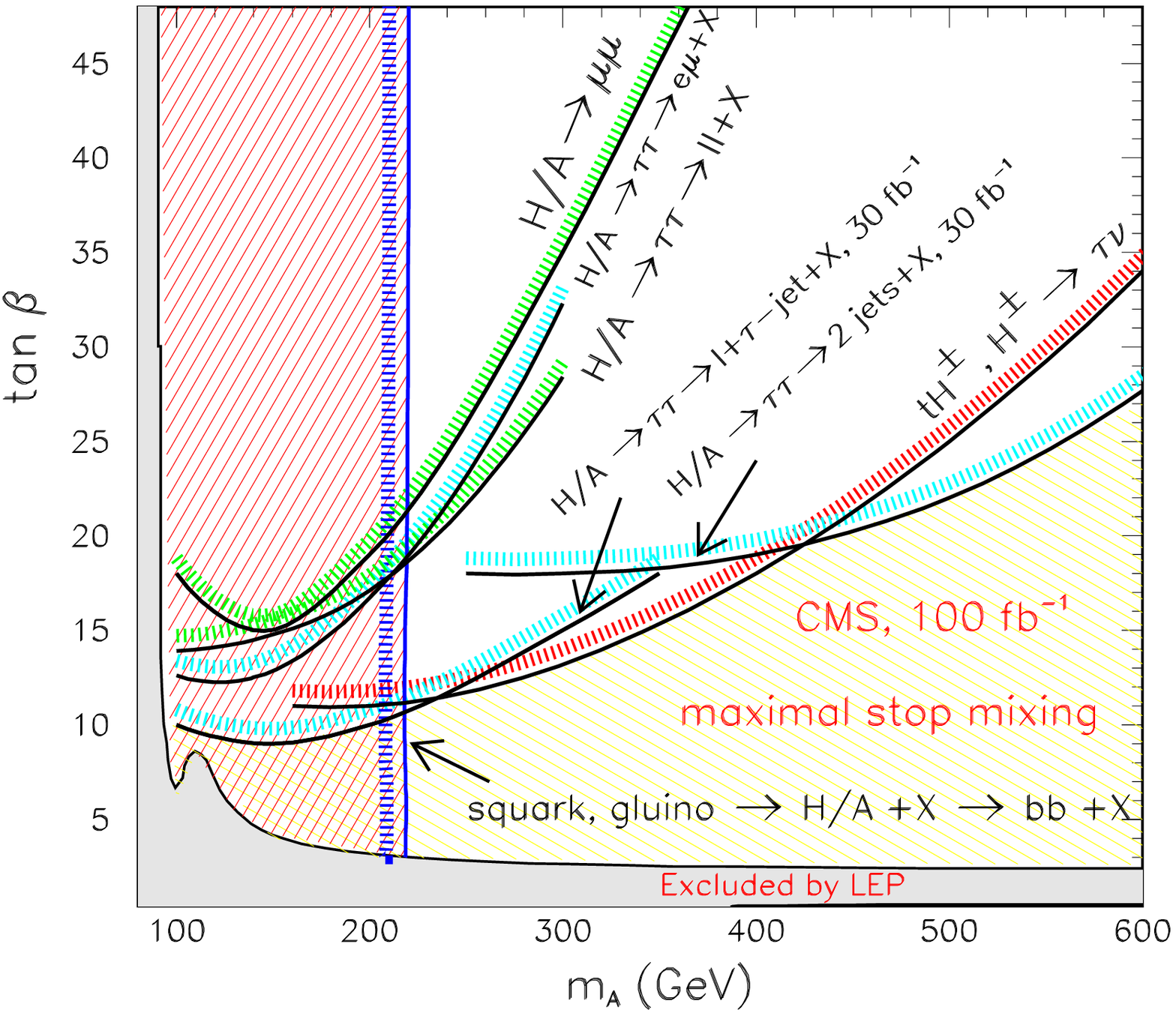,width=8cm}  
\epsfig{file=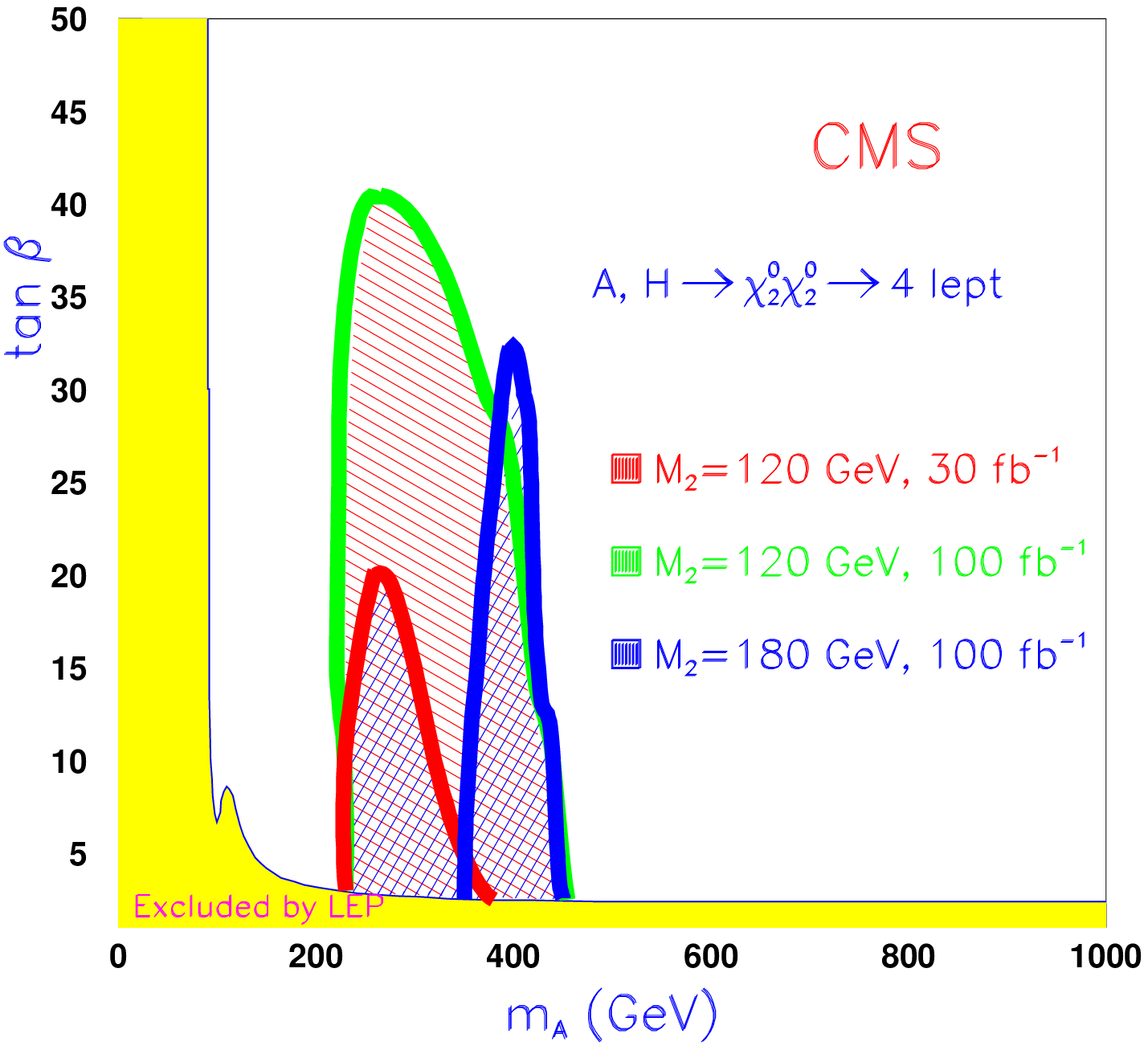,width=7.8cm} }
\end{center}
\vspace*{-.6cm}
\caption{Areas in the $[M_A, \tb]$ parameter space where the MSSM Higgs bosons 
can be discovered at the LHC with 100 fb$^{-1}$ data in cascades of SUSY 
particles \cite{cascade} (left) and in $A/H \ra \chi^0_2 \chi^0_2\ra 4\ell^\pm
+X$ decays (right) and for a given set of the MSSM parameters \cite{SUSY-Filip}.} 
\vspace*{-2.mm}
\end{figure}

\subsection{Measurements of parameters in the MSSM Higgs sector}

In the decoupling regime when the pseudoscalar $A$ boson is very heavy, only the
lighter MSSM boson with SM--like properties will be accessible. In this case,
the measurements which can be performed  for the SM Higgs boson with a mass
$\!\lsim\! 140$ GeV will also be possible. The $h$ mass can be measured with
a very good accuracy, $\Delta M_h / M_h\! \sim\! 0.1\%$,  in the $h\!\to\!
\gamma \gamma$ decay \cite{CMSTDR,ATLASTDR} which incidentally, verifies the
spin--zero nature of the particle. However, the total decay width is very small
and it cannot be resolved experimentally. The parity quantum numbers will be
very challenging to probe \cite{VVHspin}, in particular since the $h\!\to\!ZZ^*
\!\to\!4\ell^\pm$ decay in which some correlations between the final state leptons
can characterize a $J^{\rm PC}\!=\!0^{++}$ particle, might be very rare. This
will be also the case of the trilinear Higgs--self coupling which needs
extremely high luminosities \cite{Htrilin}.\s

Nevertheless, combinations of Higgs production cross sections and decay
branching ratios can be measured with a relatively good accuracy
\cite{Houches,Dieter}. The Higgs couplings to fermions and gauge bosons can be
then determined from a fit to all available data. However, while in the SM one
could make reasonable theoretical assumptions to improve the accuracy of the
measurements, in the MSSM the situation is made more complicated by several
features, such as  the possibility of invisible decay modes, the radiative
corrections in the Higgs sector which can be different for $b,\tau$ and $W/Z$
couplings, $etc.$..   Under some assumptions and with 300 fb$^{-1}$ data, one 
can distinguish an MSSM from a SM Higgs particle at the $3\sigma$ level for $A$
masses up to $M_A=$300--400 GeV \cite{Dieter}.  \s

The heavier Higgs particles $H,A$ and $H^\pm$ are accessible mainly in the $gg
\to b \bar b+ H/A$ and $gb  \to H^\pm t$ production channels for large $\tb$
values, the main decay modes being $H/A \to b\bar b, \tau^+ \tau^-$ and $H^+ \to
t \bar b, \tau^+ \nu$. The Higgs masses cannot be determined with a very good
accuracy as a result of the poor resolution. However, for $M_A \lsim 300$ GeV
and with high luminosities, the $H/A$ masses can be measured with a reasonable
accuracy by considering the rare decays $H/A \to \mu^+ \mu^-$ as the resolution
on the muon pairs is much better \cite{intense,CMSTDR}. The discrimination
between $H$ and $A$ is nevertheless  difficult as the masses are close in
general and the total decay widths large \cite{intense}. The Higgs spin--parity
quantum numbers  cannot be probed in these fermionic decays, too. \s

There is, however, one very important measurement which can be performed in
these channels. As the production cross sections above are all proportional to
$\tan^2\beta$ and, since the ratios of the most important decays fractions are
practically independent of $\tb$ for large enough values [when higher--order
effects are ignored], one has an almost direct access to this parameter.  In
Ref.~\cite{Sasha-tb}, a detailed simulation of the two production channels $gb
\to H^-t$ and $q\bar q/gg \to H/A+b\bar b$ at CMS has been performed.  At a
luminosity of 30 fb$^{-1}$ and  if only the statistical errors are taken into
account, one can make a rather precise measurement, $\Delta \tb/\tb \lsim  10\%$
for $M_A \lsim 400$ GeV. However,  there are also  systematical errors from e.g.
the luminosity measurement and theoretical errors due to the uncertainties on
the PDFs \cite{Samir-PDFs} and  higher--order effects in the production cross
sections and decay rates \cite{bbH,bg-bH}. The theoretical errors are estimated
to be $\sim 20\%$ for the production cross section and $\sim 5\%$ for the decay
branching ratio.  The total accuracy of the measurement worsens then to the
level of $\sim 30\%$  for $M_A \sim 400$ GeV, $\tb=20$ with 30 fb$^{-1}$ data. 
\s

Note that in the anti--decoupling regime, it is the heavier CP--even $H$ boson
which is SM--like and for which the previously discussed measurements for a SM
Higgs particle apply. In this case, the $h$ boson is degenerate in mass with the
pseudoscalar Higgs boson and both can be detected in the decays $h/A \to \mu^+
\mu^-$ for large enough values of $\tb$ and $M_A \gsim 110$ GeV.  In the
intense--coupling regime, as discussed earlier, the three Higgs bosons will be
difficult to disentangle and the situation will be somewhat confusing
\cite{intense}. In the intermediate--coupling regime, there will be a hope to
measure the trilinear $Hhh$ coupling and to have a direct access to part of the
scalar potential  which breaks the electroweak symmetry. Finally, light SUSY
particles would give us the hope to access some important parameters which enter
both the Higgs and sparticle sectors.

\subsection{The Higgs bosons beyond the CP--conserving MSSM}

In the CP--violating MSSM, the production processes of the neutral and charged
Higgs particles are the same as in the CP--conserving case once the couplings
have been properly adapted. All neutral Higgs particles can be produced in the
four dominant processes of Fig.~12. However, in the Higgs--strahlung and vector
boson fusion processes, only the CP--even components of the couplings $g_{H_i
VV}$ will be projected out. The final rates will then simply depend on the
masses and couplings of the  states $H_i$. For the charged Higgs boson, the
cross sections are the same  as in the CP--conserving MSSM.  To illustrate the
impact of these CP--violating phases,  a benchmark scenario  called CPX
\cite{cpxbenchmark} has been defined using the set of input   parameters  
$\mu = 2 |A_t| =2|A_b| = 4 M_S$, while the two basic parameters of the Higgs
sector, $\tb$ and $M_{H^\pm}$, are allowed to vary. For the CP violating phases,
one can assume  that the phases of $\mu$ and $M_3$ are zero, as these
parameters do not play the leading  role, while the  phases of the trilinear
couplings $A_t$ and $A_b$ are set to a  common value  $\Phi_A$.  In this CPX
scenario \cite{cpxbenchmark}, for given  $\tan\beta$  and $M_{H^\pm}$, there
are  values of the argument $\Phi_A$ for which the mass of  the lighter $H_1$
boson becomes very small, $M_{H_1} \lsim 50$ GeV,  and at the same time its
coupling to the gauge bosons negligible.  The other neutral Higgs bosons $H_2$
and $H_3$ have masses substantially larger  than $M_{H_1}$ and their couplings
to gauge bosons (as well as the trilinear self--couplings) can be substantial as
a result of the sum rule $\sum_i g_{H_iVV}^2=  g_{H_{\rm SM}VV}^2$.\s

\begin{figure}[!h]
\hspace*{-2.cm}
\vspace*{-.2cm}
\begin{center}
\begin{tabular}{ll}
\begin{minipage}{8cm}
\epsfig{file=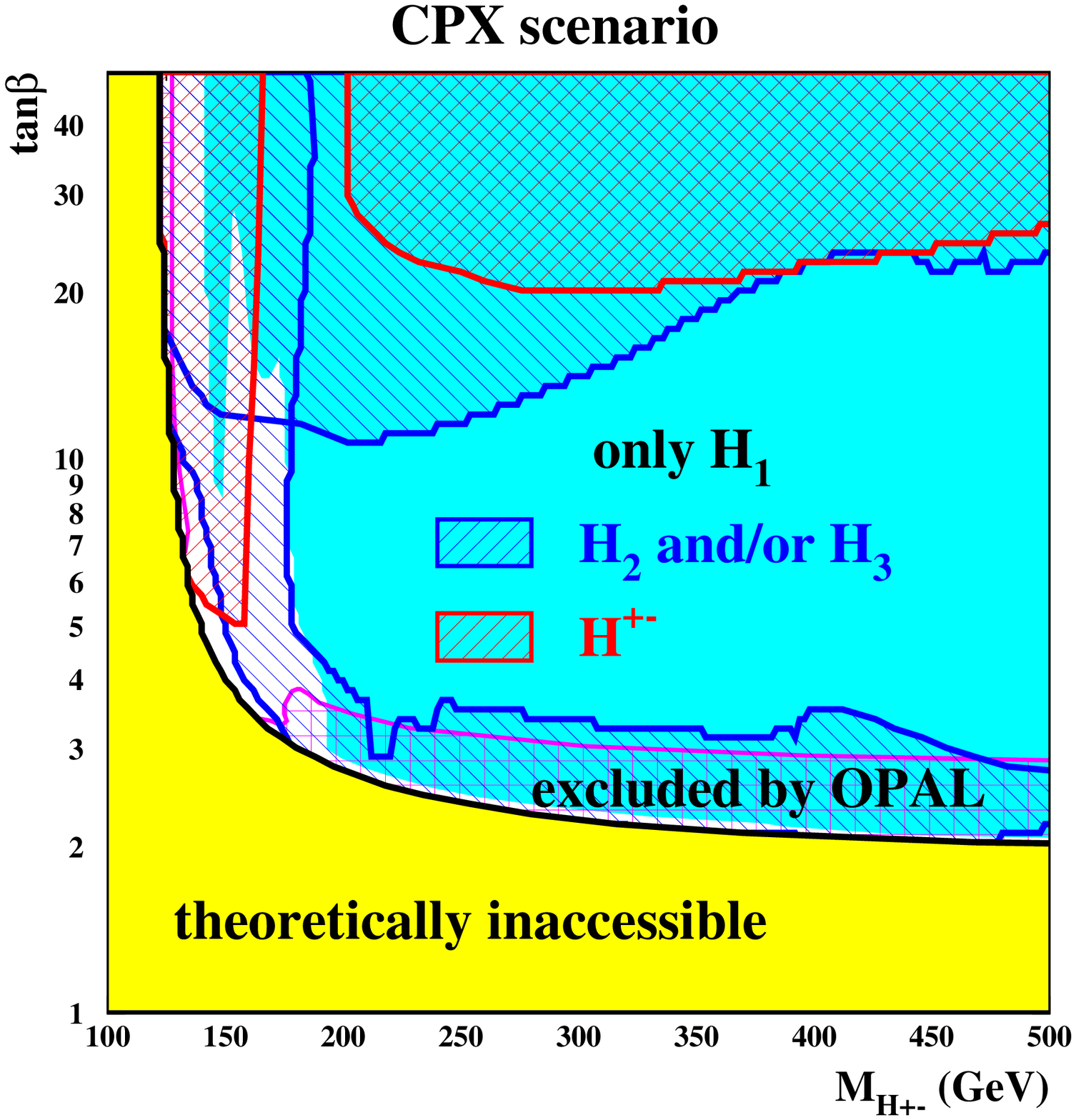,width=8.cm,height=8.cm}
\end{minipage}
& \hspace{5mm}
\begin{minipage}{8cm}
\epsfig{file=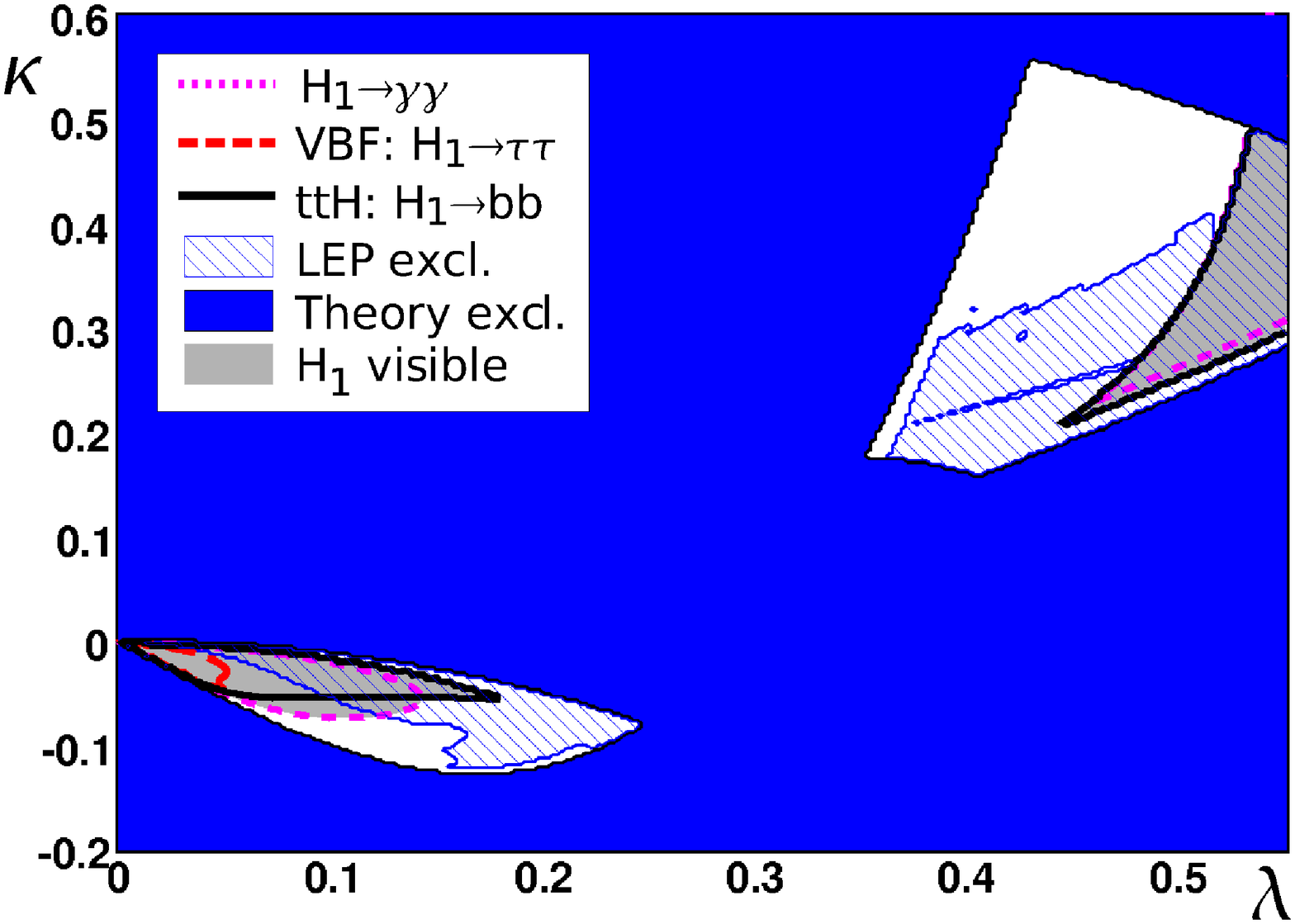,width=8.cm,height=7cm}
\end{minipage}
\end{tabular}
\end{center}
\vspace*{-.5cm}
\caption{Left: the overall discovery potential for Higgs bosons in ATLAS in a
CP--violating scenario after collecting 300 fb$^{-1}$ of data, with the white
region indicating the area where no Higgs boson can be found; from
\cite{CP-schumacher}. Right: regions of the NMSSM parameter space $[\lambda,
\kappa$] in which a light pseudoscalar Higgs boson can be detected in an
ATLAS simulation \cite{Houches-last}.}
\vspace*{-.2cm}
\label{Fextended}
\end{figure}

In this scenario, the lighter $H_1$ state cannot be observed at the LHC as the
cross sections for vector boson fusion  $qq \to qq H_1$  and Higgs--strahlung 
$q\bar q \to H_1V$ are strongly  suppressed as a result of the small $g_{H_1
VV}$ coupling. This is also the case for the production cross section in the 
gluon--gluon fusion and associated production with top quark pairs: besides the 
fact that  the $ttH_1$ coupling is also suppressed, the QCD background events 
for the dominant $H_1 \to b\bar b$ decays is too large. In turn, since the state
$H_2$ has couplings that are similar to that of the SM Higgs boson, the rates
are substantial in the four production mechanisms and, in particular, in the
gluon--gluon $gg \to  H_2$ and  vector boson $qq \to H_2 qq$ fusion channels.  
However, the $H_2$ state  will decay mostly into two $H_1$ sates with a 
branching fraction BR$(H_2 \to H_1 H_1) \gsim 80\%$, and the latter will
subsequently decay into $b\bar b$ pairs with a branching ratio of BR($H_1 \to
b\bar b) \sim 90\%$.   This leads to final state topologies with four $b$ quarks
that are subject to a huge QCD background and which will be extremely difficult
to detect.  Note also that for moderate values of $\tan\beta$, the cross
sections for the production of the heavier neutral $H_3$ and the charged $H^\pm$
Higgs bosons are also too small and no Higgs particle will be thus accessible at
the LHC. This is exemplified in the left-hand side of Fig.~\ref{Fextended} where
the result of an ATLAS simulation show the [$\tb,M_{H^\pm}]$ regions of the
CP--violating MSSM  parameter space that are accessible with 300 fb$^{-1}$ data
\cite{CP-schumacher}. \s

In the NMSSM, where a complex iso-scalar field is introduced, leading to an
additional pair of scalar and pseudoscalar Higgs particles,  the axion--type or
singlino  character of the pseudoscalar $A_1$ boson  makes it preferentially
light and decaying into $b$ quarks or $\tau$ leptons \cite{NMSSMb,Houches-last}.
Therefore, in some areas of the NMSSM parameter space, the lightest CP--even
Higgs boson may dominantly decay into a pair of light pseudoscalar $A_1$ bosons
generating four $b$ quarks or $\tau$ leptons  in the final state, $H_1 \to A_1
A_1 \to 4b, 2b2\tau, 4\tau$. In fact, it is also possible that  $H_1$ is very
light with small $VV$ couplings, while $H_2$ is not too heavy and plays the role
of the SM--like  Higgs particle; the decays $H_2\to H_1 H_1$ can also be
substantial and will give the same signature as above.\s

This situation, similar to the CPX scenario discussed above, is very challenging
at the LHC. Indeed, all the production mechanisms of the light $A_1$  or $H_1$
singlino--like state will have small cross sections as both couplings to vector
bosons and top quarks are tiny. The SM--like Higgs $H_1$ or $H_2$ will have 
reasonable production rates but the dominant decay channels into $4b, 2\tau 2b$
and $4\tau$ will be swamped by the QCD background. Nevertheless, in the case  of
very light $A_1$ bosons with masses smaller than 10 GeV and, therefore decaying
almost exclusively into $\tau^+ \tau^-$ pairs, the  $H_1 \rightarrow A_1 A_1
\rightarrow 4\tau \rightarrow 4\mu + 4\nu_\mu +4\nu_\tau$ final state with the
$H_1$ boson dominantly produced in vector boson fusion can be isolated in some
cases. This is exemplified in the right-hand side of Fig.~\ref{Fextended} where
the result of a simulation of this process by members of the ATLAS collaboration
is shown in  the parameter space formed by the trilinear NMSSM couplings 
$\lambda$ and $\kappa$. While there are regions in which the final state can be
detected, there are other regions in which the light $H_1$ and $A_1$states
remain invisible even for the high luminosity which has been assumed.\s

In the most general  SUSY model, with an arbitrary number of singlet and doublet
fields and an extended  matter content to allows for the unification of the
gauge couplings, a Higgs boson should have a mass smaller than 200 GeV and
significant couplings to gauge bosons and top quarks; this particle can be thus
searched for in the $gg$ and $VV$ fusion channels with the signature $WW \to
\ell  \ell \nu \nu$ which should not be missed. Furthermore, in scenarios with 
spontaneously broken R--parity, the Higgs particles could decay dominantly into
escaping Majorons, $H_i \to JJ$ and the searches would also be more complicated
than in the usual MSSM. However, invisible decays  could be isolated in vector
boson fusion or in associated production with a $Z$ boson, albeit with some
efforts as the final state is very challenging \cite{Hinvisble}. In turn, decays
of the pseudoscalar Higgs $A_i \to H_j Z \to Z$ and missing energy could be
detected if the cross sections for $A_i$ production are large enough.\s

Other SUSY scenarios can also be probed at the LHC.  In GUT theories which lead 
to the presence of an extra neutral gauge boson at low energies, the $Z'$  boson
decays $Z' \to Zh$ which occur via $Z$--$Z'$ mixing could have non--negligible
rates and would lead to a detectable $\ell \ell b\bar b$ signature
\cite{Zprime,ppZprime}; the $Z'$ production cross section would be large enough 
for $M_{Z'} \lsim 2$ TeV \cite{ppZprime-nous} to compensate for the tiny mixing
and hence, the small $Z+$Higgs branching ratio.  If relatively light doubly
charged Higgs bosons exist, they can be produced in the Drell--Yan process
$q\bar q \to H^{++} H^{--}$   \cite{H++LHC} and, if their leptonic decays   
$H^{--} \to \ell \ell$ are not too suppressed,  they  would lead to a
spectacular 4--lepton final state that cannot be missed. \s

Hence, many SUSY scenarios beyond the MSSM  might lead to an interesting 
phenomenology which could be probed at the LHC.

\section{SUSY Higgs bosons at the ILC}
\newcommand{\eei}{e^+e^-} 

\subsection{Higgs production in the SM}

\begin{figure}[!h]
\vspace{-2.5cm}
\begin{center}
\hspace*{-4.5cm}
\includegraphics[width=0.5\linewidth,bb=73 500 600 745]{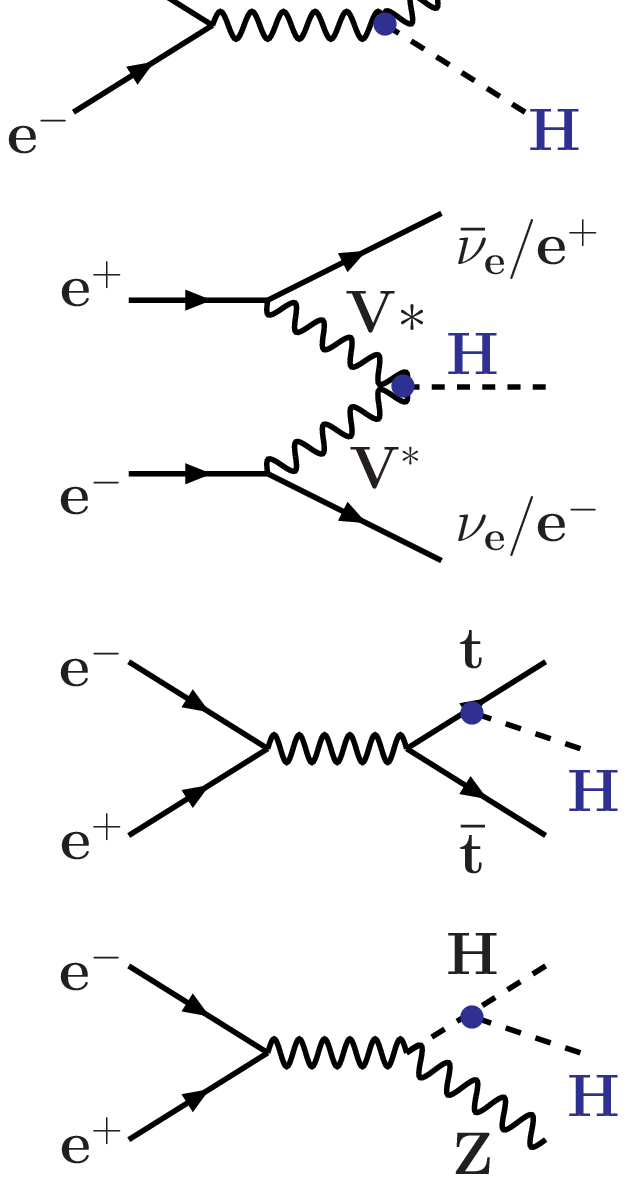} 
\hspace*{-4.3cm}
\includegraphics[width=0.62\linewidth,bb=73 420 600 745]{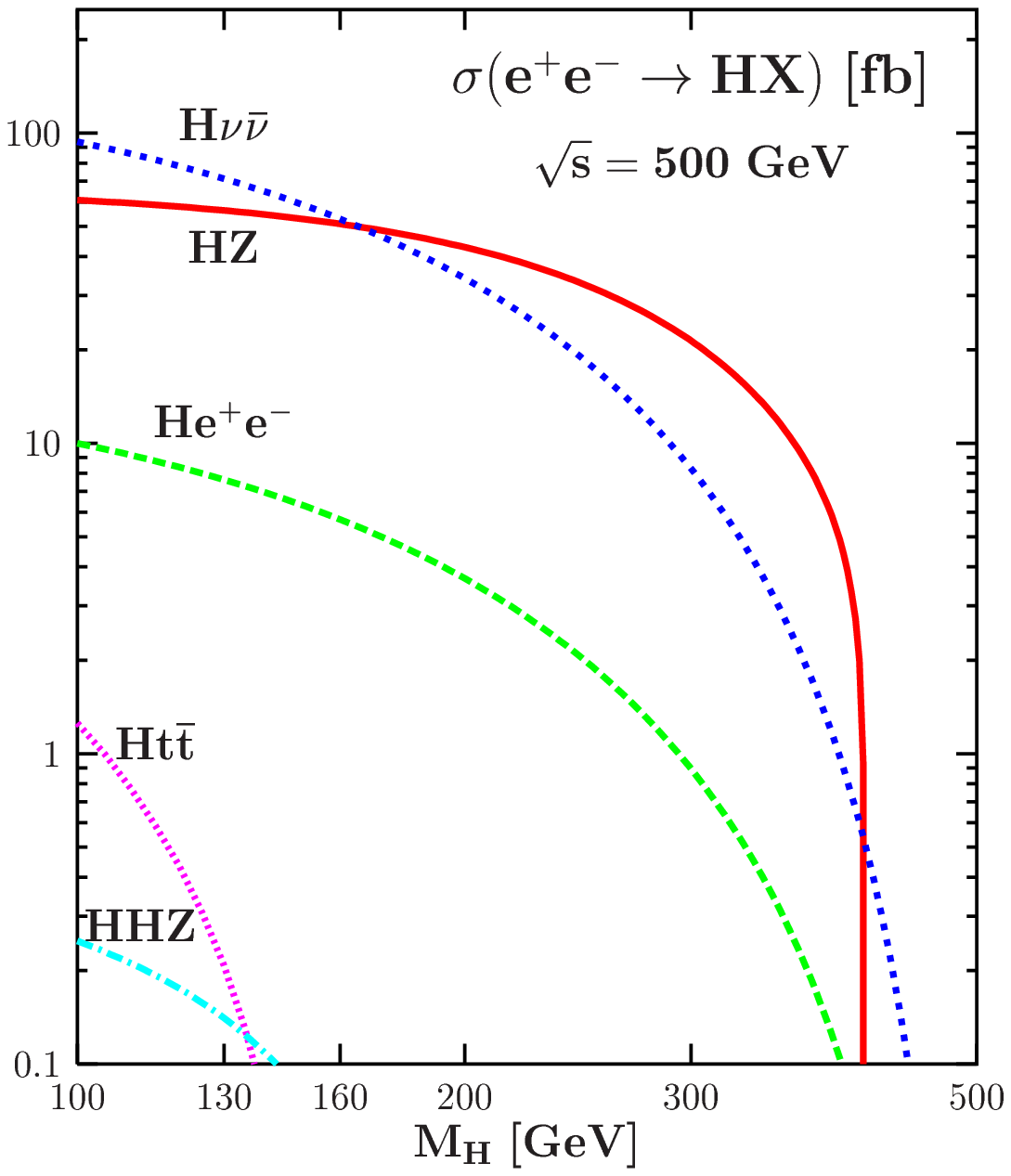} 
\hspace*{-2.3cm}
\includegraphics[width=0.4\linewidth,bb=73 50 600 745]{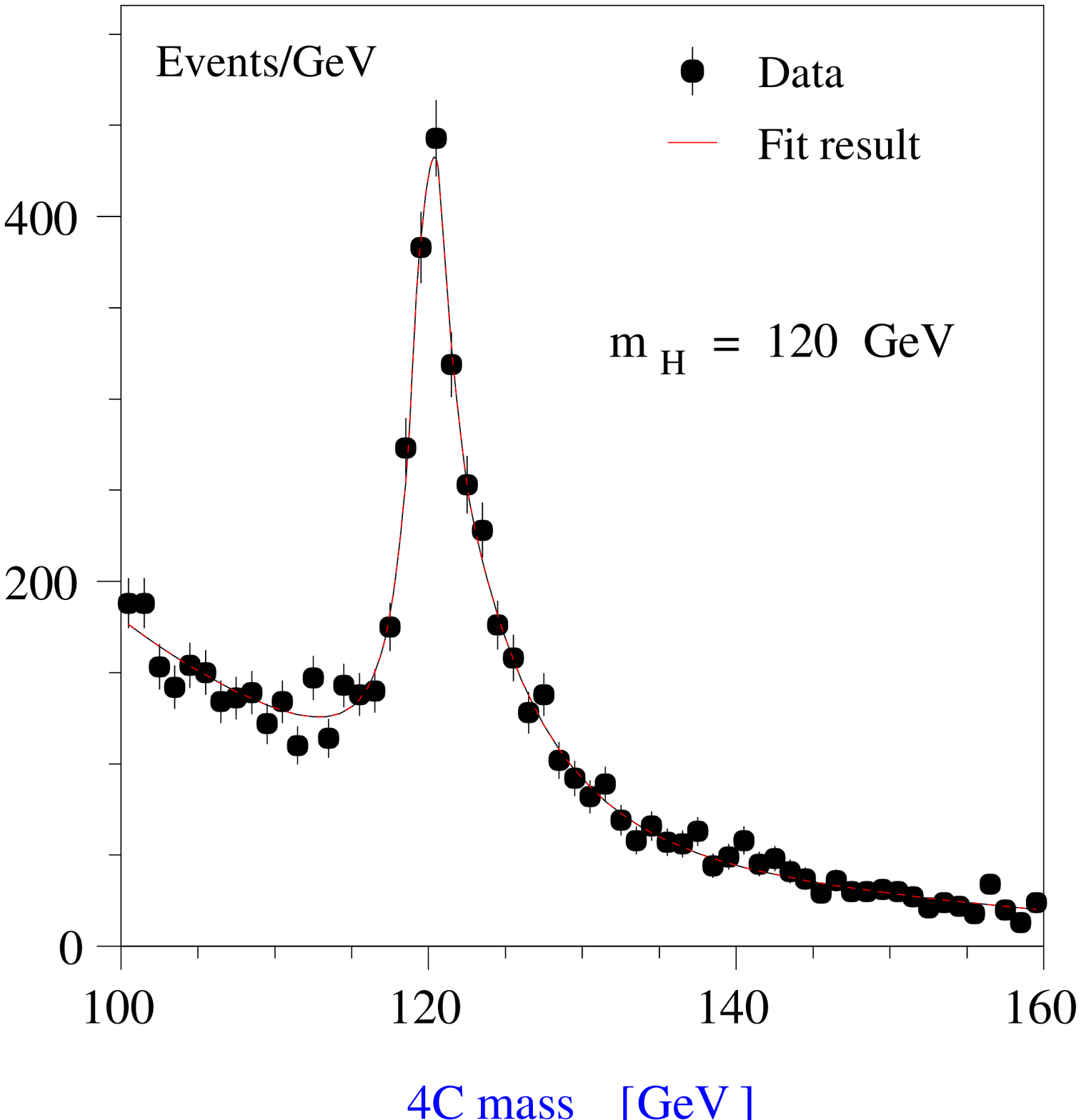} 
\hspace*{-2.5cm}
\end{center}
\caption[Production cross sections of the SM Higgs boson at ILC]{Production 
mechanisms (left), the total cross sections as a function of $M_H$ 
\cite{Anatomy1} (center) and the detection for $M_H=120$ GeV \cite{TESLA} 
(right) of the SM Higgs boson at the ILC.}
\vspace{-.2cm}
\label{Hfig:xs}
\end{figure}

In $\eei$ collisions \cite{Weiglein:2004hn,TESLA,H-Desch,DCR}, the main
production mechanisms for the SM Higgs particles are the Higgs--strahlung
\cite{H-LQT,Higgs:R1} and the $WW$ fusion  \cite{P1:VVH,Higgs:R2} processes
$\eei \to ZH \to f\bar f H$ and $\eei \to \bar{\nu}_e  \nu_e H$; see
Fig.~\ref{Hfig:xs} (left). The final state $H\nu \bar \nu$ is generated in both
the fusion and Higgs--strahlung processes.  Besides the $ZZ$ fusion mechanism
\cite{P1:VVH,Higgs:R2} $\eei \to \eei H$ which is similar to $WW$ fusion but
with an order of magnitude smaller cross section, sub--leading Higgs production
channels are associated production with top quarks $\eei \to t\bar{t}H$ 
\cite{Higgs:R3} and double Higgs production  \cite{Higgs:R4, Higgs:R4b} in the
Higgs--strahlung $\eei \to ZHH$ and fusion  $\eei \to \bar{\nu} \nu HH$
processes. Despite the smaller production rates,  the latter mechanisms are very
useful when it comes to the study of the Higgs fundamental properties. The
production  rates for all these processes are shown in Fig.~\ref{Hfig:xs}
(center)  at a c.m. energy of $\sqrt s\!=\!500$ GeV  as a function of $M_H$. 

The cross section for Higgs--strahlung scales as $1/s$ and therefore dominates
at low energies, while the one of the $WW$ fusion mechanism rises like
$\log(s/M_H^2)$ and becomes more important at high energies. The electroweak
radiative corrections to both processes are known and are under control 
\cite{HO-HZ,HO-VVH}. At $\sqrt{s} \sim 500$ GeV, the two processes have
approximately the same cross sections, ${\cal O} (50~{\rm fb})$ for the
interesting Higgs mass range 115 GeV$\,\lsim M_H \lsim\,$200 GeV favored by
high--precision data.  For the expected ILC integrated luminosity ${\cal L} \sim
500$ fb$^{-1}$, about  35000  events can be collected in the $\eei \to HZ$ and
$\eei \to \nu \bar \nu H$ channels for $M_H \sim 120$ GeV, which is  more than
enough to observe the Higgs particle and to study its properties in great
detail.\s

Turning to the sub--leading processes, the $ZZ$ fusion mechanism $\eei \to
H\eei$ is similar to $WW$ fusion but has a cross section that is one order of
magnitude smaller; however, the full final state can be reconstructed. The
associated production with top quarks has a very small cross section at $\sqrt s
=500$ GeV due to phase space suppression but, at $\sqrt s=800$ GeV, it can reach
the level of a few fbs.  The $t\bar{t}H$ final state is generated almost
exclusively through radiation off top quarks,  thus allowing an unambiguous
determination of the $g_{Htt}$ Yukawa coupling; the process  is also very
sensitive to the spin--parity of the $H$ boson \cite{Hspin}. The electroweak and
QCD corrections are moderate \cite{HO-Htt}, except near  threshold where large
coulombic corrections occur and double the production rate. For $M_H \lsim 140$
GeV, the main signal $t \bar t H \to W^+W^- b\bar{b} b \bar{b}$ is spectacular
and $b$--tagging as well as the reconstruction of the $M_H$ peak are essential
to suppress the large backgrounds. For $M_H \gsim 140$ GeV, the process leads
mainly to $H t \bar t \to 4W b\bar b$ final states which give rise to ten jets
if all $W$ bosons are allowed to decay hadronically to increase the
statistics.\s

The cross section for double Higgs production in the strahlung process, $\eei
\to HHZ$, is at the level of $\sim \frac12$ fb at $\sqrt{s}=500$ GeV for  a
light Higgs boson, $M_H \sim 120$ GeV, and is smaller at higher energies
\cite{Higgs:R4b}. It is rather sensitive to the trilinear Higgs--self coupling
$\lambda_{HHH}$: for $\sqrt{s}\!=\!500$ GeV and $M_H\!=\!120$ GeV for instance,
it varies by about 20\% for a 50\% variation of $\lambda_{HHH}$.  The
electroweak corrections to the process have been shown to be moderate
\cite{HO-HHH}.  The characteristic signal  for $M_H \lsim 140$ GeV consists of
four $b$--quarks to be tagged and a $Z$ boson which needs to be reconstructed in
both leptonic and hadronic final states to increase the statistics.  For higher
Higgs masses, the dominant signature is $Z+4W$ leading to multi--jet (up to 10)
and/or multi--lepton final states.  The rate for double Higgs production in 
$WW$ fusion, $\eei \to \nu_e \bar{\nu}_e HH$, is extremely small at $\sqrt{s}
=500$ GeV but increases with energy to reach the level of $\frac12$ fb at 1 TeV.
\s

Finally, future linear colliders can be turned to $\gamma \gamma$ colliders, in
which the photon beams are generated by Compton back--scattering of laser light
with c.m.~energies and integrated luminosities only slightly lower than that of
the original $\eei$ collider. Tuning the maximum of the $\gamma \gamma$ spectrum
to the value of $M_H$, the Higgs can be formed as $s$--channel resonances,
$\gamma \gamma \to H$, decaying mostly into $b\bar{b}$  and/or $WW^*,ZZ^*$
final states. This allows precise measurement of the Higgs couplings to photons
as well as the  CP nature of the Higgs particle \cite{Higgs:R5}. The $e^- e^-$
option is also possible.\s

In Higgs--strahlung, the recoiling $Z$ boson is mono--energetic and the Higgs
mass can be derived from the $Z$ energy when the initial $e^\pm$ beam energies
are sharp  (the effects of beamstrahlung must be
thus suppressed as strongly as possible).  The $Z$ boson can be tagged through
its clean $\ell^+ \ell^-$ decays ($\ell\!=\!e,\mu$) but also through decays into
quarks which have a much larger statistics. Therefore, it will be easy to
separate the signal from the backgrounds. In the low
mass range, $M_H\!\lsim\!140$ GeV, the process leads to $b\bar{b}q\bar{q}$ and
$b\bar{b}\ell \ell$ final states, with the $b$ quarks being efficiently tagged
by  micro--vertex detectors. For $M_H\!\gsim\!140$ GeV where the decay $H \to
WW^*$ dominates, the Higgs boson can be reconstructed by looking at the $\ell
\ell + \,$4--jet or 6--jet final states, and using the kinematical constraints
on the fermion invariant masses which peak at $M_W$ and $M_H$, the backgrounds
are efficiently suppressed. Also the $\ell \ell q\bar q  \ell \nu$ and $q\bar q
q\bar q  \ell \nu$ channels are easily accessible.\s

It has been shown in detailed simulations \cite{TESLA,H-Desch} that only a few
fb$^{-1}$ data are needed to obtain a 5$\sigma$ signal for a Higgs boson with a
mass $M_H \sim 120$ GeV at a 350 GeV collider; see Fig.~\ref{Hfig:xs} (right)
with 500 fb$^{-1}$ data. In fact, for such small masses, it is better to move to
lower energies where the Higgs--strahlung cross section is larger and the
reconstruction of the $Z$ boson is better \cite{H:Richard}. 
Moving to higher energies, Higgs
bosons with masses up to $M_H\sim 400$ GeV can be discovered in the
Higgs--strahlung  process at an energy of 500 GeV and with a luminosity of 500
fb$^{-1}$. For even larger masses, one needs to increase the c.m. energy of the
collider and, as a rule of thumb, Higgs masses up to $\sim 80$\% $\sqrt{s}$ can
be probed. This means that a 1 TeV collider can probe the entire Higgs mass
range that is theoretically allowed in the SM, $M_H \lsim 700$ GeV.\s

The $WW$ fusion mechanism offers a complementary production channel. For low
$M_H$  where the decay $H\to b\bar{b}$ is dominant, flavor tagging plays an
important role to suppress the background.  The $\eei \to H\bar{\nu}\nu \to
b\bar{b}\bar{\nu}\nu$ final state can be separated  from the corresponding one
in the process, $\eei \to HZ \to b\bar{b}\bar{\nu}\nu$, by exploiting their
different characteristics in the $\nu \bar{\nu}$ invariant mass \cite{TESLA}.
The polarization of the $e^\pm$  beams, which allows tuning of the $WW$ fusion
contribution, can be very useful to control the systematic uncertainties.  For
larger $M_H$, when the decays $H \to WW^{(*)},ZZ^{(*)}$ and even $t\bar t$ are
dominant, the backgrounds can be suppressed using kinematical constraints from
the reconstruction of the Higgs mass  peak and exploiting the signal
characteristics.

\subsection{Higgs production in the MSSM}

\begin{figure}[hbtp]
\vspace*{-.1mm}
\begin{center}
\includegraphics[width=0.85\linewidth,bb=73 430 600 745]{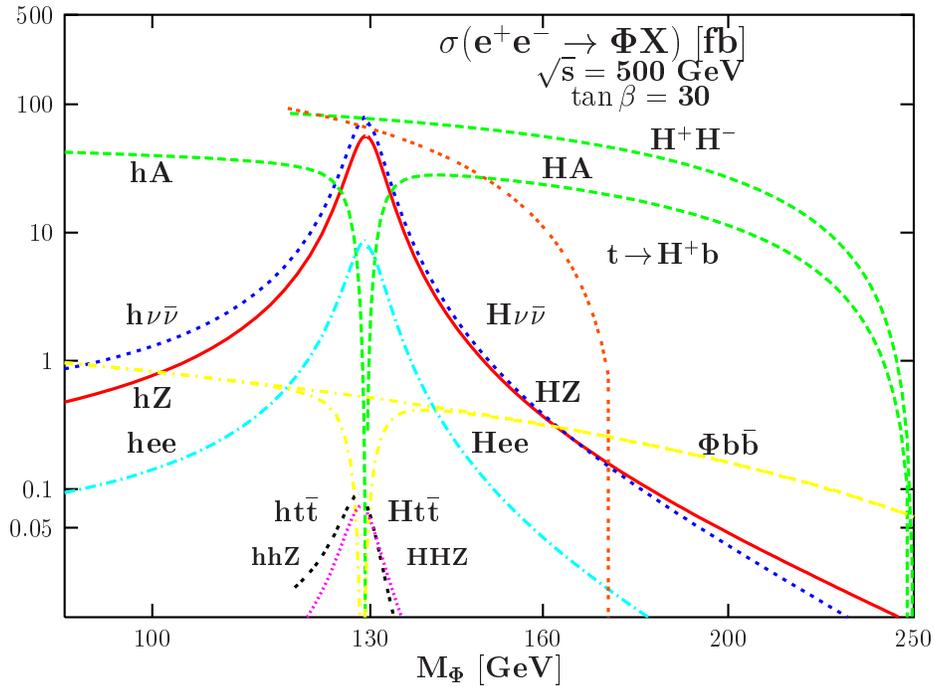} 
\end{center}
\vspace*{-5mm}
\caption[Production cross sections of the MSSM Higgs bosons at a 500 GeV ILC]
{Production cross sections of the MSSM Higgs bosons in $e^+ e^-$ collisions
as functions of the masses for $\tb=30$ and $\sqrt{s}=500$ GeV; from
Ref.~\cite{Anatomy2}.}
\label{Hfig:ee-MSSM}
\vspace*{-1mm}
\end{figure}

At the ILC, besides the usual Higgs--strahlung and fusion processes for $h$ and
$H$ production, the neutral Higgs particles can also be produced pairwise: $\eei
\to A + h/H$ \cite{H:eeSUSY}. The cross sections for the Higgs--strahlung and
the pair production as well as the cross sections for the production of $h$ and
$H$ are mutually complementary, coming either with a coefficient $\sin^2(\beta-
\alpha)$ or $\cos^2(\beta -\alpha)$; Fig.~\ref{Hfig:ee-MSSM}.  The cross section
for $hZ$ production is large for large values of $M_h$, being of ${\cal O}(100$
fb) at $\sqrt{s}=500$ GeV; by contrast, the cross section for $HZ$ is large for
light $h$ (implying small $M_H$). In major parts of the parameter space, the
signals consist of a $Z$ boson and $b\bar{b}$ or $\tau^+ \tau^-$ pairs, which is
easy to separate from the backgrounds with flavor tagging. For associated
production, the situation is opposite: the cross section for $Ah$ is large for
light $h$ whereas $AH$ production is preferred in the complementary region.  The
signals consists mostly of four final $b$ quarks, requiring efficient $b$--quark
tagging; mass constraints help to eliminate the QCD jets and $ZZ$ backgrounds.
The CP--even Higgs particles can also be searched for in the $WW$ and $ZZ$
fusion mechanisms.\s

In $\eei$ collisions, charged Higgs bosons can be produced pairwise, $\eei \to
H^+H^-$, through $\gamma,Z$ exchange. The cross section depends only on the
charged Higgs mass; it is large almost up to $M_{H^\pm} \sim \frac12 \sqrt{s}$.
$H^\pm$ bosons can also be produced in top decays; in the range $ 1 < \tb <
m_t/m_b$, the $t \to H^+b$ branching ratio and the $t\bar{t}$ production cross
sections are large enough to allow for their detection in this mode. [$H^\pm$ 
can also be pair--produced in $\gamma \gamma$ collisions with large rates].\s

The discussion of SUSY Higgs production at ILC can be summarized in the
following  points.\s

\begin{figure}[ht!]
\vspace*{4mm}
\begin{center}
\begin{tabular}{ccc}
\epsfig{file=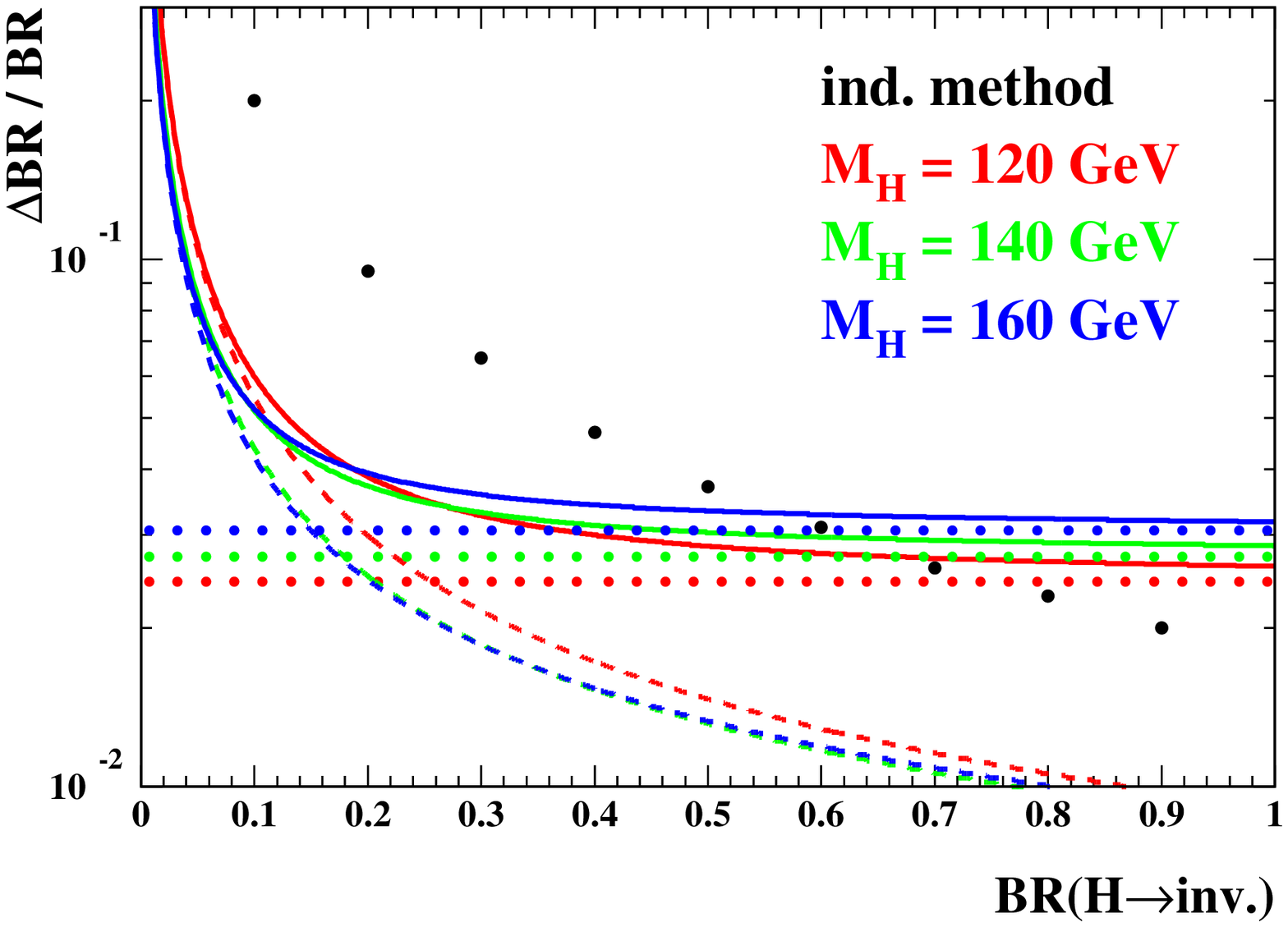,height=7cm,width=5.4cm}
&
\epsfig{file=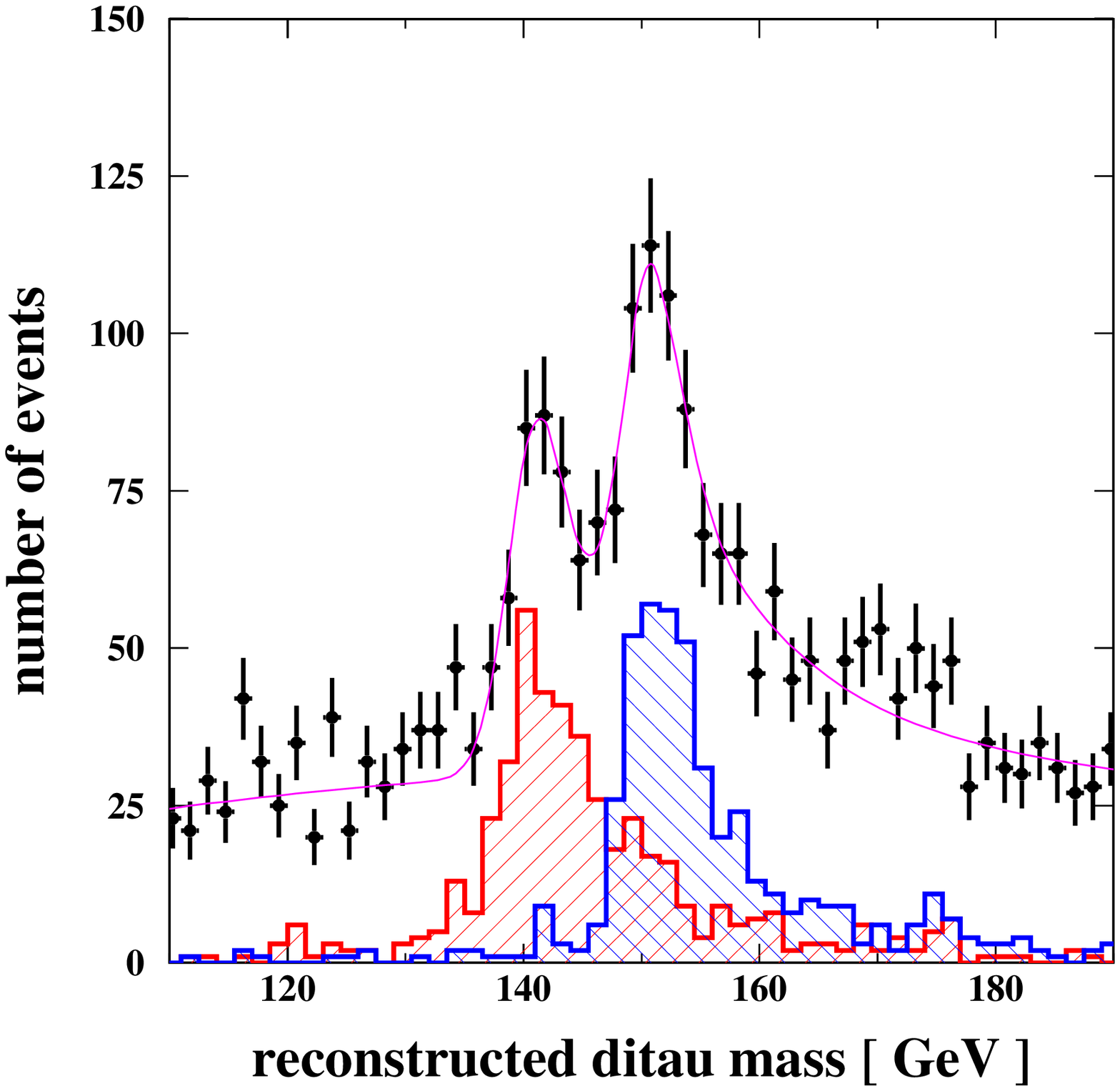,height=7cm,width=5.4cm,clip}
&
\epsfig{file=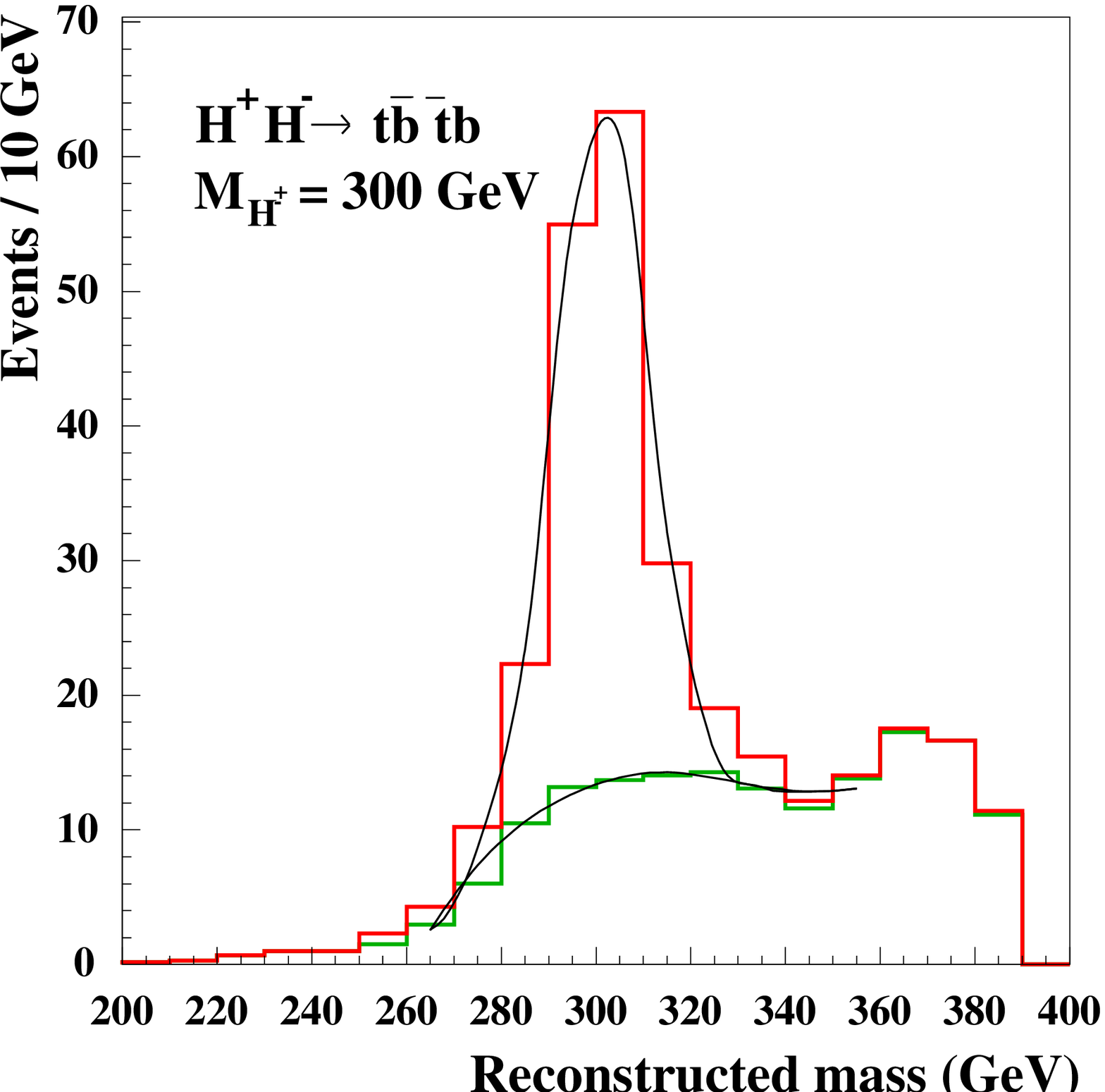,height=7cm,width=5.4cm,clip}
\end{tabular}
\end{center}
\vspace*{-4mm}
\caption[] {The expected accuracy on the invisible decay rate as a function of
the  branching ratio at $\sqrt {s}=350$ GeV  in full lines; the other lines are
from measurement of the invisible rate (dashed), the total cross section 
(dotted) and an indirect method (large dots) \cite{Schumacher:2003ss}
(left). The reconstructed $\tau\tau$ invariant mass from a kinematic fit in
 $\eei\! \to\! HA\! \to\! b \bar b\tau^+ \tau^-$ for $M_A\!=\!140$ 
GeV and $M_H\!=\! 150$ GeV at $\sqrt{s}\!=\!500$ GeV \cite{HAee-desch} (center).
The di--jet invariant mass distribution for the $e^+ e^-\! \to\! H^+H^-\! \to\!
t \bar b \bar t b$  process for $M_{H^\pm}\!=\!300$ GeV
after final state constraints at $\sqrt{s}\!=\! 800$ GeV
\cite{TESLA} (right).  In all cases, a luminosity of 500 fb$^{-1}$  is assumed.}
\vspace*{-2mm}
\label{Hfig:MSSM-heavy}
\end{figure}

-- The Higgs boson $h$ can be detected in the entire range of the MSSM parameter
space, either through the Higgs--strahlung (and $WW$ fusion) process or
associated production with the pseudoscalar $A$ boson.  In fact, this conclusion
holds true even at a c.m. energy of 250 GeV and with a luminosity of a few
fb$^{-1}$.  Even if the decay modes of the $h$ boson are very complicated,
missing mass techniques allow for their detection. For instance, the branching 
ratios for the invisible $h$ boson decays into the LSP neutralinos can be
measured at the percent level; see Fig.~\ref{Hfig:MSSM-heavy} (left). The
accuracy can be substantially improved by running at lower c.m. energies
\cite{H:Richard}. The same very detailed tests and precision measurements for
the SM Higgs boson (see later) can be performed for the MSSM $h$ boson, in
particular in the decoupling limit, thus complementing  LHC analyses
\cite{Weiglein:2004hn}.\s

-- All SUSY Higgs bosons can be discovered at an $\eei$ collider if the $H,A$
and $H^{\pm}$ masses are less than the beam energy; for higher masses, one
simply has to increase the c.m. energy, $\sqrt{s} \gsim 2M_A$.  Several 
channels might be observable depending on the value of $\tan{\beta}$. The
dominant processes will be however Higgs pair production $\ee \to HA$ and $H^+
H^-$ for which the cross sections are not suppressed by mixing factors (for $M_A
\gsim M_h^{\rm max}$ in the case of $HA$ production). This is exemplified in the
central and right--handed panels of Fig.~\ref{Hfig:MSSM-heavy}.  Note that the
additional associated neutral Higgs production processes with  $t\bar{t}$ and
$b\bar{b}$ allow for the measurement of the Yukawa couplings. In particular,
$\eei \to b\bar{b}+h/H/A$ for high $\tb$ values allow for the determination of 
the important $\tb$ parameter for low $M_A$ values.\s

-- If the energy is not high enough to open the $HA$ pair production threshold,
the photon collider option may become the discovery machine for the heavy Higgs
bosons \cite{Higgs:R5,Asner:2001ia}. Since the $A,H$ bosons are produced as
$s$--channel resonances, the mass reach at a photon collider is extended
compared to the $e^+e^-$ mode and masses up to 80\% of the original c.m. energy
can be probed.  It has been shown in Ref. \cite{Asner:2001ia} that the whole
medium $\tan{\beta}$ region up to about 500 GeV, where only one light Higgs
boson can be found at the LHC (the so--called wedge region of the LHC), can be
covered by the photon collider option with three years of operation with an
$e^-e^-$ c.m.~energy of 630 GeV. The photon collider mode is also important to
determine the CP properties of the heavy Higgs bosons, either by studying
angular correlation of Higgs decay products or by using initial beam
polarization. The discrimination between the scalar and pseudoscalar particles
can be performed and CP violation  can be unambiguously probed
\cite{Choi:2004kq}  .

\subsection{High--precision measurements of the Higgs properties}

The profile of the lighter Higgs boson can be entirely  determined. This is
particularly the case close to the decoupling regime where the $h$ boson behaves
like the SM Higgs particle but with a mass below $M_h \sim 140$ GeV.  This is,
in fact, the most favorable mass range for precision measurements as the Higgs
boson has many decay channels that are accessible in this case.  A short summary
of the measurements which can be performed is as follows; see
Refs.~\cite{TESLA,H-Desch,DCR} for details and references.\s

$\bullet$ The measurement of the recoil $f\bar{f}$ mass in the Higgs--strahlung
process, $\ee \ra  h f\bar{f}$ allows a very good determination of the Higgs 
mass: at $\sqrt{s}=350$ GeV and with $ 500$ fb$^{-1}$ data, a precision of $
\Delta M_h \sim 50$ MeV can be reached for $M_h \sim 120$ GeV.  [Accuracies 
$\Delta M_H \sim 80$ MeV can also be reached for $M_H=150$ and 180 GeV when the
heavier Higgs decays mostly into gauge bosons.]  \s

$\bullet$ The angular distribution of the $Z/h$ in the strahlung process,  $\sim
\sin^2\theta$ at high energy, characterizes the production of a  $J^P=0^+$
particle. The Higgs spin--parity quantum numbers can also be checked by looking
at correlations in the production $\ee \ra hZ \ra 4f$ or decay $h \ra WW^* \ra
4f$ processes, as well as in the channel $h \ra \tau^+ \tau^-$. An unambiguous
test of the CP nature of the $h$ boson can be made in threshold and polarization
analyses in the process $\ee \ra t \bar{t}h$ [or at laser photon colliders
in the  loop--induced process $\gamma \gamma \ra h$].\s

$\bullet$ The Higgs couplings to $ZZ/WW$ bosons, which are predicted to be
proportional to the masses, can be directly determined by measuring the
production cross sections in the strahlung and the fusion processes.  In the
$\ee \ra \ell^+ \ell^-+h$ and $\nu \bar{\nu}+h$ processes, the total cross 
section can be measured with a precision less than $\sim$ 3\% at $\sqrt{s}\sim
500$ GeV  with 500 fb$^{-1}$ integrated luminosity if $h$ is SM--like. This
leads to an accuracy of  less than 1.5\% on the $hVV$ couplings.\s

$\bullet$ The measurement of the Higgs branching ratios  is of utmost
importance. Since $M_h \lsim 130$ GeV, a large variety of branching ratios can
be measured: the $b\bar{b}, c\bar{c}$  and $\tau^+ \tau^-$ branching ratios
allow us to derive the relative Higgs--fermion couplings and to check the
prediction that they are proportional to the masses. The gluonic branching ratio
is sensitive to the $t\bar{t}h$ Yukawa coupling and to new strongly interacting
particles, such as stops in the MSSM. The branching ratio into $W$ bosons allows
a measurement of the $hWW$ coupling, while the branching ratio of the
loop--induced $\gamma \gamma$ decay is also very important since it is sensitive
to new particles.  \s

$\bullet$ The Higgs coupling to top quarks, which is the largest coupling in the
theory, is directly accessible in the process where the Higgs boson is radiated
off top quarks, $\ee \ra t\bar{t}h$. For $M_h \lsim 130$ GeV, the Yukawa
coupling  can be measured with a precision of less than 5\% at $\sqrt{s}\sim
800$ GeV with a luminosity of ${\cal L} \sim 1$ ab$^{-1}$. \s

$\bullet$ The total width of the SM Higgs boson, for masses less than $\sim 200$
GeV, is so small that it cannot be resolved experimentally. However, the
measurement of BR($h \ra WW$) allows an indirect determination of $\Gamma_h$,
since the $hWW$ coupling can be determined from the measurement of the Higgs
cross section in the $WW$ fusion process. [$\Gamma_{\rm tot}$ can also be
derived by measuring the $\gamma \gamma \to h$ cross section at a 
$\gamma\gamma$ collider or the branching ratio of $h \to \gamma \gamma$ in 
$\ee$ collisions].\s

$\bullet$ Finally, the measurement of the trilinear Higgs self--coupling, which
is the first non--trivial test of the Higgs potential, is accessible  in the
double Higgs production processes $\ee \ra Zhh$ [and in the $\ee \ra \nu
\bar{\nu}hh$ process at high energies]. Despite its smallness, the cross 
sections can be determined with an accuracy of the order of 20\% at a 500 GeV 
collider if a high luminosity, ${\cal L} \sim 1$ ab$^{-1}$, is available. [For
not too large $M_H$, the coupling $\lambda_{Hhh}$ can also be accessed.] \s

An illustration of the experimental accuracies that can be achieved in the
determination of the mass, CP--nature, total decay width and the various
couplings of a SM--like Higgs boson for the two masses $M_h=120$ and 140 GeV is
shown in Table 1 for $\sqrt{s}=350$ GeV [for $M_h$ and the CP nature] and $500$
GeV [for $\Gamma_{\rm tot}$ and all couplings except for $g_{htt}$] and for
$\int {\cal L}=500$ fb$^{-1}$ [except for $g_{htt}$ where $\sqrt{s}=1$ TeV and
$\int {\cal L}=1$ ab$^{-1}$ are assumed]. The achievable accuracy is
impressive.\s

\begin{table}[htbp] 
\vspace*{1mm}
\hspace*{-.7cm} 
\renewcommand{\arraystretch}{2.}
\hskip3pc
\vbox{\columnwidth=26pc
\begin{tabular}{|c|c|c|c||c|c|c|c|c|c|c|c|}\hline 
$M_h$ (GeV) & $\Delta M_h$ & $\Delta {\rm CP}$ & $\ \Gamma_{\rm tot}\ $ & $g_{hWW}$
& $g_{hZZ}$ & $g_{htt}$ & $g_{hbb}$ & $g_{hcc}$ & $g_{h\tau \tau}$ & $g_{hhh}$ 
\\ \hline 
$120$ & $\pm 0.033$ & $\pm 3.8$ & $\pm 6.1$ & $\pm 1.2$ & $\pm 1.2$ & $\pm 3.0$
& $\pm 2.2$ & $\pm 3.7$ & $\pm 3.3$ & $\pm 17$  \\ \hline 
$140$ & $\pm 0.05$ & $-$ & $\pm 4.5$ & $\pm 2.0$ & $\pm 1.3$ & $\pm 6.1$ & 
$\pm 2.2$ & $\pm 10$ & $\pm 4.8$ & $\pm 23$  \\ \hline 
\end{tabular} } 
\vspace*{-.1mm} 
\caption{Relative accuracies (in \%) on the SM--like Higgs boson mass, width 
and couplings obtained at the ILC with $\sqrt{s}=350,500$ GeV and $\int {\cal 
L}=500$ fb$^{-1}$  (except for top); Ref.~\cite{TESLA}. }
\end{table} 

A number of very important measurements can be performed at the ILC in the MSSM
heavier Higgs sector. If the $H,A$ and $H^\pm$ states are kinematically
accessible, one can measure their masses and cross sections times decay
branching ratios with a relatively good accuracy. In the pair production process
$\eei \to HA$, a precision of the order of $0.2\%$ can be achieved on the $H$
and $A$ masses, while a measurement of the cross sections can be made at the
level of a few percent in the $b\bar b b\bar b$ and ten percent in the $b\bar b
\tau^+ \tau^-$ channels. For the charged Higgs boson, statistical uncertainties
of less than 1 GeV on its mass and less than 15\% on its production cross
section times branching ratio can be achieved in the channel $e^+e^- \to H^+H^-
\to t \bar b \bar t b$ for $M_{H^\pm} \sim 300$ GeV with high enough energy and
luminosity. \s

 These measurements allow the determination of the most important branching
ratios, $b\bar b$ and $\tau^+\tau^-$ for the $H/A$ and $tb$ and $\tau \nu$ for
the $H^\pm$ particles, as well as the total decay widths which can be turned
into a determination of the value of $\tb$, with an accuracy of 10\% or less. 
The spin--zero nature of the particles can be easily checked by looking at the
angular distributions which should go as $\sin^2\theta$. Several other
measurements, such as the spin--parity of the Higgs particles in $H/A \to \tau^+
\tau^-$ decays and, in favorable regions of the parameter space, some trilinear
Higgs couplings such as $\lambda_{Hhh}$, can be made.\s

The high--precision achievable at the ILC in the SUSY Higgs sector would allow
to determine two very important parameters, $\tb$ and $M_A$, which can be used
as inputs in the extrapolation of low energy scenarios to the GUT scale to
reconstruct the fundamental SUSY theory.

\subsection{Global analyses and LHC--ILC complementarity}

\begin{figure}[h!]
\begin{center}
\mbox{
\epsfig{file=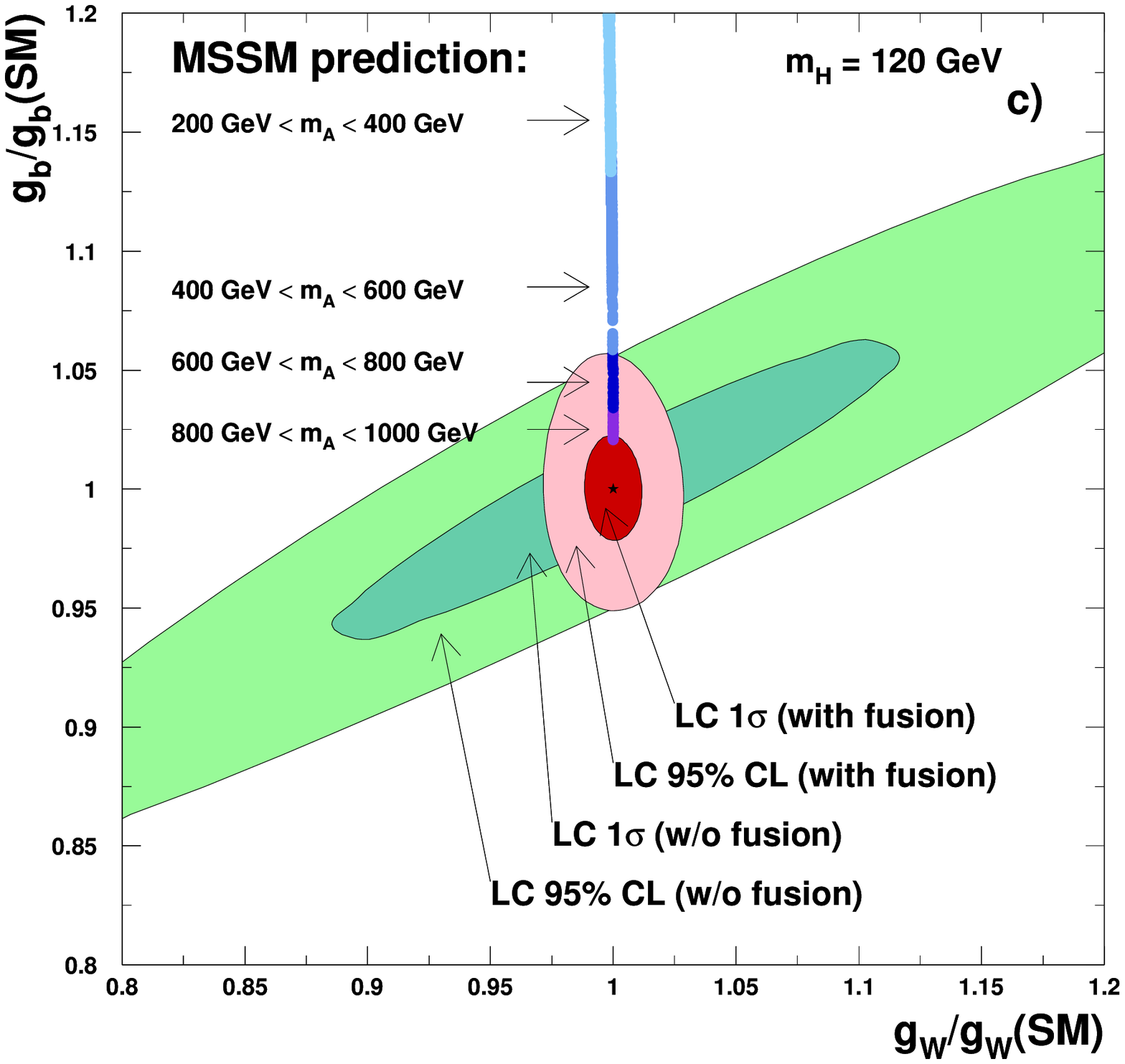,width=0.35\linewidth}\hspace*{-4mm}
\epsfig{file=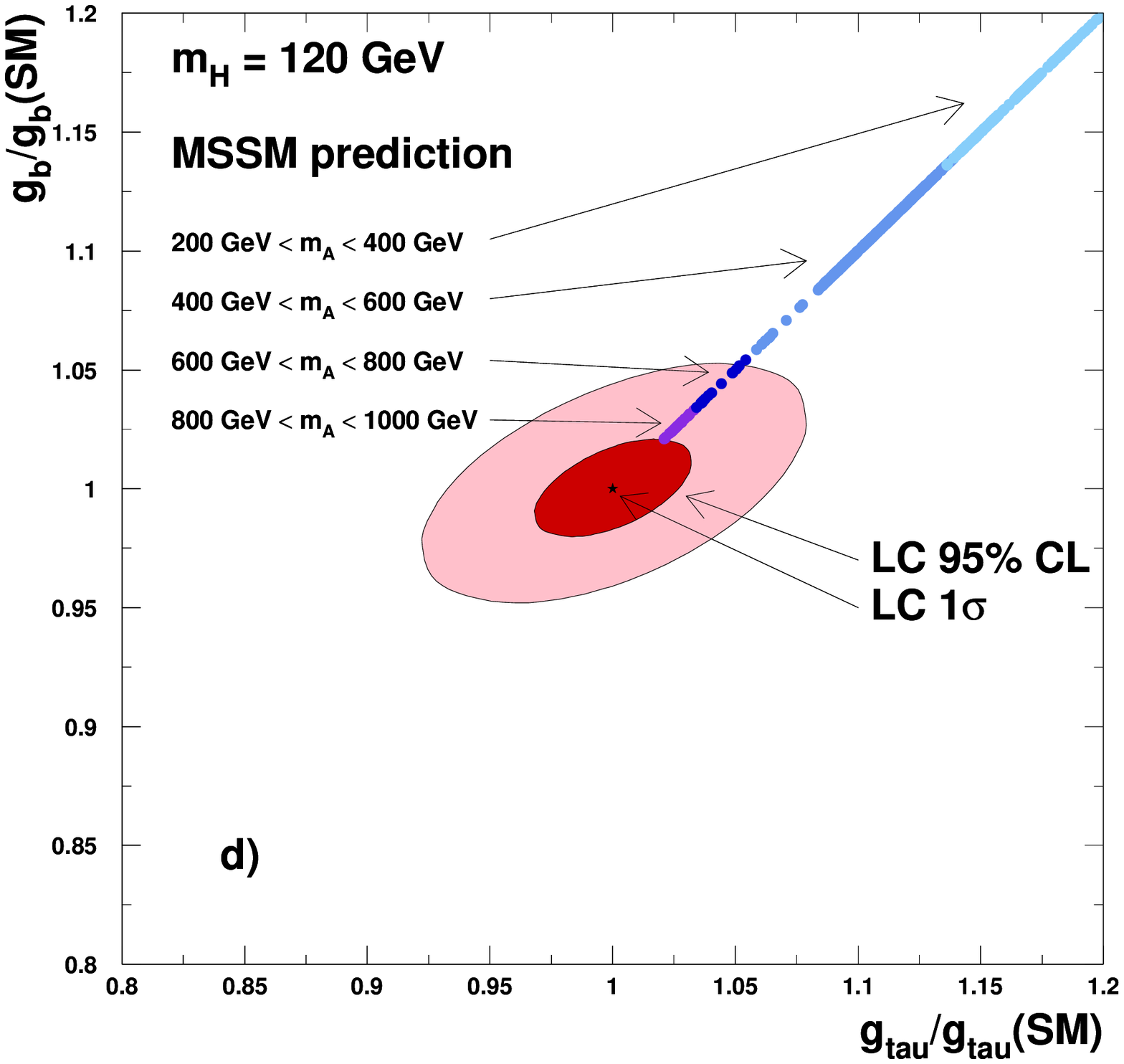,width=0.35\linewidth}\hspace*{-4mm}
\epsfig{file=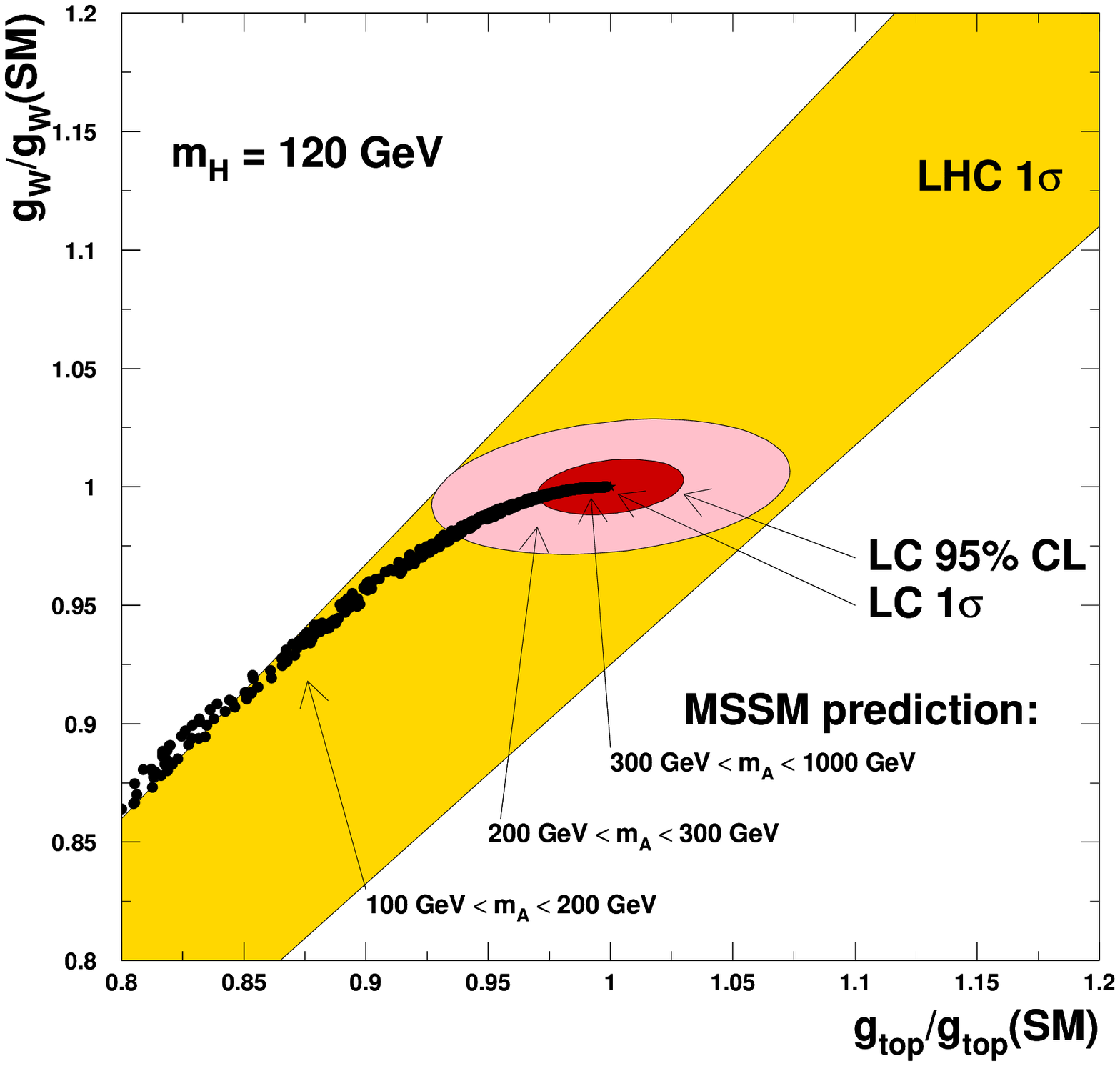,width=0.32\linewidth}
}
\end{center}
\vspace*{-.6cm}
\caption[Determination of the couplings of a SM--like Higgs and MSSM interpretation]
{Determination of the couplings of a SM--like Higgs boson at the ILC
and the interpretation within the MSSM. The contours are the couplings of
a 120 GeV Higgs boson as measured with 500\,fb$^{-1}$ data at $\sqrt{s}=350$
GeV except for $g_{Htt}$ which uses 800 GeV (here the expectation at the
LHC is also shown); from Ref.~\cite{TESLA}.}
\vspace*{-.1cm}
\label{Hfig:cpl-TESLA}
\end{figure}

A detailed analysis of the deviations of the couplings of the $h$ boson with a
mass $M_h=120$ GeV, from the predictions in the SM  has been performed in
Ref.~\cite{TESLA} using a complete scan of the MSSM $[M_A,
\tb]$ parameter space, including radiative corrections. In
Fig.~\ref{Hfig:cpl-TESLA}, shown are the  1$\sigma$ and 95\% confidence level
contours for the fitted values of various pairs of ratios of couplings, assuming
the experimental accuracies at the ILC discussed in the previous section
and summarized in Tab.~1.\s

From a $\chi^2$ test which compares the deviations, the MSSM  can be
distinguished from the SM case at the 95\% confidence level for $M_A \lsim 600$
GeV (and only at the 68\% confidence level for $M_A \lsim 750$ GeV). In some
cases, one is sensitive to MSSM effects even for masses $M_A\!\sim\!1\;$TeV,
i.e. beyond the LHC mass reach. If the deviations compared to the SM are large,
these precision measurements would also allow for an indirect determination of
$M_A$; for instance, in the mass range $M_A=300$--600 GeV an accuracy of 70--100
GeV is possible on the $A$ mass. \s

This type of indirect determination cannot be made in a convincing way at the
LHC as the experimental errors in the various measurements are worse than at the
ILC; see Fig.~\ref{Hfig:cpl-TESLA} (right) where the $g_{hWW}$ and $g_{htt}$
contours are displayed. While at the ILC, MSSM effects can be probed for  masses
close to $M_A=1$ TeV, there is practically no sensitivity at the LHC. However,
the precision measurements at the ILC can gain enormously from other
measurements that can be performed only at the LHC. \s

Indeed, the various Higgs couplings are not only sensitive to the tree--level
inputs $M_A$ and $\tb$ but also, on parameters that enter through radiative
corrections such as the stop and sbottom masses which could be accessible only
at the LHC. If, in addition, the $A$ boson is seen at the LHC (which means that
$\tb$ is large, $\tb \gsim 10$) and its mass is measured at the level of 10\%,
the only other important parameter entering the Higgs sector at one--loop is the
trilinear coupling $A_t$ (and to a lesser extent, $A_b$ and $\mu$) which will be
only loosely constrained at the LHC. Nevertheless, using this knowledge and the
fact that the top mass (the uncertainty of which generates the largest error as
the corrections are $\propto m_t^4$) can be measured with a precision of 100 MeV
at the ILC, one can vastly improve the tests of the MSSM Higgs sector that can
be performed at the LHC or at the ILC alone. This is one example of the possible
complementarity between the LHC and the ILC;  for more discussions  and examples
see Ref.~\cite{Weiglein:2004hn}.

\subsection{Extended Higgs sectors at the ILC}

In the CP--violating MSSM where the three neutral Higgs bosons $H_1, H_2, H_3$
are mixtures of CP--even and CP--odd states, because of the sum rule for the
Higgs couplings to gauge bosons, $\sum_i g_{H_iVV}^2 =g^2_{H_{ SM}}$, the
production cross sections in the Higgs--strahlung and $WW$ fusion processes
should be large for at least one of the particles and there is a complementarity
between $H_i$ single and $H_j H_k$ pair production.  In fact, similarly to the
usual MSSM, the  normalized couplings are such that $|g_{H_1VV}| = |g_{H_2H_3V}|
\sim 1$ in the decoupling limit $M_{H^\pm} \gsim 200$ GeV and at least $H_1$ is
accessible for $\sqrt s \gsim 300$ GeV,  since $M_{H_1}\lsim 130$ GeV.  If two
or the three Higgs particles  are very close in mass, the excellent energy and
momentum resolution on the recoiling $Z$ boson in the Higgs--strahlung process
would allow to resolve  the coupled Higgs systems, e.g. from an analysis of the
lineshape (this is in fact similar to the MSSM in the intense  coupling
regime).  . The presence of CP--violation can be   unambiguously checked by
studying the spin--spin correlations in Higgs decays into tau lepton pairs or
controlling the  beam polarization of the colliding photon beams at the $\gamma
\gamma$ option of the ILC; see Ref.~\cite{Choi:2004kq} for instance. \s

The ILC will also be very useful in probing the Higgs sector of  the NMSSM with
the additional CP--even and CP--odd Higgs particles. As seen previously, 
Higgs--strahlung, $\ee \to Z H_i$,   allows for the detection of CP--even Higgs
particles independently of their decay modes and thus, even if they decay  into
the singlino--like light $A_1$ or $H_1$ states. This is possible provided that
their couplings to the $Z$ boson are substantial, as it always occurs for at
least one CP--even Higgs boson. In fact,  thanks to the usual  sum rule which
relates the CP--even Higgs couplings to the those of the SM Higgs boson, a
``no--lose theorem" for discovering at least one Higgs state has been
established for ILC while for LHC, as discussed in the previous subsection, the
situation is presently less clear and all Higgs particles could escape
detection.\s

In the general SUSY scenario with an arbitrary number of singlet and doublet
fields, one Higgs particle has significant $ZZ$ coupling and a mass smaller than
200 GeV.  This particle should be therefore kinematically accessible at the ILC
with a c.m. energy $\sqrt s \gsim 350$ GeV. It can be detected in the
Higgs--strahlung process independently of its (visible or invisible) decay
modes. If its mass happens to be in the high range, $M_h \sim 200$ GeV, at least
its couplings to $W,Z$ bosons and $b$--quarks (eventually $t$--quarks at high
energies and luminosities), as well as the total decay widths and the
spin--parity quantum numbers can be  determined. \s

We should stress again that even in scenarios with invisible Higgs decays, as
would be the case for instance of spontaneously broken R--parity scenarios in
which the Higgs particles could decay dominantly into invisible Majorons, $H_i
\to JJ$, at least one CP--even Higgs boson is light and has sizable couplings to
the gauge bosons and should be observed by studying the recoil mass spectrum
against the $Z$ boson in the Higgs--strahlung process. Furthermore, if doubly
charged Higgs bosons of left--right symmetric models occur \cite{H++th,H:higheR}
and if kinematically accessible at the ILC, they can be pair produced in $\ee$
collisions, $\ee \to H^{++} H^{--}$ with large rates  [also in $\gamma \gamma$
collisions where, because of the large electric charge, the rates are more than
an order of magnitude larger than for singly charged $H^\pm$ bosons]; they can 
also be singly produced  in $e^- e^-$ collisions, and thus with a much more
favorable phase space,  if the Yukawa couplings are not too small \cite{ee-H++}.
Finally, in the presence of a new $Z'$ boson \cite{Zprime,eeZprime}, the
Higgs--strahlung process would receive additional contributions from the virtual
exchange of this new particle, $\ee \to Z, Z'\to  hZ$, and  thanks to the
high--luminosity and to  the clean environment, the expectedly small deviations
of the production cross section from the SM or MSSM cases could be detected.  \s

Thus, from the previous discussions, one can thus conclude that the ILC is an
ideal  machine for the SUSY Higgs sector, whatever  scenario nature has chosen.

\section{Conclusion}

The LHC will soon provide us with the  answer to the question that particle
physicists are asking themselves since the seminal paper of Julius Wess and
Bruno Zumino,  almost four decades ago: is low--energy Supersymmetry realized in
Nature? The answer might first come from the Higgs sector, as a generic
prediction of low--energy SUSY is  the existence of at least one light Higgs
particle with a mass below $\sim 200$ GeV.  If the answer to the question is
positive, a new continent will be open to experimental investigation as well as
theoretical development. While the LHC will make the pioneering  exploration of
the new continent, the ILC will  be needed to fully chart it.\s


We are anxiously waiting for these breathtaking times.  Much to our regret,
Julius Wess will not be among us during these times. \bigskip

\nn {\bf Acknowledgments:} I thank J. Bartels and S. Bethke  for inviting me to
write this review and for their patience while waiting for the manuscript and
P.M. Zerwas  for comments on the draft. Support from  the Alexander von--Humbold Foundation
(Bonn, Germany) is acknowledged.

\newpage

\section*{Appendix}

In this appendix we sketch the derivation of the scalar Higgs potential 
eq.~(2) from the SUSY Superpotential and its soft--SUSY breaking counterpart.\s

The most general globally supersymmetric superpotential, compatible with 
gauge invariance, renormalizability and $R$--parity conservation can be written 
in terms of (hatted) superfields, as 
\beq
{\cal W}=\sum_{i,j=\rm generation} - Y^u_{ij} \, {\widehat {u}}_{Ri} \widehat{H_2} \! 
\cdot  \! {\widehat{ Q}}_j+ Y^d_{ij} \, {\widehat{ d}}_{Ri} \widehat{H}_1  
\! \cdot  \! {\widehat{ Q}}_j+Y^\ell_{ij} \,{\widehat{\ell}}_{Ri} \widehat{H}_1
 \! \cdot  \! {\widehat{ L}}_j+ \mu \widehat{H}_2  \! \cdot  \! \widehat{H}_1 
 \non
\label{defW}
\eeq
The product between SU(2)$_{\rm L}$ doublets for Higgses, quarks and leptons
reads $H\cdot Q \equiv \epsilon_{a b} H^a Q^b$; etc... where $a, b$ are
SU(2)$_{\rm L}$ indices and $ \epsilon_{12}=1 = - \epsilon_{21}$; 
$Y^{u,d,\ell}_{ij}$ denote the Yukawa couplings among generations. The first
three terms are nothing else but a superspace generalization of the Yukawa
interaction in the SM, while the last term is a globally supersymmetric Higgs
mass term. \s

The supersymmetric part of the tree--level scalar potential  is  the sum of
the   F-- and D--terms, where the F--terms  come from the superpotential
through derivatives with respect to all scalar fields $S_i$  and the D--terms 
correspond to the quartic scalar interactions under the SM gauge group
\beq
V_{F}={\sum_{i}  |W^{i}|^2} \ \ {\rm with} \ W^{i} = \partial{\cal W}/\partial{ 
S_i}  \ , \ 
V_{D}= \frac{1}{2}  \sum_{a=1}^{3}  \left(\sum_{i} g_a S_i^* T^a S_i
\right)^2 \non
\eeq

One then adds a set of terms which break  SUSY explicitely but softly: mass
terms for the gauginos $\sum_i \frac12 M_i V_i^\mu V_{i\mu}$, mass terms for
the  sfermions $\sum_i m^2_{{\tilde {F}}_i}
{\tilde{F}}_i^{\dagger}{\tilde{F}}_i$ as well as mass and bilinear terms for the
Higgs bosons and  trilinear couplings between sfermions and Higgs bosons:  
\beq
-{\cal L}_{\rm Higgs} &=& m^2_{H_2} H_2^{\dagger} H_2+m^2_{H_1}  H_1^{\dagger} 
H_1 + B \mu (H_2 \! \cdot  \! H_1 + {\rm h.c.} ) \non \\
&+& 
{\sum_{i,j=gen} { \left[ A^u_{ij} Y^u_{ij}  {\tilde{u}}^*_{R_i} H_2  \! \cdot 
\! {\tilde{Q}}_j+ A^d_{ij} Y^d_{ij}  {\tilde{d}}^*_{R_i} H_1  \! \cdot  \! 
{\tilde{Q}}_j +A^l_{ij} Y^\ell_{ij} {\tilde{\ell}}^*_{R_i} H_1  \! \cdot 
{\tilde{L}}_j \ + \ {\rm h.c.} \right] }} \non
\eeq

The terms contributing to the scalar Higgs potential $V_H$ come  from three
different sources: \s

\nn $i)$ The $D$ terms: for the two Higgs fields $H_1$ and $H_2$ with $Y=-1$ and 
$+1$, they are given by
\beq
V_D= {g_2^2\over 8}\bigg[ 4| H_1^\dagger\! \cdot\! H_2|^2 -2 |H_1|^2|H_2|^2
+ (|H_1|^2)^2+(|H_2|^2)^2 \bigg] +{g_1^{2}\over 8} (|H_2|^2 
-|H_1|^2)^2  \non
\eeq
\nn $ii)$ The $F$ term: from the term $W \sim \mu \hat{H}_1 \!\cdot\!\hat{H}_2$
of the Superpotential, one obtains the component 
\beq
V_{F}= \mu^2( |H_1|^2+|H_2|^2) \non
\eeq
\nn $iii)$ The soft SUSY--breaking scalar Higgs mass terms and the  bilinear 
term in ${\cal L}_{\rm Higgs}$ which give
\beq
V_{\rm soft}=  m^2_{H_1} H_1^{\dagger} H_1+m^2_{H_2}  H_2^{\dagger} H_2 + B \mu 
(H_2 \! \cdot\! H_1 + {\rm h.c.} ) \non
\eeq
The full scalar potential involving the Higgs fields, eq.~(2), is then the sum
of these terms:
\beq
V_H= V_D+V_F+V_{\rm soft} \non
\eeq

Note that in the NMSSM, with an additional singlet superfield $\widehat S$, 
the superpotential writes
\beq
{\cal W} = \sum_{i,j=gen} - Y^u_{ij} \, {\widehat {u}}_{Ri} \widehat{H_2} \! 
\cdot  \! {\widehat{ Q}}_j+ Y^d_{ij} \, {\widehat{ d}}_{Ri} \widehat{H}_1  
\! \cdot  \! {\widehat{ Q}}_j+Y^\ell_{ij} \,{\widehat{\ell}}_{Ri} \widehat{H}_1
 \! \cdot  \! {\widehat{ L}}_j +
\lambda \widehat{S} \widehat{H}_2 \widehat{H}_1 +
\frac{\kappa}{3} \, \widehat{S}^3 \non 
\eeq
and the soft--SUSY breaking potential has additional terms besides those of the
MSSM 
\beq
-{\cal L}_\mathrm{Higgs}= -{\cal L}_\mathrm{Higgs}^{\rm MSSM}+  
m_{S}^2| S |^2 + \lambda A_\lambda H_2 H_1 S + \frac13 \kappa  A_\kappa S^3 
\non
\eeq
An effective $\mu$ value is then generated when the additional field $S$ 
acquires a vev,  $\mu_{\rm eff} = \lambda \langle S \rangle$. 



\end{document}